\documentclass[a4paper,fleqn,usenatbib,useAMS]{mnras}
\usepackage[T1]{fontenc}
\usepackage{ae,aecompl}

\usepackage{graphicx}	% Including figure files
\usepackage{amsmath}	% Advanced maths commands
\usepackage{amssymb}	% Extra maths symbols
\usepackage{bm}		% Bold maths symbols, including upright Greek
\usepackage{pdflscape}	% Landscape pages

\usepackage{ulem}
\usepackage{times}
\usepackage{txfonts}
\usepackage{color}
\newcommand{\fracb}[2]{\left(\frac{#1}{#2}\right)}

\newcommand{\blue}[1]{\textcolor{blue}{#1}}

\usepackage[dvipsnames]{xcolor}
\definecolor{blazeorange}{rgb}{1.0, 0.4, 0.0}
\definecolor{seagreen}{rgb}{0.18, 0.55, 0.34}
\definecolor{rufous}{rgb}{0.66, 0.11, 0.03}
\definecolor{royalfuchsia}{rgb}{0.79, 0.17, 0.57}
\definecolor{scarlet}{rgb}{1.0, 0.13, 0.0}
\definecolor{royalpurple}{rgb}{0.47, 0.32, 0.66}

\title[Temporal Evolution of Prompt GRB Polarization]
{Temporal Evolution of Prompt GRB Polarization}

\author[Gill \& Granot]{
Ramandeep Gill$^{1,2,3}$\thanks{E-mail: rsgill.rg@gmail.com},
Jonathan Granot$^{2,3,1}$\thanks{E-mail: granot@openu.ac.il}
\\
$^{1}$Department of Physics, The George Washington University, Washington, DC 20052, USA\\
$^{2}$Department of Natural Sciences, The Open University of Israel, P.O Box 808, Ra'anana 43537, Israel \\
$^{3}$Astrophysics Research Center of the Open university (ARCO), The Open University of Israel, P.O Box 808, Ra'anana 43537, Israel
}

\date{Accepted XXX. Received YYY; in original form ZZZ}

\pubyear{2020}

\begin{document}
\label{firstpage}
\pagerange{\pageref{firstpage}--\pageref{lastpage}}
\maketitle

\begin{abstract}
The dominant radiation mechanism 
that produces the prompt emission in gamma-ray bursts (GRBs) remains a major open question. Spectral information alone has proven insufficient in elucidating its nature. Time-resolved linear polarization has 
the potential to distinguish between popular emission mechanisms, e.g., synchrotron radiation from electrons with a
power-law energy distribution or inverse Compton scattering of soft seed thermal photons, which can yield the 
typical GRB spectrum but produce different levels of polarization. Furthermore, it can be 
used to learn about the outflow's composition (i.e. whether it is kinetic-energy-dominated or Poynting-flux-dominated) and angular structure. 
For synchrotron emission it is a powerful probe of 
the magnetic field geometry. Here we consider synchrotron emission from a thin ultrarelativistic outflow, 
with bulk Lorentz factor $\Gamma(R)=\Gamma_0(R/R_0)^{-m/2}\gg1$, that radiates a Band-function spectrum in a 
single (multiple) pulse(s) over a range of radii, $R_0\leq R\leq R_0+\Delta R$. Pulse profiles and polarization 
evolution at a given energy are presented for a coasting ($m=0$) and accelerating ($m=-2/3$) thin spherical 
shell and for different viewing angles for a top-hat jet with sharp as well as smooth edges in emissivity. Four different magnetic field 
configurations are considered, such as a locally ordered field coherent over angular scales $\theta_B\gtrsim1/\Gamma$, 
a tangled field ($B_\perp$) in the plane transverse to the radial direction, an ordered field ($B_\parallel$) 
aligned in the radial direction, and a globally ordered toroidal field ($B_{\rm tor}$). All field 
configurations produce distinct polarization evolution with single (for $B_\perp$ and $B_\parallel$) and 
double (for $B_{\rm tor}$) $90^\circ$ changes in the polarization position angle. 
\end{abstract}


%%%%%%%%%%%%%%%%%%%%%%%%%%%%%%%%%%%%
%========================================
\section{Introduction}
%========================================

% Main Goals:
%  \begin{enumerate}
%         \item Include temporal variation of the emissivity and spectrum in a more realistic pulse model that assumes a particular dissipation 
%         mechanism, e.g. internal shocks or magnetic reconnection. The comoving spectral luminosity can be parameterized to vary with radius
%         \begin{equation}
%             L_{\nu'}'(R) = L'_0\fracb{R}{R_0}^a S\fracb{\nu'}{\nu_p'},\quad\quad \nu_p'(R) = \nu_0'\fracb{R}{R_0}^d
%         \end{equation}
%         where for simplicity we can assume a band-function spectrum in $S(\nu')$ for which the peak of the spectrum would vary with $R$.
%         \item Show spectro-polarimetric correlation in a given frequency band, e.g. how polarization $\Pi(E,t)$ would change as the peak 
%         of the spectrum crosses the POLAR-II energy window.
%         \item Include integration over multiple pulses to show how $\Pi(E,t)$ will change for a given emission model when dim to moderately 
%         bright bursts are detected that will not allow time-resolved polarization analysis and model comparison.
%         \item Consider synchrotron emission with different B-field structures, namely $B_\perp$ (random field in the plane transverse to 
%         the shock normal), $B_\parallel$ (ordered field in the radial direction), and $B_{\rm tor}$ (a globally-ordered toroidal field).
       
%         \item Consider both a top-hat jet and a structured jet model, e.g. a power law jet.
%     \end{enumerate}

The non-thermal prompt gamma-ray burst (GRB) spectrum is generally described by the empirical Band function \citep{Band+93} 
that features a smoothly broken power law. In rare cases deviations from the Band spectrum have also been found where the 
spectrum, e.g., appears to be quasi-thermal \citep[e.g.,][]{Ryde-04,Ryde-05}, features a thermal component in addition to 
the non-thermal Band component \citep[e.g.,][]{Guiriec+11,Guiriec+17}, shows a low-energy spectral break 
\citep{Oganesyan+17,Ravasio+18,Ravasio+19}, and a high-energy break usually interpreted as due to $\gamma\gamma$-annihilation 
\citep[e.g.,][]{Ackermann+13,Tang+15,Vianello+18}. The $\nu F_\nu$ spectrum peaks at a mean energy of $\langle E_{\rm pk}\rangle\simeq 250\,$keV, 
around which most of the energy in the burst comes out. Theoretical efforts focused 
on understanding the origin of this spectrum over the last few decades \citep[see reviews by][]{Piran-04,Kumar-Zhang-15} have narrowed it down to 
two popular radiative mechanisms -- synchrotron emission from power law electrons \citep{Sari-Piran-97,Daigne-Mochkovitch-98} and inverse-Compton scattering by 
mildly relativistic electrons on soft seed thermal photons operating in both optically thick and thin emission regions 
\citep{Thompson-94,Ghisellini-Celotti-99,Giannios-06,Thompson-Gill-14,Gill-Thompson-14,Vurm-Beloborodov-16}. 

Both mechanisms are able to explain 
the typical prompt GRB spectrum, although there are exceptional cases where some spectral features are more naturally explained by one or the other. 
For example, the canonical optically-thin synchrotron emission model is disfavoured in bursts that show low-energy spectral slopes harder than the 
so-called synchrotron line-of-death, with $d\ln F_{\nu}/d\ln\nu>1/3$ \citep[e.g.,][]{Crider+97,Preece+98,Ghirlanda+03}. Harder low-energy spectral 
slopes have been accommodated using this model if the magnetic field gradually declines in the emission region as the flow expands \citep{Uhm-Zhang-14,Geng+18}. 
Alternatively, these can be explained with sub-photospheric dissipation models that feature a quasi-thermal spectral peak broadened by Comptonization 
\citep[e.g.][]{Vurm-Beloborodov-16,Beloborodov-Meszaros-17}. Sub-photospheric dissipation models sometimes also produce double-hump spectra 
\citep{Guiriec+11,Guiriec+17} where the quasi-thermal component produces the $\nu F_\nu$-peak and the non-thermal component dominates the spectrum at 
energies above and below $E_{\rm pk}$; in some very rare cases more than two spectral components have also been shown to fit the prompt GRB spectrum 
\citep{Guiriec+15a}. The double-hump spectrum has been shown \citep{Gill+20} to arise in a sub-photospheric dissipation model  by power-law electrons 
emitting synchrotron photons \citep{Beniamini-Giannios-17} or gradually heated mono-energetic electrons cooling on soft seed thermal photons via 
Comptonization \citep[e.g.,][]{Giannios-08}. This degeneracy between the two emission mechanisms has lead 
to a deadlock hindering further progress on understanding, e.g., the composition -- kinetic energy dominated or Poynting flux dominated -- of 
the ultrarelativistic outflow as well as the particle acceleration/heating mechanisms since both of these ultimately dictate the dominant 
radiation process.
%\jg{there is no clear one-to-one correspondence between the dominant radiation mechanism and the outflow composition -- these are two important open questions with some links but solving one doesn't automatically solve the other}
%\rg{Agreed, however, ruling out one of the radiation mechanisms will undoubtedly allow one to constrain models of particle heating, as shown in our recent work on gradual dissipation, and in many cases it may favor one kind of flow composition over the other. So, while it may not remove the degeneracy altogether, it will lead to stricter constraints on the composition.}

One promising way to break this degeneracy is the measurement of linear polarization ($\Pi$) which would strongly favour synchrotron emission if 
$\Pi\gtrsim20\%$ is measured in most GRBs \citep{Gill+20}. In general, negligible polarization is expected from a Comptonized spectrum since multiple 
scatterings tend to wash out any preferred direction of the polarization vector. However, if an axisymmetric flow has an angular structure, 
in particular with a steep gradient in the bulk Lorentz factor $\Gamma(\theta)$ with polar angle $\theta$ measured from the jet symmetry axis, 
Comptonized emission can yield $\Pi\lesssim20\%$ \citep[e.g.,][]{Ito+14,Lundman+14,Parsotan+20}. Therefore, apart from helping in determining the dominant 
radiation mechanism for the prompt emission, linear polarization is also a valuable tool for understanding the jet's angular structure. 
What is more, in the case of synchrotron emission, 
since different magnetic field configurations produce different levels of polarization for a given jet geometry, observer's line-of-sight (LOS), and 
spectral index, linear polarization measurements can potentially constrain the magnetic field geometry in the emission region (e.g. as was done for 
the afterglow phase in \citealt{Gill-Granot-20}). For example, even a single 
statistically significant measurement of polarization at the level of 50\%$\,\lesssim\Pi\lesssim\,$60\% will definitively indicate the presence 
of an ordered magnetic field, such as a globally ordered toroidal field \citep{Gill+20}. 

Thus far, linear polarization measurements of prompt GRB emission (see, e.g., Table 1 of \citealt{Gill+20}) have not been able to settle this issue. 
Even though several such measurements exist, they are of low statistical significance ($\lesssim3\sigma$) and the results are inconsistent when compared 
between different detectors. Measurements by IKAROS-GAP \citep{Yonetoku+11,Yonetoku+12} and AstroSat-CZTI 
\citep{,Chand+18,Chattopadhyay+19,Chand+19,Sharma+19} tentatively find high levels of polarization with $\Pi\gtrsim50\%$. However, time-integrated 
measurements obtained by POLAR \citep{Zhang+19,Burgess+19,Kole+20} yield low-levels of polarization with $\Pi\lesssim20\%$ instead, with 
most GRBs consistent with being unpolarized within the 2$\sigma$ confidence level. On the other hand, their time-resolved analysis did find moderate 
polarization with a time-varying polarization position angle (PA) for GRB 170114A. 
% \jg{any comment on the significance of this measurement?}
% \rg{I'll have to check with J. Michael for this as it's not clear from their paper.}
% ֿ\jg{J. Michael is a Bayesian priest and therefore he tries to avoid almost at all cost quoting the significance of a detection.}
If this is indeed the case, then a time-integrated analysis would indeed yield 
very low polarization since it is being averaged out by the variation of the PA. This calls for a more careful time-resolved polarization analysis of 
bright GRBs and comparison of the measured polarization with time-resolved theoretical estimates.

Detailed treatments of energy-independent linear polarization from synchrotron emission in ultrarelativistic GRB jets have appeared in several works 
\citep{Ghisellini-Lazzati-99,Gruzinov-99,Sari-99,Granot-Konigl-03,Rossi+04,Granot-Taylor-05}. 
Pulse-integrated prompt GRB polarization from synchrotron emission has been calculated for a locally ordered magnetic field in a thin spherical shell 
as well as for different field configurations, such as a tangled field in the plane transverse to the radial direction ($B_\perp$) and an ordered 
field aligned with the radial direction ($B_\parallel$), or ordered in the tangential direction ($B_{\rm ord}$), in a top-hat jet in \citet{Granot-03}. 
The same for a globally ordered toroidal field in a  spherical shell were presented in \citet{Lyutikov+03}, where in the context of afterglow polarization 
it was also calculated for a structured power-law jet \citep{Lazzati+04a} and for a top-hat jet \citep{Granot-Taylor-05}. Key results for pulse-integrated 
polarization from synchrotron emission and from different magnetic field configurations in a top-hat jet and their statistical incidence in a large sample 
of GRBs are summarized in \citet{Toma+09} and \citet{Gill+20}. The latter also discussed polarization from structured jets and alternative radiation 
mechanisms to synchrotron, such as non-dissipative photospheric emission \citep{Beloborodov-11,Lundman+14} and Compton drag \citep{Lazzati+04}. 

While most works have presented pulse-integrated and energy-independent polarization results, the need for time-resolved and energy-resolved polarization 
models has become important only recently. Time-resolved polarization was reported for GRB 170114A \citep{Burgess+19}, a bright single pulsed GRB jointly 
observed by POLAR and \textit{Fermi}-GBM, with increasing level of polarization towards the pulse peak that reached $\Pi\sim30\%$ at the peak. This 
was accompanied by a time-varying PA. Time-resolved treatment of polarization of synchrotron emission from an ordered toroidal field 
in a coasting top-hat jet is presented in \citet{Cheng+20}, where the magnetic field is assumed to decay with radius as a power law in a prescribed way. 
The power-law electron distribution, as it cools due to adiabatic, synchrotron, and synchrotron self-Compton cooling, is obtained by numerically solving 
the continuity equation in energy space. \citet{Lan+20} consider a wide variety of magnetic field configurations involving both ordered and random field 
components and for which they present time-integrated but energy-resolved polarization in a coasting top-hat jet.

In this work, we address the question of time-resolved prompt GRB polarization of synchrotron emission arising from physically motivated magnetic field 
configurations and outflow dynamics as well as angular structure. 
First, in \S\ref{sec:pulse-spectral-model}, we  calculate the polarization evolution over a single pulse using a general framework of an ultrarelativistic thin radiating shell. The comoving spectrum is described by the Band function whose $\nu F_\nu$ spectral peak energy and normalization each 
evolve as a power law with radius. 

In \S\ref{sec:scalings} synchrotron emission is assumed to be the dominant radiation mechanism for which we calculate the physically motivated scalings 
needed to describe the radial evolution of the spectral emissivity. Two different types of outflows are considered: (i) a kinetic-energy-dominated 
(KED; \S\ref{sec:KED}) coasting (constant bulk $\Gamma$) flow  in which internal shocks efficiently dissipate the baryonic kinetic energy, and (ii) a 
Poynting-flux-dominated (PFD; \S\ref{sec:PFD}) accelerating flow in which magnetic reconnection,  MHD instabilities or multiple weak shocks in a variable 
outflow tap the magnetic field energy.

In \S\ref{sec:formalism} we present the formalism to calculate the time-resolved polarization over a single pulse and integrated over the observed image
of the outflow on the plane of the sky. We consider four different magnetic field configurations, namely a locally ordered field ($B_{\rm ord}$) with 
angular coherence length 
on the order of the angular size of the beaming cone ($\theta_B\gtrsim1/\Gamma$), a tangled field ($B_\perp$) that lies entirely in the plane transverse to 
the radial direction, an ordered field aligned with the radial direction ($B_\parallel$) at each point of the outflow, and a globally ordered toroidal field 
($B_{\rm tor}$) that is axisymmetric around the jet symmetry axis. Results for a thin spherical shell are presented first in 
\S\ref{sec:Orderd-B-field-Spherical-Shell} both for a KED and PFD flow. This case is relevant when the beaming cone doesn't include the edge of the jet in 
which case the emission appears as if it's coming from a spherical flow. Time-resolved polarization curves for different viewing angles from a 
top-hat jet (\S\ref{sec:top-hat-jet-pol}) and a smooth top-hat jet (\S\ref{sec:smooth-top-hat-jet-pol}), with a uniform core 
and with either exponential or power-law wings, are presented next.

For GRBs that are not exceptionally bright and lack clear single pulses an integration over multiple pulses is generally performed that can yield polarization 
results different from those expected for a single isolated pulse. We address this point in \S\ref{sec:multi-pulse} and present results for integration over 
multiple pulses. Finally, we summarize this work (\S\ref{sec:Summary-Discussion}) and discuss time variation of the PA (\S\ref{sec:Time-Dependent-PA}) and 
energy dependence of polarization (\S\ref{sec:Energy-Dependent-Pol}) at the end.

\begin{table}
    \centering
    \begin{tabular}{|c|l|}
        Symbol & Definition \\
        \hline
        $R_0$ & Radius at which emission turns on \\
        $\Delta R$ & Radial distance over which shell emits continuously \\
        $R_f$ & Radius at which emission turns off: $R_f = R_0 + \Delta R$ \\
        $\hat R$ & Normalized radius: $\hat R = R/R_0$ \\
        $\Gamma_0$ & Bulk-$\Gamma$ of emission region at $R_0$ \\
        $t_0$ & Arrival time of first photons emitted along the LOS at $R_0$ \\
        $\nu_0$ & Observed $\nu F_\nu$ peak frequency of first photons emitted \\
        & from $R_0$ along the LOS that arrived at time $t_0$\\
        $m$ & Bulk Lorentz factor PL index: $\Gamma^2\propto R^{-m}$ \\
        $\ell$ & Lab-frame B-field PL index: $B\propto R^\ell$ \\
        $s$ & Minimum particle Lorentz factor PL index: $\gamma_m\propto R^s$ \\
        $a$ & Spectral luminosity PL index: $L_{\nu'}'\propto R^a$ \\
        $d$ & Peak frequency PL index: $\nu_{\rm pk}'\propto R^d$ \\
        % $x_0$ & Normalized observation frequency: $x_0 = \nu/\nu_0$ \\
        $b_1$, $b_2$ & Asymptotic Band-function spectral indices: \\
        & $d\ln F_\nu/d\ln\nu$ \\
        $\tilde t$ & Normalized apparent time: $\tilde t = t/t_0$ \\
        $\tilde t_{\rm pk}$ & Normalized pulse peak time: $\tilde t_{\rm pk}=t_{\rm pk}/t_0$ \\
        $\tilde t_{\rm cross}$ & Normalized crossing time of the Band-function break \\
        & frequency $x_b = (b_1-b_2)/(1+b_1)$ across $x_0 = \nu/\nu_0$ \\
        $\tilde\theta$ & Polar angle measured from the LOS \\
        $\tilde\xi$, $\tilde\xi_0$, $\xi_j$, $\xi_{0,j}$ & $(\Gamma\tilde\theta)^2$, $(\Gamma_0\tilde\theta)^2$, 
        $(\Gamma\theta_j)^2$, $(\Gamma_0\theta_j)^2$ \\
        $q$ & $\theta_{\rm obs}/\theta_j$ \\
        $\Delta$ & Smoothing parameter for a top-hat jet with exponential \\ 
        & wings in $L_{\nu'}'$ \\
        $\delta$ & Smoothing parameter for a top-hat jet with power-law \\ 
        & wings in $L_{\nu'}'$ \\
        on-beam & When emission is received from within the $1/\Gamma$ beaming \\ 
        & cone of the emitting material \\
        off-beam & When emission is received from outside of the $1/\Gamma$ \\
        & beaming cone of the emitting material
    \end{tabular}
    \caption{Various symbols or terms and their definitions}
    \label{tab:symbols}
\end{table}

%%%%%%%%%%%%%%%%%%%%%%%%%%%%%%%%%%%
\section{Pulse and Spectral Model}\label{sec:pulse-spectral-model}
%%%%%%%%%%%%%%%%%%%%%%%%%%%%%%%%%%%

We consider an ultrarelativistic ($\Gamma\gg1$) thin shell with lab-frame width much smaller than the causal size at that radius, such that 
$\Delta \ll R/\Gamma^2$. This assumption is valid when the bulk of the prompt emission arises from a very thin layer, which is guaranteed when 
the (comoving) cooling time of electrons is much shorter than the dynamical time, $t_{\rm cool}'\ll t_{\rm dyn}'=R/\Gamma c$. To study the 
temporal evolution of the polarization we follow the treatment of \citet{Genet-Granot-09} (also see, e.g., \citealt{Uhm-Zhang-15,Uhm-Zhang-16}) 
and construct a simple pulse model in which the thin shell 
starts to radiate at $R=R_0$ and continues to radiate until $R=R_f=R_0+\Delta R$, where the emission is assumed to be switched off abruptly. The 
analysis can be easily extended to more complex pulse profiles where the emission, e.g., switches off gradually 
\citep[see Appendix C of][or \citealt{Beniamini-Granot-16}]{Genet-Granot-09}, which would make the pulse peaks rounder and less spiky, but it will not alter the main results in a 
significant way. The radial evolution of the bulk Lorentz factor (LF) of the shell can be written generally as $\Gamma^2(R)=\Gamma_0^2(R/R_0)^{-m}$, with 
$\Gamma_0=\Gamma(R_0)$, which allows for a coasting flow ($m=0$), an accelerating flow ($m<0$), and a decelerating flow ($m>0$).

As the flow expands, the observed flux density, $F_\nu\propto t^{-\alpha}\nu^{-\beta}$, changes with the apparent time $t$ due in part to the radially 
evolving comoving (all comoving quantities henceforth are marked with a prime) isotropic-equivalent spectral luminosity and peak energy that scale as 
a power-law with radius,
%\jg{This is valid for a spherical shell. Since here we also consider axi-symmetric jets, there should also be some normalized angular function, $f(\theta)$ where $f(0)=1$, so that this equation gives the local isotropic equivalent $L'_{\nu'}(R,\theta)$.}
\begin{equation}\label{eq:L-and-nupk-scaling}
    L'_{\nu'}(R,\theta) = L_0'\fracb{R}{R_0}^a  S\fracb{\nu'}{\nu'_{\rm pk}}\,f(\theta)\quad{\rm with}\quad \nu'_{\rm pk} = \nu_0'\fracb{R}{R_0}^d\,,
\end{equation}
where $L_0' = L'_{\nu_{\rm pk}'}(R_0)$ and $\nu_0' = \nu_{\rm pk}'(R_0)$ are normalizations of the spectral luminosity and peak frequency at $R=R_0$. 
The factor $f(\theta)$ represents the angular structure of the emissivity normalized to unity at $\theta=0$ with $f(0)=1$, where $\theta$ is the polar 
angle measured from the jet symmetry axis. This means that $f(\theta)=1$ for a spherical flow and $f(\theta) = \mathcal{H}(\theta_0-\theta)$ for a top-hat 
jet with jet half-opening angle $\theta_0$ and $\mathcal{H}(x)$ being the Heaviside function. Here we make the explicit assumption that the emission 
is isotropic in the comoving frame as well as uniform over the entire shell, and thus depends only on $R$. We define the normalized frequency, in terms 
of the peak frequency, as
\begin{equation}
    x \equiv \frac{\nu'}{\nu_{\rm pk}'} = \frac{\nu'}{\nu_0'}\fracb{R}{R_0}^{-d} 
    = \frac{\delta_D^{-1}}{(2\Gamma_0)^{-1}}\frac{\nu}{\nu_0}\fracb{R}{R_0}^{-d} 
    = \frac{2\Gamma_0}{\delta_D}x_0\fracb{R}{R_0}^{-d}\,.
\end{equation}
Here we fix $\nu_0'$ using the peak frequency of the first photons emitted along the observer's LOS from radius $R_0$ and that were received at time 
$t=t_0$, so that $\nu_0\equiv\nu_{\rm pk}(t_0)=\nu/x_0$ and $\nu_0'=(1+z)\nu_0/2\Gamma_0$, where we made use of the Lorentz transformation, $\nu=\delta_D\nu'/(1+z)$ 
where $\delta_D$ is the Doppler factor (see \S\ref{sec:formalism}), and the fact that $\delta_D=2\Gamma(R)=2\Gamma_0$ for emission along the LOS. 
The comoving spectrum, $S(x)$, is described by the Band-function
\begin{equation}
    S(x) = e^{1+b_1} \left\{
    \begin{tabular}{c|c}
        $x^{b_1} e^{-(1+b_1)x}$\,, & $x\leq x_b$ \\
        $x^{b_2}x_b^{b_1-b_2}e^{-(b_1-b_2)}$\,, & $x\geq x_b$
    \end{tabular}\right.\,,
\end{equation}
where the break energy $x_b = (b_1-b_2)/(1+b_1)>1$ when $b_2<-1$. The local spectral index is given by $\beta\equiv-d\ln S(x)/d\ln x= x(1+b_1)-b_1$ 
for $x\leq x_b$ and $\beta=-b_2$ for $x>x_b$.
From 
the definition of $x$, the peak of the $\nu F_\nu$ spectrum occurs at $x=1$ for which $S(x=1)=1$, and therefore $xS(x=1)=1$ at the spectral peak.

The various symbols, along with their definitions, that appear in most expressions presented in this work are collected in Table~\ref{tab:symbols} 
for convenience.

%%%%%%%%%%%%%%%%%%%%%%%%%%%%%%%%
\section{Scalings for Synchrotron Emission and Outflow Dynamics}\label{sec:scalings}
%%%%%%%%%%%%%%%%%%%%%%%%%%%%%%%%

The dynamics of the relativistic outflow, in particular, the radial dependence of bulk $\Gamma$ can vary between different models that 
prescribe different compositions. Here we do not consider pair enrichment and instead assume a proton-electron plasma where the outflow 
composition describes the division of energy between the particles and the electromagnetic field. The composition can be characterized using 
the magnetization\footnote{More generally, the specific enthalpy $h=w'/\rho' c^2$ appears in the denominator, but here we assume a 
cold plasma where $h\cong1$.} $\sigma = B'^2/4\pi n'm_pc^2$, which is the ratio of the (proper) enthalpy density of the comoving magnetic 
field ($B'^2/4\pi$), with strength $B'=B/\Gamma\propto R^{\ell+m/2}$ where $B\propto R^\ell$ is the lab-frame magnetic field, to that of 
the cold baryons ($w'\cong n'm_pc^2$) with comoving number density $n'$. Here $m_p$ is the proton mass and $c$ is the speed of light. 
The comoving number density of electrons in the ejecta shell scales 
as\footnote{More generally, $V'\propto R^2\Delta'$ where $\Delta'=\Gamma\Delta$ and $\Delta\approx\max(\Delta_0,R/\Gamma^2)$, so the 
expression here holds only before the spreading radius, $R<R_\Delta\approx\Delta_0\Gamma^2(R_\Delta)$, where $\Delta_0\approx cT_{\rm GRB}/(1+z)$ 
is the radial width of the outflow and $T_{\rm GRB}$ is the total GRB duration. The dissipation radius is typically 
$R_{\rm dis}\lesssim \Gamma^2(R_{\rm dis})ct_v/(1+z)$ where $t_v<T_{\rm GRB}$ is the variability time, so that 
$[R_{\rm dis}/\Gamma^2(R_{\rm dis})]/[R_\Delta/\Gamma^2(R_\Delta)]\lesssim t_v/T_{\rm GRB}\ll1$ and indeed $R_{\rm dis}\ll R_{s}$ since 
$R/\Gamma^2$ typically increases with $R$ (i.e. $m>-1$).} $n'\propto V'^{-1}\propto (R^2\Gamma)^{-1}\propto R^{(m-4)/2}$. Therefore, 
the magnetization of the outflow scales as $\sigma\propto R^{(4+4\ell+m)/2}$. When $\sigma\ll1$, most of the energy resides in the kinetic 
energy of the baryons, part of which is dissipated in internal shocks and then part of that is radiated. Alternatively, when $\sigma>1$, 
the outflow is Poynting flux dominated and the main energy reservoir is the magnetic field. The dynamics of the flow are different in both scenarios. 

Here we assume that synchrotron emission, arising from power-law electrons gyrating in the shock-generated magnetic field or that advected from the base 
of the flow, forms the dominant radiation component that produces the Band-like spectra of GRBs. In order to ensure a high radiative efficiency, 
the electrons must be in the fast-cooling regime \citep{Sari+98}, so that $\nu_c'<\nu_m'$ where 
%$\nu_c' = 9\pi em_ec/\sigma_T^2B'^3t_{\rm dyn}'^2$ 
$\nu_c' \propto (B'^3t_{\rm dyn}'^2)^{-1}\propto R^{-(5m+6\ell+4)/2}$ is the characteristic cooling break frequency, 
with $t_{\rm dyn}'\sim R/\Gamma c\propto R^{(m+2)/2}$ 
being the comoving dynamical time, and 
%$\nu_m' = eB'\gamma_m^2/2\pi m_ec^2$ 
$\nu_m' \propto B'\gamma_m^2\propto R^{(2\ell+m+4s)/2}$ is the synchrotron radiation frequency of minimal energy electrons with LF $\gamma_m\propto R^s$. 
The $\nu F_\nu$ synchrotron spectrum peaks at $\nu_{\rm pk}\propto\Gamma\nu_m'\propto R^{\ell+2s}$ in the central engine frame, making it the characteristic 
frequency at which most of the energy of the burst comes out. The number of electrons emitting synchrotron photons at the peak luminosity and occupying 
the causal volume $\tilde V' \propto R^3/\Gamma\propto R^{(6+m)/2}$ are $N_e=n'\tilde V'\propto R^{1+m}$. 
The comoving spectral luminosity at the peak frequency is obtained from $L_{\nu_{\rm pk}'}' = L_{\nu_{\rm max}'}'(\nu_m'/\nu_c')^{-1/2}\propto R^{-(\ell+s)}$, 
where the maximum spectral luminosity can be approximated by using the synchrotron power emitted by the electron, $P_{\rm syn}'$, at the characteristic 
synchrotron frequency, $L_{\nu_{\max}'}' \sim N_eP'_{\rm syn}/\nu \propto N_eB'\propto R^{(2+2\ell+3m)/2}$. From the above discussion we find the power-law 
indexes of the spectral luminosity and peak energy in Eq.~(\ref{eq:L-and-nupk-scaling}) to be
\begin{equation}
    a=-(\ell+s)\quad\quad{\rm and}\quad\quad d=\ell+2s+m/2\,.
\end{equation}

%%%%%%%%%%%%%%%%%%%%%%%%%%%%%%%%%%%%%%%%%%%%%%%%%%%%%%%%%%%%
\subsection{Kinetic Energy Dominated Flow: Internal Shocks}\label{sec:KED}
%%%%%%%%%%%%%%%%%%%%%%%%%%%%%%%%%%%%%%%%%%%%%%%%%%%%%%%%%%%%
The temporal variability of GRB lightcurves, with variability timescale $t_v\sim10^{-3}-1\,$s \citep{Fishman-Meegan-95}, can be understood 
using the internal shocks model \citep{Rees-Meszaros-94,Paczynski-Xu-94,Sari-Piran-97,Daigne-Mochkovitch-98}, which posits that the observed 
variability reflects that of the central engine. Variability in long-soft GRBs can also be embedded in the flow as the jet traverses the stellar 
envelope due to pressure confinement, mixing, and shocks \citep[e.g.][]{Matzner-03}. A similar situation can arise in the prompt emission of short-hard 
GRBs where the jet breaks out of the circum-merger ejecta. Here we consider the canonical scenario where the central engine accretes intermittently and ejects 
shells of matter that are initially separated by length scale $\sim ct_v/(1+z)$ and have fluctuations in bulk LFs of order 
$\Delta\Gamma\sim\Gamma$, the mean bulk LF of the unsteady flow. As a result, after the outflow acceleration saturates with 
$\Gamma(R)=\Gamma_\infty\propto R^0$, and therefore $m=0$, faster moving shells catch up from behind with slower ones and collide with each other to dissipate 
their kinetic energy at internal shocks that occur at the dissipation radius 
$R_{\rm dis}=2\Gamma_\infty^2ct_v/(1+z) = 6\times10^{13}(1+z)^{-1}\Gamma_{\infty,2}^2t_{v,-1}\,$cm for a source at redshift $z$. 

In each collision between two adjecent shells a double shock structure forms, with forward and reverse shocks going into the slower and faster shells, 
respectively, which shock heat a fraction $\xi_e$ of the electrons into a power-law energy distribution, $dN_e/d\gamma_e\propto\gamma_e^{-p}$ for 
$\gamma_e>\gamma_m$, which holds a fraction $\epsilon_e$ of the total internal energy density behind the shock. Here 
$\gamma_m = [(p-2)/(p-1)](\epsilon_e/\xi_e)(m_p/m_e)(\Gamma_{\rm ud}-1)$ is the LF of minimal energy electrons, 
where $m_p$ and $m_e$ are the proton and electron masses. The strength of the two shocks is characterized by the relative upstream to downstream LF \footnote{$\Gamma_{\rm ud} \cong \frac{1}{2}(\Gamma_u/\Gamma_d + \Gamma_d/\Gamma_u)$ is the LF of the upstream (unshocked) material 
moving with LF $\Gamma_u\gg1$ as measured in the frame of the downstream (shocked) material moving with LF $\Gamma_d\gg1$.}, 
$\Gamma_{\rm ud}$, which is expected to be roughly 
constant with radius for the simplest case of two uniform shells, as considered here. In this case, the LF of minimal energy electrons, 
$\gamma_m\propto R^0$, remains independent of radius, and therefore $s=0$. 

In a coasting flow, the radial size of a given fluid element remains 
constant, but its transverse size increases with radius. Consequently, magnetic flux conservation yields the scaling $B_r\propto R^{-2}$ for the radial 
component of the magnetic field and $B_{\theta,\phi}\propto R^{-1}$ for the transverse components. Energy is dissipated at a radial distance much larger 
than that where the flow bulk-$\Gamma$ saturates, such that $R_{\rm dis}/R_s = 2\Gamma_\infty ct_v/R_\ell\gg1$, where $R_s = \Gamma_\infty R_\ell$ is the 
saturation radius for a fireball expanding under its own pressure and $R_\ell$ is the jet launching radius. A similar situation can arise even in initially 
Poynting-flux-dominated highly variable flows \citep{Granot+11,Granot-12,Komissarov12} for which the magnetization declines with radius, 
$\sigma(R)\propto R^{-1/3}$. Once the flow transitions to being weakly magnetized, with $\sigma<1$ at $R>R_s\sim\Gamma_\infty^2 ct_v\sim R_{\rm dis}$ 
where the saturation radius in this scenario is defined below, internal shocks again become efficient at dissipating energy.
%\jg{This assumes thermal acceleration by the radiation pressure, i.e. the fireball model, but internal shocks can also arise in other circumstances, such as in a variable flow that is initially Poynting-flux dominated, in which case this expression for $R_s$ no longer holds. It is improtant to stress that $\sigma=\sigma(R)$ can naturally evolve with radius and an initially Poynting-flux dominated outflow can eventually become kinetic-energy dominated.} %\rg{Done.}
In both scenarios, the transverse component of the magnetic field dominates, which yields $B\propto R^{-1}$ and $\ell=-1$.

For the internal shock model in a KED flow, we find that $a=1$ and $d=-1$.

%%%%%%%%%%%%%%%%%%%%%%%%%%%%%%%%%%%%%%%%%%%%%%%%%%%%%%%%%%%%%%%%%%
\subsection{Poynting Flux Dominated Flow: Magnetic Reconnection}\label{sec:PFD}
%%%%%%%%%%%%%%%%%%%%%%%%%%%%%%%%%%%%%%%%%%%%%%%%%%%%%%%%%%%%%%%%%%
An attractive alternative to internal shocks is the possibility that the relativistic outflow is permeated by strong magnetic fields advected from 
the base of the flow at $R=R_\ell$ \citep[e.g.,][]{Thompson-94,Lyutikov-Blandford-03}. Since $\sigma>1$ in this scenario, internal shocks are rendered 
inefficient in dissipating any kinetic energy of the flow. Instead, the main energy reservoir is the magnetic field, which is dissipated due to magnetic 
reconnection and/or MHD instabilities, e.g. the Kruskal-Schwarzchild instability \citep{Lyubarsky-10,Gill+18} which is the magnetic analog of the 
Rayleigh-Taylor instability. A popular model of a PFD outflow is that of a striped-wind 
\citep{Lyubarsky-Kirk-01,Spruit+01,Drenkhahn-02,Drenkhahn-Spruit-02,Begue+17} where the magnetic field lines 
reverse polarity over a characteristic length scale $\lambda\sim\pi R_L = \pi c/\Omega=cP/2=1.5\times10^7P_{-3}\,$cm. Here $R_L$ is the light cylinder radius, $\Omega=2\pi/P$ is the central engine's 
rotational angular frequency, and $P=10^{-3}P_{-3}\,$s is its spin period. While this parameterization is relevant for a millisecond magnetar 
\citep[e.g.,][]{Metzger+11}, more generally, stochastic polarity flips in the outflow launched by a BH have $\lambda\gtrsim R_L$ 
\citep{Mckinney-Uzdensky-12,Parfrey+15}. Magnetic energy is dissipated as opposite polarity 
field lines are brought together at the inflow plasma velocity $v_{\rm in}=\epsilon v_A$, where $\epsilon\sim0.1$ and $v_A\simeq c$ is the Alfv\'{e}n 
speed which approaches the speed of light $c$ when $\sigma\gg1$, and undergo reconnection. A significant fraction of the dissipated energy goes 
towards accelerating the flow with $\Gamma\propto R^{1/3}~(m=-2/3)$ for $R_A<R<R_s$, where $R_A\sim{\rm few}\times R_L$ is the Alfv\'{e}n radius and 
$R_s = \Gamma_\infty^2\lambda/6\epsilon = 1.7\times10^{13}\Gamma_\infty^2(\lambda/\epsilon)_8\,$cm is the saturation radius. The latter defines the 
radial distance beyond which no magnetic dissipation occurs and the flow starts to coast at $\Gamma=\Gamma_\infty$. 

Broadly similar flow dynamics and energy dissipation is obtained in a highly variable magnetized outflow \citep{Granot+11,Granot-12,Komissarov12} that 
doesn't require magnetic field polarity reversals to dissipate energy at reconnection sites. Instead, energy is dissipated in multiple weak internal 
shocks that operate at $R\ll R_s$ when $\sigma\gg1$. These shocks gradually become more efficient as the magnetization declines and become strongest 
and most efficient when $\sigma\lesssim1$ for $R\gtrsim R_s$. In this case $\lambda/c$ is the variability time of the outflow emanating from the central source.

For an axisymmetric magnetic field, such as a globally toroidal field centered on the jet symmetry axis, the poloidal component declines faster, 
with $B_p\propto R^{-2}$, as compared to the toroidal component that scales as $B_\phi\propto R^{-1}$. Therefore, at large distances from the central 
engine, where dissipation occurs and detectable non-thermal emission is produced, the toroidal component dominates, which again yields $\ell=-1$.

To determine the scaling of $\gamma_m$ with radius, we consider the mean energy per baryon which cannot exceed $\sigma m_pc^2$ as this is the total 
dissipated energy per baryon-electron for complete magnetic dissipation. A fraction of the dissipated energy is deposited in the electrons and their 
mean energy per particle, $\langle\gamma_e\rangle$, scales with that of the protons. Since 
$\gamma_m\propto\langle\gamma_e\rangle\propto\sigma$, which yields $s=(4+4\ell+m)/2 = -1/3$.
This is strictly valid for $p>2$ for which most of the energy resides near $\gamma_m$. However, many works find that the power-law index $p=p(\sigma)$ 
depends sensitively on $\sigma$, where the dependence can be approximately expressed as $p=4\sigma^{-0.3}$ \citep{Sironi-Spitkovsky-14,Guo+15,Kagan+15,Werner+16}. 
As a result, $p<2$ for $\sigma>10$ in which case most of the energy in particles starts shifting towards larger particle LFs, $\gamma_e\gg\gamma_m$. 
What's relevant here is the value of $\sigma$ when the flow becomes optically thin so that radiation can stream out. \citet{Gill+20b} carried out numerical 
simulations of a PFD flow over a range of dissipation radii and in all cases of interest here $\sigma<10$ in the optically thin parts of the flow. Therefore, 
we find that the scaling of $\gamma_m$ is reasonably justified.

For the PFD flow with a striped wind magnetic field structure, we find that $a=4/3$ and $d=-2$.

%%%%%%%%%%%%%%%%%%%%%%%%%%%%%%%%%%%%%%%%%%%%%%%%%%%%%%%%%%%%%%%%%%%%%%
\section{Linear Polarization Over a Single Pulse}\label{sec:formalism}
%%%%%%%%%%%%%%%%%%%%%%%%%%%%%%%%%%%%%%%%%%%%%%%%%%%%%%%%%%%%%%%%%%%%%%

Synchrotron emission, in general, is partially linearly polarized. Its anisotropic emissivity and local (for a given point on the surface of the outflow) 
polarization depends on the direction of the magnetic field with respect to the observer's LOS in the comoving frame. The direction of the polarization 
vector, $\hat\Pi' = (\hat n\times\hat B)/|\hat n\times\hat B|$, is always transverse to both $\hat n$, the unit vector pointing along the LOS 
(the direction of a photon that reaches the observer), and $\hat B$, the unit 
vector pointing in the local direction of the magnetic field, regardless of the Lorentz frame. While the magnitude of 
local polarization remains Lorentz invariant, its direction in the observer frame is obtained with appropriate Lorentz transformations. Since synchrotron 
emission arising from different fluid elements is incoherent, the total polarization is obtained by taking a ratio of the total polarized intensity to the 
total intensity, where both are integrated over the entire image of the outflow as observed on the plane of the sky.

Here we first present a formalism valid for a spherical shell. We later generalize it to cover other outflow structures, namely a top-hat 
jet and a uniform jet with smooth edges in emissivity.

The flux density measured by a distant observer in the direction of the unit vector $\hat n$ from an infinitely `thin-shell' at an apparent 
time $t$ for a source at redshift $z$ with luminosity distance $d_L(z)$ is given by \citep{Granot-05}
\begin{equation}
    F_\nu(t) = \frac{(1+z)}{16\pi^2d_L^2}\int\delta_D^3L'_{\nu'}d\tilde\Omega\,,
\end{equation}
where $\delta_D=[\Gamma(1-\vec\beta\cdot\hat n)]^{-1}=[\Gamma(1-\beta\tilde\mu)]^{-1}$ is the Doppler factor, 
$\hat n\cdot\hat\beta\equiv\tilde\mu=\cos\tilde\theta$ with $\tilde{\theta}=\tilde{\theta}(\theta,\phi)$ being the polar angle measured from the 
LOS, and $d\tilde\Omega=d\tilde\varphi d\tilde\mu$ is the solid angle. For an 
ultra-relativistic flow $\Gamma\gg1$ for which $\delta_D\approx2\Gamma/(1+\tilde\xi)$ with $\tilde\xi\equiv(\Gamma\tilde\theta)^2$. The \textit{anisotropic} 
comoving spectral luminosity for synchrotron emission can be expressed as \citep[e.g.,][]{Rybicki-Lightman-79}
\begin{equation}
    L'_{\nu'} = L'_{\nu'}(R)(\sin\chi')^\epsilon = L'_{\nu'}(R)[1-(\hat n'\cdot \hat B')^2]^{\epsilon/2}\,,
\end{equation}
where $\chi'$ is the angle between the direction of local magnetic field and that of the emitted photon in the comoving frame. 
Since synchrotron radiation is highly beamed in the forward 
direction of the electron's motion, $\chi'$ is also the pitch angle between the electron's velocity vector and the magnetic field. When the power-law 
electrons' energy distribution is independent of their pitch angles, $\epsilon=1+\alpha$ where $\alpha=-d\log F_\nu/d\log\nu$ is the spectral index \citep{Laing-80,Granot-03}.

\begin{figure*}
    \centering
    \includegraphics[width=0.48\textwidth]{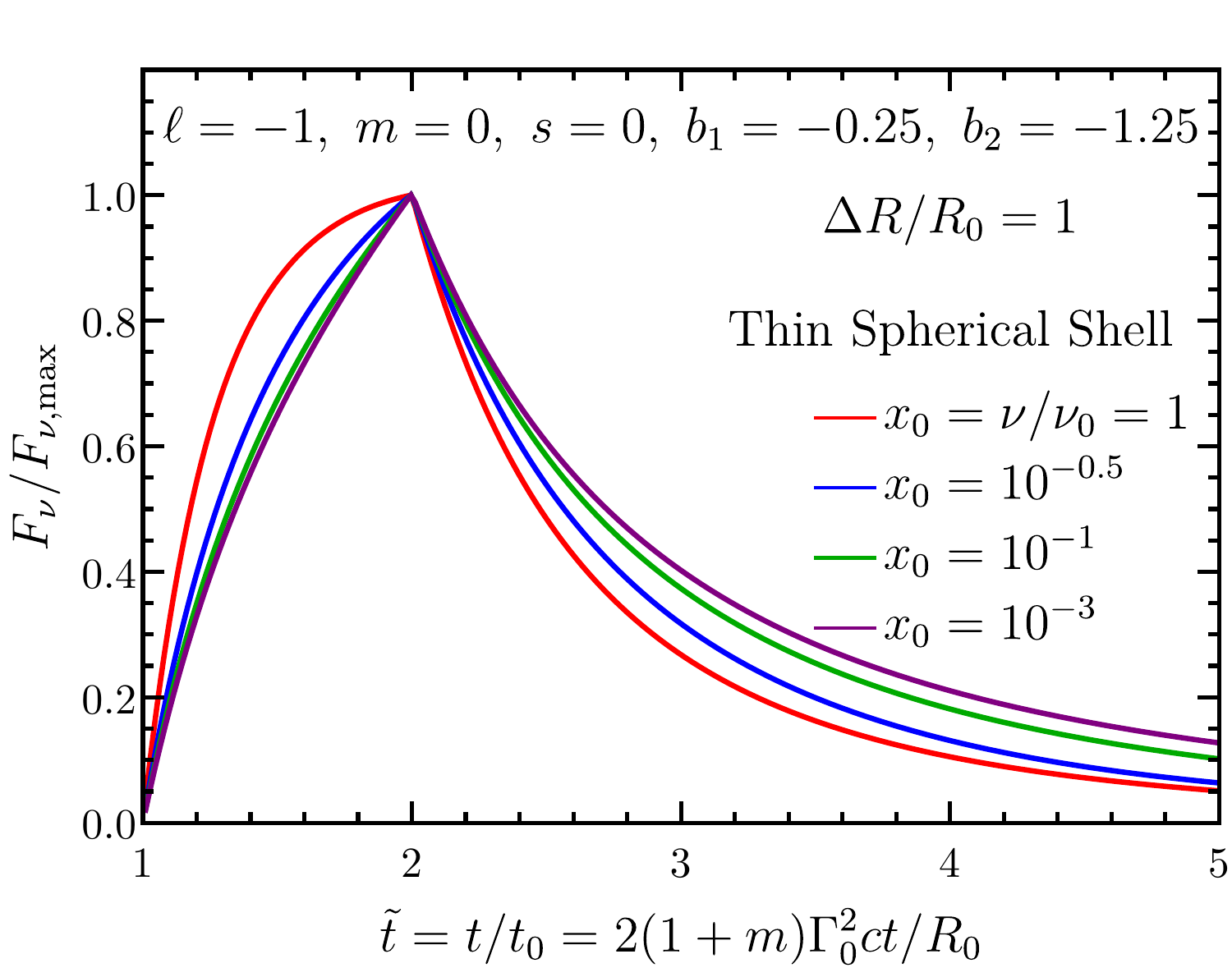}\quad\quad
    \includegraphics[width=0.48\textwidth]{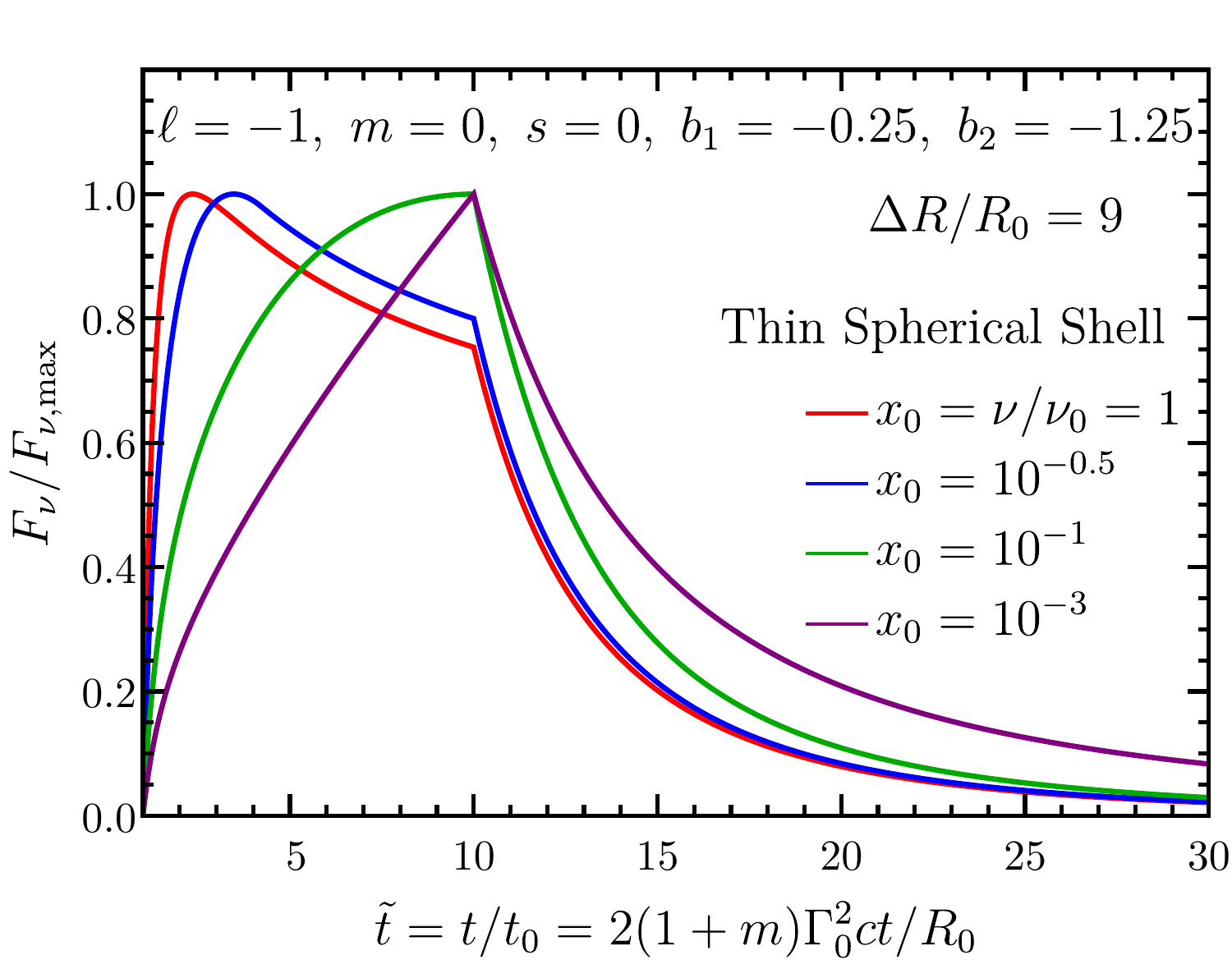}
    \includegraphics[width=0.48\textwidth]{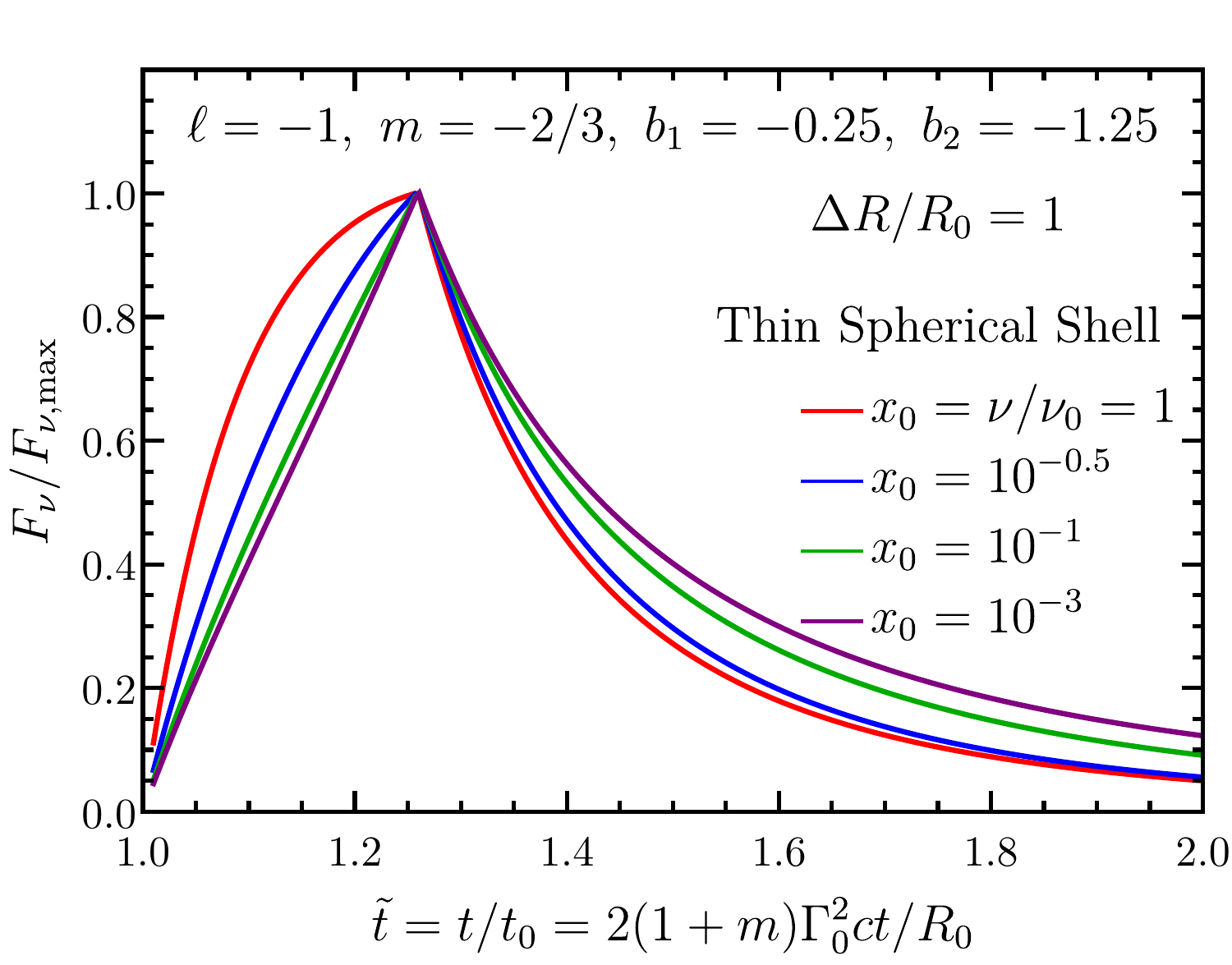}\quad\quad
    \includegraphics[width=0.48\textwidth]{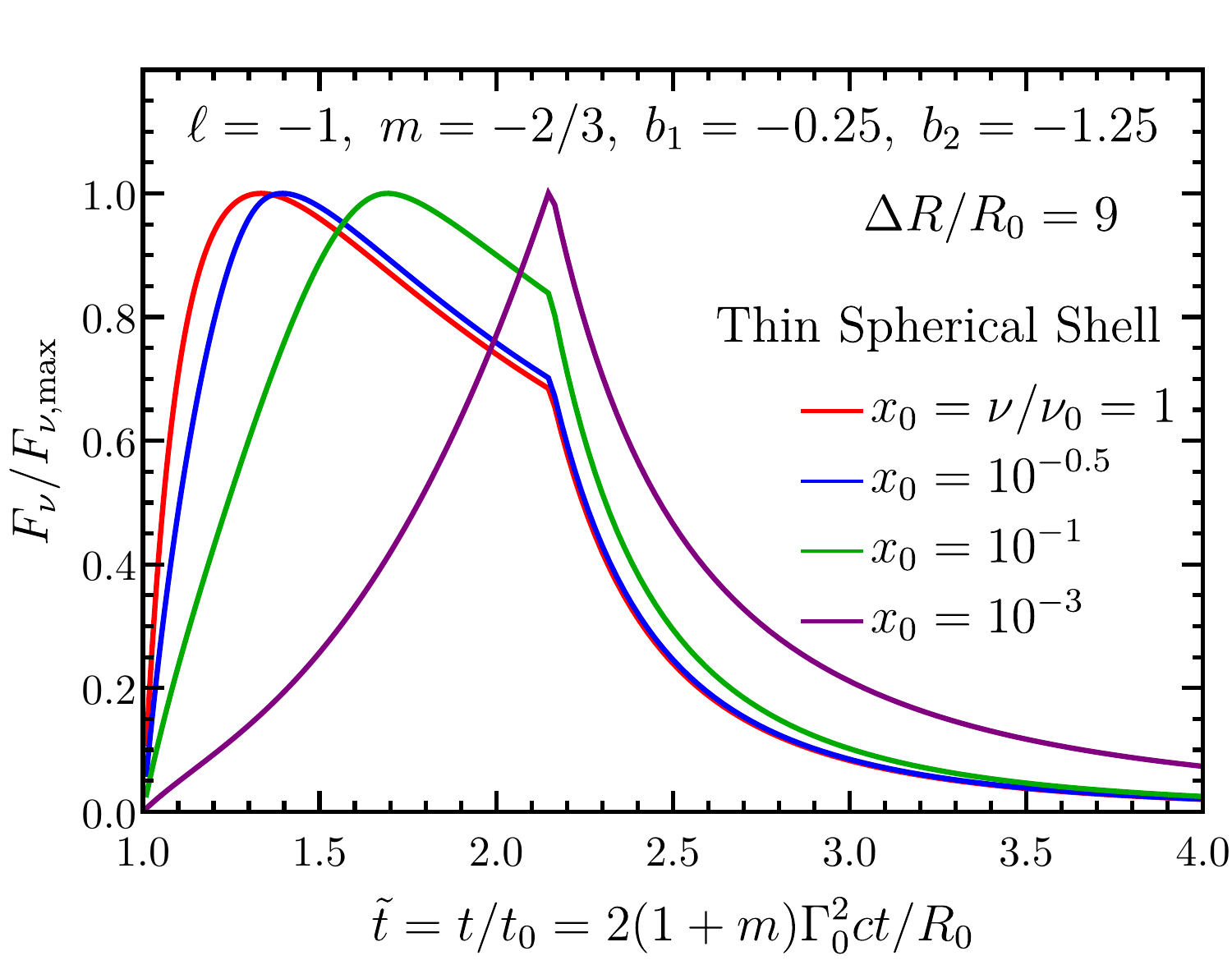}
    \caption{Pulse profiles for an ultrarelativistic thin spherical shell that is kinetic energy dominated (top; $a=1$ and $d=-1$) 
    or Poynting flux dominated (bottom; $a=4/3$ and $d=-2$), shown as a function of the apparent time $t$ normalized by the arrival 
    time of the first photons emitted from radius $R_0$ along the LOS. The shell starts radiating at radius $R_0$ and continues to 
    radiate over a radial distance of $\Delta R$. The different curves show the trend at different observed 
    frequencies $\nu=x_0\nu_0$, where $\nu_0$ is the spectral peak frequency of the photons that arrive at time $t_0$. }
    \label{fig:Sph-shell-pulse}
\end{figure*}

\begin{figure}
    \centering
    \includegraphics[width=0.48\textwidth]{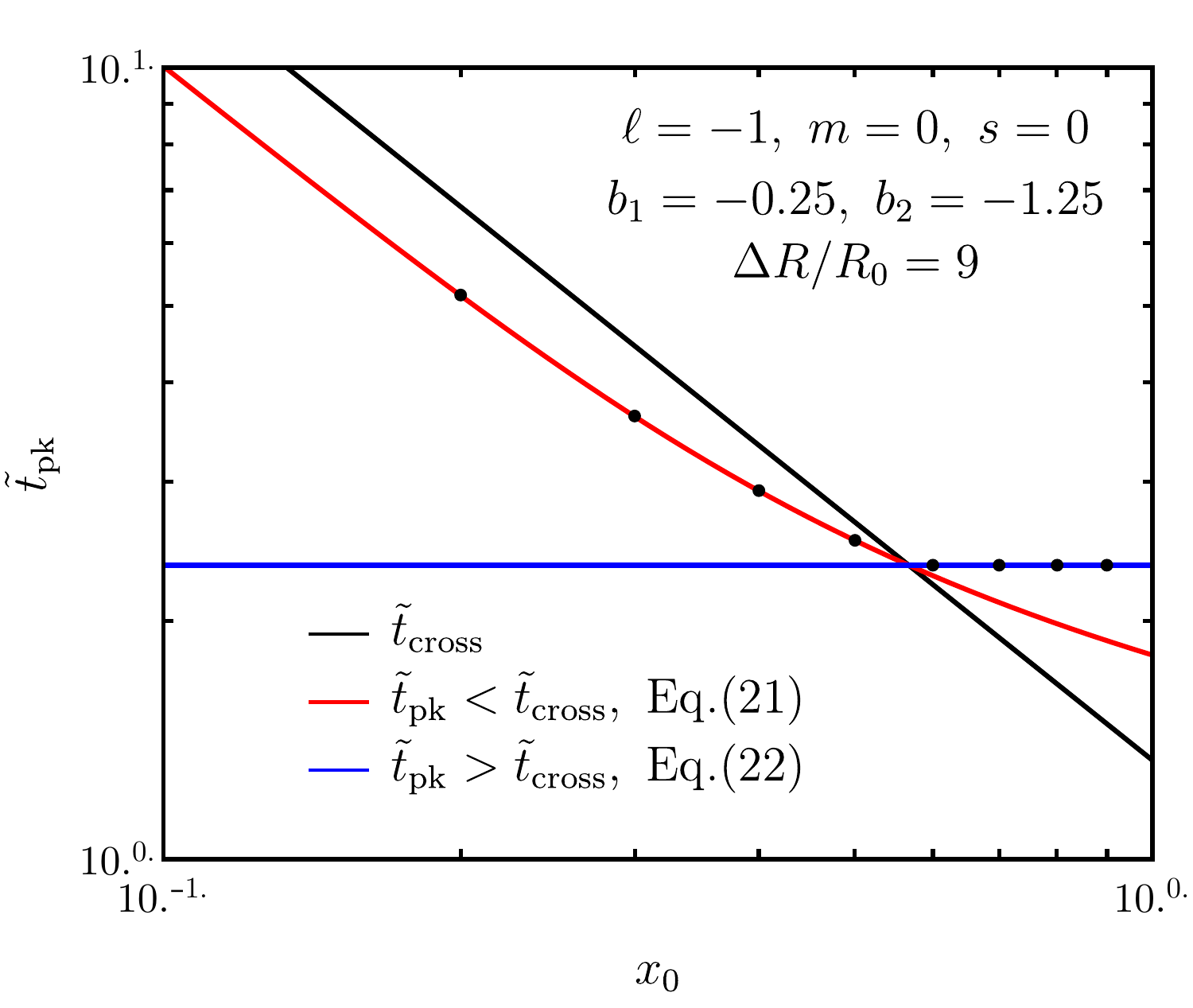}
    \caption{The peak time of pulses from the full numerical solution (Eq.~\ref{eq:Fnu}) with different $x_0$ are shown using black dots. 
    These are compared with analytic solutions obtained from $dF_\nu/d\tilde t=0$, where approximate scaling of the flux $F_\nu(\tilde t)$ 
    is given in Eq.~(\ref{eq:Flux-Scaling-General}). It's clear that the pulse peak times don't exactly follow the crossing time ($\tilde t_{\rm cross}$) 
    of the spectral break energy that can only be used as a zeroth-order approximation.
    }
    \label{fig:tpk-x0}
\end{figure}

\begin{figure*}
    \centering
    \includegraphics[width=0.48\textwidth]{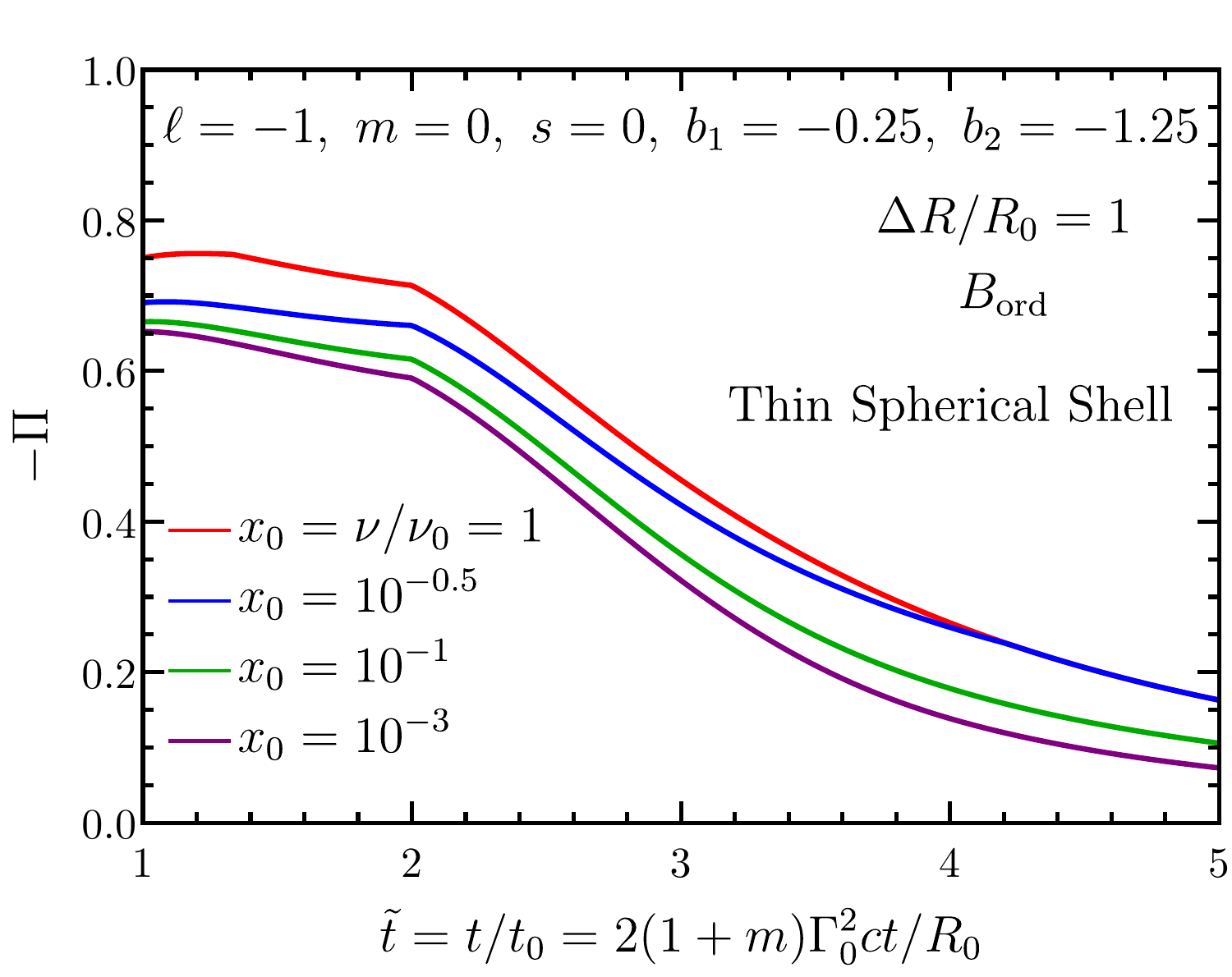}\quad\quad
    \includegraphics[width=0.48\textwidth]{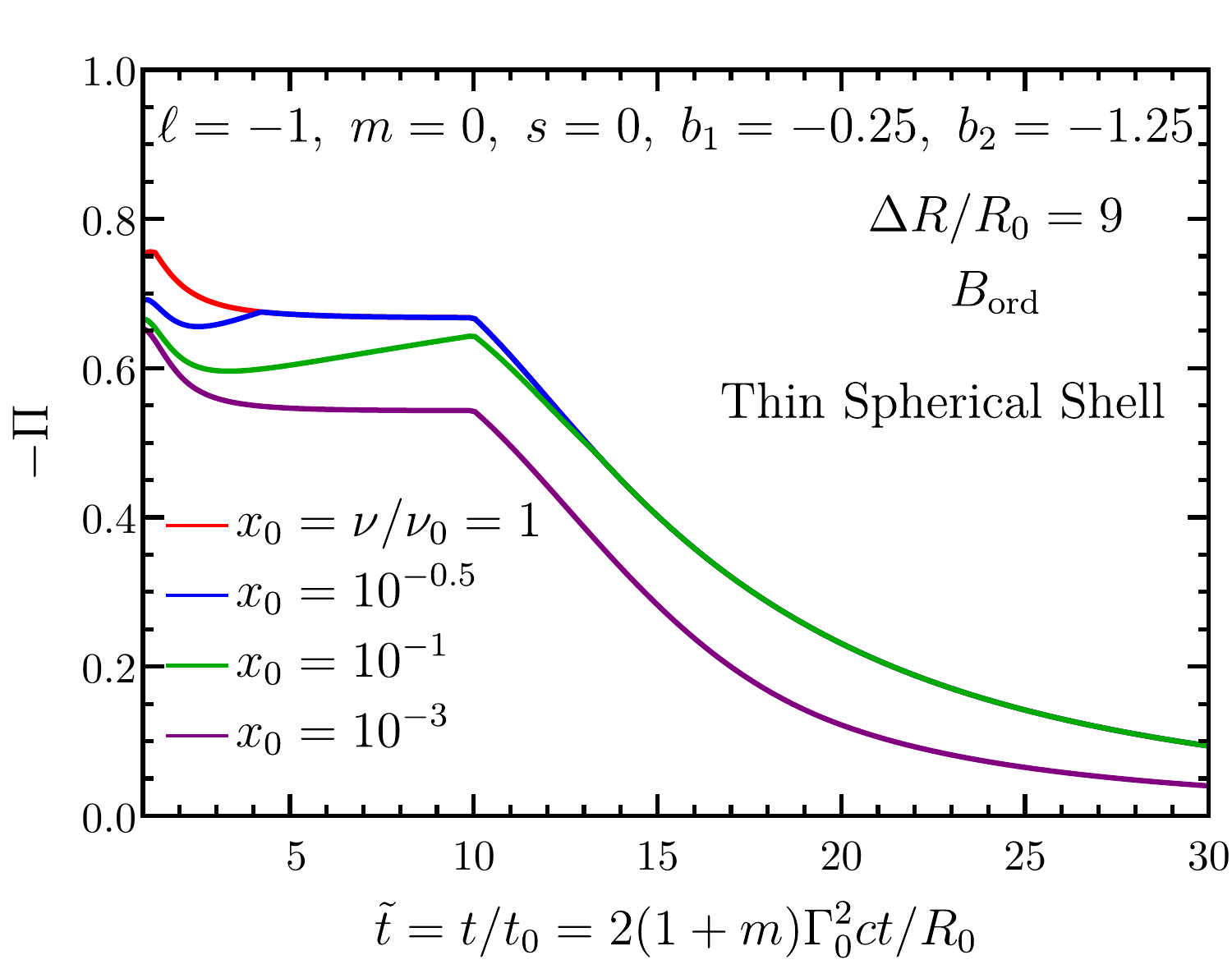}
    \includegraphics[width=0.48\textwidth]{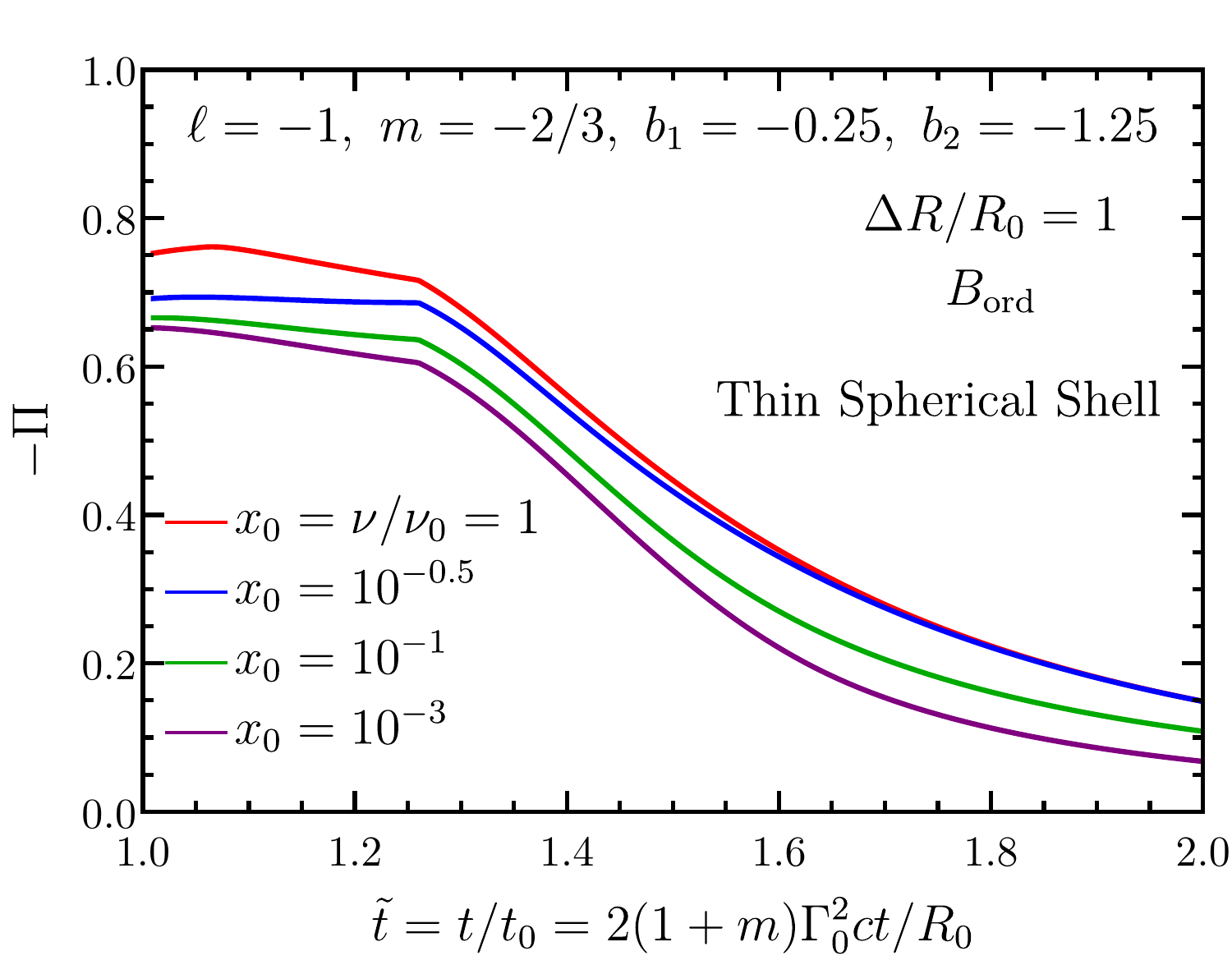}\quad\quad
    \includegraphics[width=0.48\textwidth]{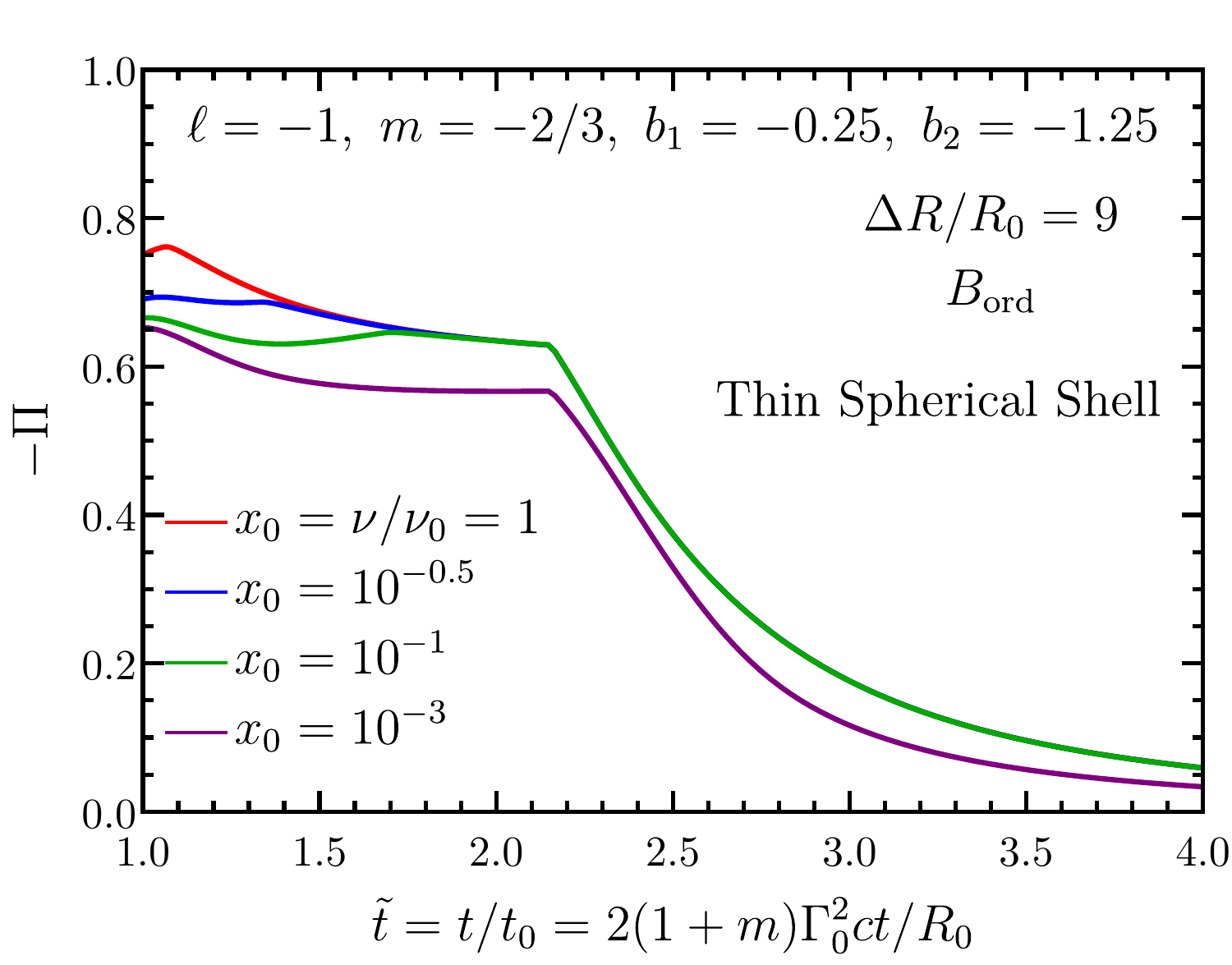}
    \caption{Temporal evolution of polarization over a pulse shown for the same flow dynamics as in Fig.~\ref{fig:Sph-shell-pulse}, 
    with KED (top; $a=1$ and $d=-1$) and PFD (bottom; $a=4/3$ and $d=-2$) flows. The magnetic field is assumed to be ordered with 
    angular coherence length, $\theta_B\gtrsim 1/\Gamma$, larger than the observed region of the flow. See Table~\ref{tab:symbols} 
    and caption of Fig.~\ref{fig:Sph-shell-pulse} for definition of various symbols.}
    \label{fig:Sph-shell-pol}
\end{figure*}

The arrival time $t$ of a photon originating at an angle $\tilde\theta$ and from a radius $R$ is given by $t_z\equiv t/(1+z) = t_{\rm lab}-R\tilde\mu/c$, 
where the lab-frame time for an ultra-relativistic thin-shell expanding with bulk LF $\Gamma(R)\propto R^{-m/2}$ is
\begin{equation}
    t_{\rm lab}(R) = \int_0^R\frac{dR'}{\beta(R')c} \approx \frac{R}{c} + \frac{R}{2(1+m)\Gamma^2c}
\end{equation}
for $m>-1$. Expressions for general $m\leq-1$ are straightforward. This yields the equal arrival time surface (EATS) for a given apparent time 
\citep[e.g.][]{Granot+08}
\begin{equation}\label{eq:EATS}
    t_z(R,\tilde\mu) = \frac{R}{c}\left[1-\tilde\mu+\frac{1}{2(1+m)\Gamma^2}\right]
    \approx \frac{R}{2\Gamma^2c}\left(\tilde\xi + \frac{1}{1+m}\right)\,,
\end{equation}
where we made the approximation $\tilde\mu\approx1-\tilde\theta^2/2$ for $\tilde\theta\ll1$ when $\Gamma\gg1$ and where 
$\tilde\xi = (\Gamma\tilde\theta)^2 = (\Gamma_0\tilde\theta)^2\hat R^{-m} = \tilde\xi_0\hat R^{-m}$ with $\hat R\equiv R/R_0$. 
The arrival time of the first photons originating at radius $R_0$ and along the LOS with $\tilde\mu=1~(\tilde\xi=0)$ is $t_{0,z}=R_0/2(1+m)\Gamma_0^2c$ 
(where $t_{z}=0$ corresponds to the arrival time of a hypothetical photon emitted at the central source together with the 
ejection of the expanding relativistic thin shell). Normalizing the apparent time by $t_{0,z}$ allows to express the EATS 
condition in a more convenient form
\begin{equation}\label{eq:EATS2}    \tilde t \equiv \frac{t_z}{t_{0,z}} = \tilde t_R + \tilde t_\theta = \hat R^{1+m} + (1+m)\tilde \xi_0\hat R\,,
\end{equation}
where we identify the normalized radial and angular delay times as $\tilde t_R$ and $\tilde t_\theta$. Then, for a given apparent 
time the arriving photons originate from different angles on the outflow with $-1\leq\tilde\mu\leq1$ (as well as different $0\leq\tilde\varphi\leq2\pi$ 
but $\tilde\varphi$ does not affect the arrival time as long as the emitting shell is spherical). This corresponds to different 
limiting radii\footnote{The limiting radii obtained here are strictly valid for a spherical flow and not for a top-hat jet (\S\ref{sec:top-hat-jet-pol}) 
when $\theta_{\rm obs}>\theta_j$ and when $\tilde t_{0-}<\tilde t<\tilde t_{f-}$ and $\tilde t_{0+}<\tilde t<\tilde t_{f+}$. In such cases, these 
limiting radii can be used when the angular dependence of the emissivity is taken into account, so that only emission from parts of the 
EATS that intersect the jet is included.} 
$\hat R_{\min}\leq \hat R\leq \hat R_{\max}$. The minimum radius is obtained for $\tilde\mu=-1$ with 
$\hat R_{\min}=\max[1,\hat{\mathcal R}_{\min}]$ where in general $\hat{\mathcal R}_{\min}$ is the root of the EATS equation 
\begin{equation}
    \tilde t = 4(1+m)\Gamma_0^2\hat{\mathcal R}_{\min} + \hat{\mathcal R}_{\min}^{1+m}\,.
\end{equation}
For an ultrarelativistic flow, with $\Gamma_0\gg1$, the above equation yields $\hat{\mathcal R}_{\rm min}>1$ only when $\tilde t\gg1$, which is not 
relevant for our case here, and therefore, $\hat R_{\rm min}=1$. The maximum radius is obtained for $\tilde\mu=1$ with 
$\hat R_{\max}=\min[1+\Delta R/R_0, \hat{\mathcal R}_{\max}]$ with $\hat{\mathcal R}_{\max}=\tilde t^{1/(1+m)}$. 
By using the EATS condition we can relate the integration over $\tilde\mu$ to that over $R$ with $dR=\vert d\tilde\mu/dR\vert dR$ 
when calculating the flux density, where 
\begin{eqnarray}
    R_0\frac{d\tilde\mu}{dR} = \frac{d\tilde\mu}{d\hat R} 
    &&= \frac{ct_zR_0}{R^2}\left[1+\frac{m}{2(1+m)}\frac{R}{\Gamma^2ct_z}\right] \\
    &&= \frac{1}{2(1+m)\Gamma_0^2}\frac{\tilde t}{\hat R^2}\left[1+\frac{m\hat R^{1+m}}{\tilde t}\right]
\end{eqnarray}
which yields
\begin{equation}\label{eq:Fnu}
    F_\nu(\tilde t) = \frac{(1+z)}{16\pi^2d_L^2}\int_{\hat R_{\min}}^{\hat R_{\max}}d\hat R \left\vert \frac{d\tilde\mu}{d\hat R}\right\vert \delta_D^3
    \int_0^{2\pi}d\tilde\varphi\, L'_{\nu'}\!\left[\hat{R},\theta(\hat{R},\tilde{\varphi})\right]\,,
\end{equation}
where, for an ultrarelativistic flow, the Doppler factor can be expressed as
\begin{equation}
    \delta_D\approx \frac{2\Gamma}{(1+\tilde\xi)} = 2(1+m)\Gamma_0\frac{\hat R^{-m/2}}{\hat R^{-(m+1)}\tilde t+m}\,.
\end{equation}

The energy-dependent linear polarization can be expressed using the Stokes parameters, with $\Pi_\nu=\sqrt{Q_\nu^2+U_\nu^2}/I_\nu$ 
where the specific intensity $I_\nu\propto F_\nu$ and ratio of the Stokes parameters (polarized intensities $Q_\nu$ and $U_\nu$ to the total 
intensity $I_\nu$) can be obtained from \citep{Gill+20}
\begin{equation}\label{eq:general-pol}
    \left\{\begin{aligned}
        \frac{Q_\nu(t_z)}{I_\nu(t_z)} \\
        \frac{U_\nu(t_z)}{I_\nu(t_z)}
    \end{aligned} \right\} 
    = \frac{\int_{\hat R_{\min}}^{\hat R_{\max}} d\hat R \left\vert \frac{d\tilde\mu}{d\hat R}\right\vert \delta_D^3 L'_{\nu'}(\hat R) 
    \int_0^{2\pi}d\tilde\varphi \Lambda(\tilde\xi,\tilde\varphi)\Pi'
    \left\{\begin{aligned}
        \cos(2\theta_p) \\
        \sin(2\theta_p)
    \end{aligned}\right\}
    }
    {\int_{\hat R_{\min}}^{\hat R_{\max}} d\hat R \left\vert \frac{d\tilde\mu}{d\hat R}\right\vert \delta_D^3 L'_{\nu'}(\hat R) 
    \int_0^{2\pi}d\tilde\varphi \Lambda(\tilde\xi,\tilde\varphi)}
\end{equation}
where $\Lambda(\tilde\xi,\tilde\varphi)=\langle[1-(\hat n'\cdot\hat B')^2]^{\epsilon/2}\rangle$ represents the factor relating to the pitch 
angle of electrons averaged over the local probability distribution of the comoving magnetic field $\hat B'$, and $\Pi'$ is the local 
(and not averaged over the whole observed region of the emitting shell) degree of polarization. Expressions for $\Lambda(\tilde\xi,\tilde\varphi)$, 
$\theta_p$, and $\Pi'$ for different magnetic field configurations and assuming an ultrarelativistic uniform flow were first derived in \citet{Granot-03,Granot-Konigl-03,Lyutikov+03,Granot-Taylor-05} and are summarized in \citet{Toma+09,Gill+20}. For an axisymmetric flow, $U_\nu=0$ due to symmetry, 
and the instantaneous polarization is given by $\Pi_\nu=Q_\nu/I_\nu$. For an axisymmetric flow, the net polarization can only be aligned with 
two direction, either along the line connecting the LOS and the jet symmetry axis ($\Pi<0$) on the plane of the sky or transverse to it ($\Pi>0$).

We consider different magnetic field configurations here, including tangled fields ($B_\perp$) constrained to be in the plane transverse to the local velocity 
vector, which we take to be in the radial direction, fields ($B_\parallel$) that are ordered along the direction of the local velocity vector, ordered fields 
($B_{\rm ord}$) in the plane of the ejecta with angular coherence scales $\theta_B\gtrsim1/\Gamma$, and a globally ordered toroidal field ($B_{\rm tor}$) that 
is axisymmetric around the jet symmetry axis and is ordered in the transverse direction.

%%%%%%%%%%%%%%%%%%%%%%%%%%%%%%%%%%%%%%%%%%%%%%%%%%%%%%%%%%%%%%%%%%%%%%
\subsection{Polarization from a thin spherical shell}\label{sec:Orderd-B-field-Spherical-Shell}
%%%%%%%%%%%%%%%%%%%%%%%%%%%%%%%%%%%%%%%%%%%%%%%%%%%%%%%%%%%%%%%%%%%%%%

We first consider the simplest scenario of an ultrarelativistic thin spherical shell, where the emission region is pervaded by a locally 
ordered B-field ($B_{\rm ord}$) that lies in the plane transverse to the local fluid velocity, $\vec\beta=(\varv/c)\hat R$ with $\varv$ being the fluid 
velocity, which we take to be in the radial direction with unit vector $\hat R$. This case is both instructive and relevant 
when the jet has sharp edges, with half-opening angle $\theta_j$, and the observer is on-beam (see Table~\ref{tab:symbols} and \S\ref{sec:top-hat-jet-pol} 
for definition), with viewing angle $\theta_{\rm obs}\lesssim\theta_j-1/\Gamma$, so that the emission is indistinguishable from that arising 
from a spherical flow at early times when the observer is unaware of the jet's edge. The field is ordered with an angular coherence 
length scale $\theta_B\gtrsim1/\Gamma$, where $\tilde\theta = 1/\Gamma$ is the angular size of the beaming cone.

The approximate scaling of flux density with $\tilde t$ can be derived from Eq.~(\ref{eq:Fnu}) to obtain 
\begin{equation}\label{eq:Flux-Scaling-General}
    F_\nu(\tilde t) \propto \tilde t\hat R^{a-1}\left(1+\frac{m\hat R^{1+m}}{\tilde t}\right)
    \left(\frac{\hat R^{-m/2}}{\hat R^{-(m+1)}\tilde t+m}\right)^3 S(x)\left\vert_{\hat R_{\min}}^{\hat R_{\max}}\right.\,,
\end{equation}
where this expression is evaluated at the two limiting radii, and where $S(x)\propto x^{b_1}\exp[-(1+b_1)x]$ for $x\leq x_b$ and 
$S(x)\propto x^{b_2}$ for $x> x_b$ with 
\begin{equation}\label{eq:x-t}
    x = \frac{2\Gamma_0}{\delta_D}\frac{x_0}{\hat R^d} = \frac{x_0}{1+m}\frac{\hat R^{-(m+1)}\tilde t+m}{\hat R^{(2d-m)/2}}\,.
\end{equation}
There are two important timescales at which the behavior of the flux density and polarization changes, namely (i) the crossing time, $\tilde t_{\rm cross}$, 
of the spectral break frequency ($x_b$) across a given observed frequency ($x$) for the emission from the LOS, after which point the 
observer only samples the spectrum above the spectral 
break frequency, which for the Band-function is a strict power-law, and (ii) the timescale $\tilde t_f=\hat{R}_f^{1+m}$ that corresponds to the arrival of the last photons 
emitted along the LOS from radius $R_f$ at which emission is suddenly switched off. Exact expressions for the flux density for different temporal power-law 
segments are derived in \citet{Genet-Granot-09}. 

When the spectral break crossing occurs at $\tilde t<\tilde t_f$, the emission is dominated by that along the LOS. In this case, the time of the spectral 
break crossing can be approximately obtained from Eq.~(\ref{eq:x-t}) with $\tilde t=\tilde t_R = \hat R^{1+m}$, the radial delay time for the arrival of 
the first photons along the LOS from any radius, so that 
\begin{equation}\label{eq:x-t-on-axis}
    x_b = x = x_0\tilde t^{(m-2d)/2(1+m)}\Rightarrow \tilde t_{\rm cross} \approx \fracb{x_b}{x_0}^{2(1+m)/(m-2d)}\,,\quad x_0<1.
\end{equation}
The crossing time derived above is only approximate and not exact since the observer receives emission from different radii. Although the dominant contribution 
does come from emission along the LOS, the contribution from smaller radii has a small but non-negligible effect on $\tilde t_{\rm cross}$. 
Since for the outflow dynamics considered here $m>-1$, and if $d\leq m/2$ as well, which is true for the two types of outflows considered in 
this work, then the crossing time increases with decreasing energy $x_0$. In that case, for $\tilde t>\tilde t_{\rm cross}$ the pulse profile 
for a given $x_0=x_{0,a}$ should have an exactly similar trend with $\tilde t$ as for pulses with $x_0=x_{0,b}>x_{0,a}$, if all else remains unchanged. 
Furthermore, since the degree of polarization depends on the spectral index, the polarization for $x_0=x_{0,a}$ at any given $\tilde t>\tilde t_{\rm cross}$ 
should be exactly the same as that for $x_0=x_{0,b}>x_{0,a}$. In the case of a KED flow, with $m=0$ and $d=-1$, we get
\begin{equation}\label{eq:tcross-KED}
    \tilde t_{\rm cross} \approx \frac{x_b}{x_0}\,.
\end{equation}
When the spectral break crossing occurs at $\tilde t>\tilde t_f$, there is no longer emission arriving along the LOS. Instead, the flux is dominated 
by emission from $\hat R=\hat R_f$ but from angles away from the LOS, with $\tilde \theta\geq\tilde{\theta}_f>0$ where $\tilde{\theta}_f=\Gamma_0^{-1}[(\tilde{t}-\tilde{t}_f)/(1+m)\hat{R}_f]^{1/2}$ (for $\tilde{\theta}_f\ll1$). In this case, $\tilde t_{\rm cross}$ can be approximated 
using Eq.~(\ref{eq:x-t}), now evaluated at $\hat R=\hat R_f$ for $x=x_b$, which yields
\begin{equation}
    \tilde t_{\rm cross} \approx \left[(1+m)\fracb{x_b}{x_0}\hat R_f^{(2d-m)/2}-m\right]\hat R_f^{1+m}\,,\quad x_0<1.
\end{equation}
Interestingly, for the special case of a KED flow with $m=0$ and $d=-1$, we recover Eq.~(\ref{eq:tcross-KED}).

In Fig.~\ref{fig:Sph-shell-pulse}, we show the pulse profiles for a KED (top row) and a PFD (bottom row) flows with $\Delta R/R_0=\{1, 9\}$ for 
different $x_0=\nu/\nu_0$. Similar results have also been obtained in \citet{Genet-Granot-09,Uhm-Zhang-15,Uhm-Zhang-16}. 
The pulse profiles are shown using normalized flux $F_\nu/F_{\nu,\rm max}$ as a function of $\tilde t$, where 
$F_{\nu,\rm max}$ is the maximum flux density. For $\Delta R/R_0=1$, $F_{\nu,\rm max} = F_\nu(\tilde t_f)$ where 
$\tilde t_f = \hat R_f^{1+m} = (1+\Delta R/R_0)^{1+m}$ 
gives the normalized arrival time of the last photon emitted along the LOS when the emission is switched off at $R_f=R_0+\Delta R$. Then, for 
$1\leq\tilde t\leq\tilde t_f$, the flux rises to its peak and decays abruptly at $\tilde t>\tilde t_f$ after which point high latitude emission starts 
to dominate. All the curves show temporal slopes that are dissimilar from each other, and only at late times when $\tilde t > \tilde t_{\rm cross}(x_0)$ 
the temporal slopes of curves with $x_0<1$ match with that of $x_0=1$. One of the main differences between the $m=0$ (top-left) and $m=-2/3$ (bottom-left) 
case is that there is a linear relation between $\bar{t}_f=\tilde t_f-1$ and $\Delta R/R_0$ which is not true for an accelerating ($m<0$) flow. 
The pulse profiles do show subtle differences between the two cases, e.g. when comparing the $x_0=10^{-3}$ curves when $\Delta R/R_0=1$, the $m=0$ case 
shows a curved pulse profile up to the peak whereas the $m=-2/3$ case shows a much more linear rise to the peak. These differences are even more pronounced 
when $\Delta R/R_0=9$.

For the $\Delta R/R_0=9$ ($\hat{R}_f=10$) case (right column of Fig.~\ref{fig:Sph-shell-pulse}), the pulse profiles show a peak at $\tilde t_{\rm pk}<\tilde t_f$ instead. 
The reason for this behavior is the passage of the spectral peak across the observed frequency which, despite the rise of the peak spectral power, 
$L'_{\nu',\rm max}\propto R^{-(\ell+s)}$, with radius, causes the observed flux density to decline. Such a peak occurs at $\tilde{t}_{\rm pk}<\tilde{t}_f$ 
for $x_0\gtrsim x_b\hat{R}_f^{(2d-m)/2}$ i.e. $x_0\gtrsim0.133$ for a KED flow and $x_0\gtrsim0.029$ for a PFD flow for our parameter values. 
After the pulse peak, it can be seen that all curves 
show the same temporal slope, except the green and purple curves for $m=0$ and only the purple curve for $m=-2/3$ for which $\tilde t_{\rm cross}>\tilde t_f$. 
Since the Band spectrum is smoothly broken at the spectral peak, rather than having a sharp break, the pulse peak can lie both before or after 
$\tilde t_{\rm cross}$ due to the curvature. The scaling of the flux density in the two relevant time segments, $\tilde t<\tilde t_{\rm cross} < \tilde t_f$ and 
$\tilde t_{\rm cross} < \tilde t < \tilde t_f$ can be obtained from Eq.~(\ref{eq:Flux-Scaling-General}). 
The peak of the pulse is obtained from $dF_\nu/d\tilde t=0$, which for a KED flow ($m=0,a=1,d=-1$) yields the equations
\begin{align}
    \label{eq:tpk<tcross}
    (\tilde t_{\rm pk}^{\,4}-\tilde t_{\rm pk})x_0-\tilde t_{\rm pk}^{\,3} = \frac{2-b_1}{1+b_1}, & \quad\quad \tilde t_{\rm pk} < \tilde t_{\rm cross} < \tilde t_f \\
    \tilde t_{\rm pk} = \fracb{b_2-2}{b_2+1}^{1/3}, & \quad\quad \tilde t_{\rm cross} < \tilde t_{\rm pk} < \tilde t_f\,.
\end{align}
Here Eq.~(\ref{eq:tpk<tcross}) is implicit and the peak time is obtained by root finding. We show both solutions in Fig.~\ref{fig:tpk-x0} and compare 
them with the peak times of pulses with different $x_0$. The solution from Eq.~(\ref{eq:tpk<tcross}) asymptotically becomes parallel to 
$\tilde t_{\rm cross}(x_0)$ but scaled down in magnitude. This scale factor can be derived by looking at the $x_0\ll1$ and $\tilde t_{\rm pk}\gg1$ 
limit of Eq.~(\ref{eq:tpk<tcross}) that yields $\tilde t_{\rm pk}=x_0^{-1}$, and therefore the scale factor is $\tilde t_{\rm cross}/\tilde t_{\rm pk}=x_b$.

\begin{figure*}
    \centering
    \includegraphics[width=0.8\textwidth]{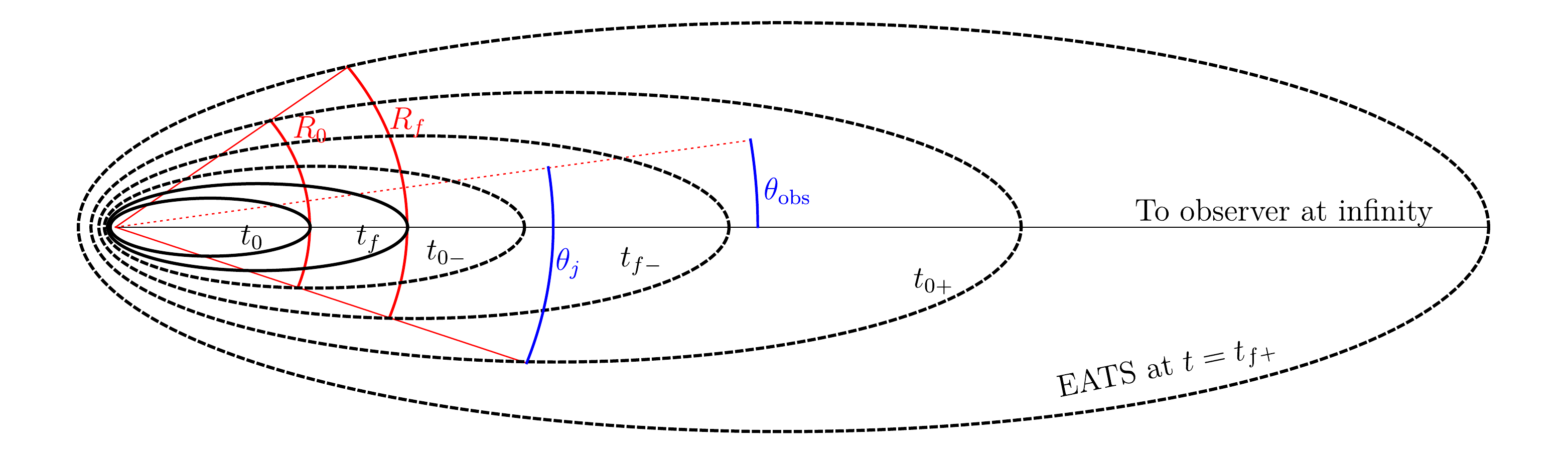}
    \includegraphics[width=0.8\textwidth]{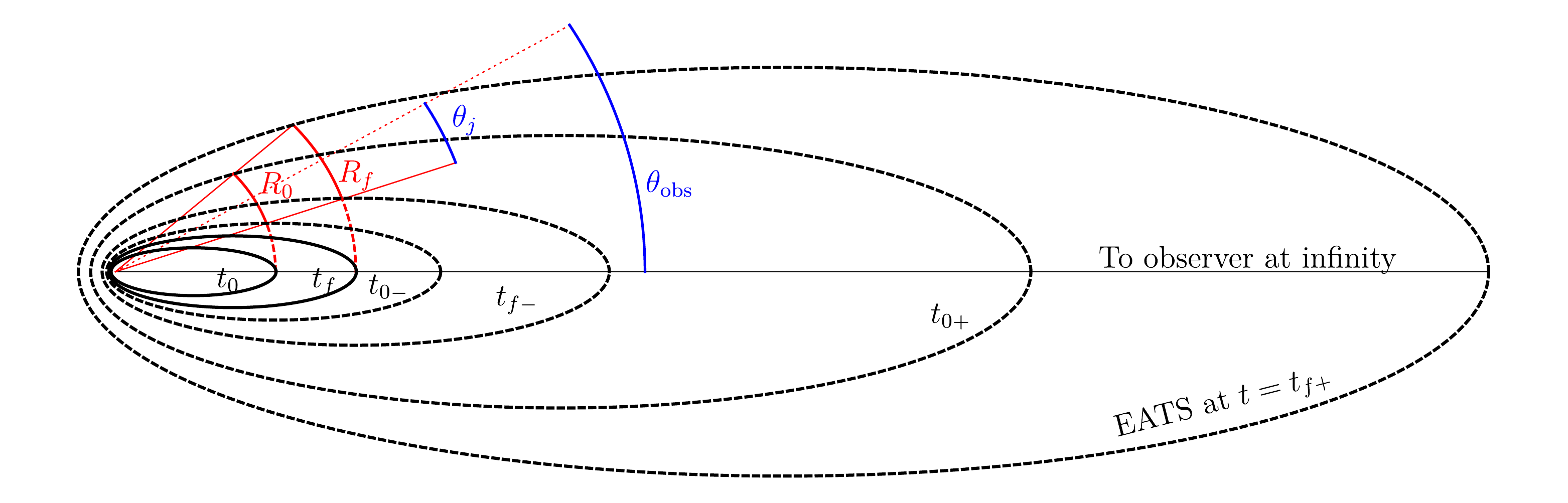}
    \caption{Two-dimensional cross-sections out of three-dimensional equal arrival time surfaces (EATSs) for an observer (with $\theta_{\rm obs}>0$) 
    receiving on-beam (top; $\theta_{\rm obs}<\theta_j+1/\Gamma$) and off-beam (bottom; $\theta_{\rm obs}>\theta_j+1/\Gamma$) emission from a top-hat jet 
    coasting at a constant $\Gamma$ ($m=0$). The spherical thin shell with sharp edges (at $\theta=\theta_j$; all 
    angles are measured from the jet symmetry axis (red dotted line)) starts to radiate at $R=R_0$ and stops at $R=R_f$. 
    \textbf{Top}: The on-beam emission arrival time of first photons emitted at $R=R_0$ along the LOS is $t=t_0$, while the same for photons emitted at 
    $R=R_f$ is $t=t_f$. At $t>t_f$ high latitude emission (dashed EATS) starts to dominate the observed flux and the photon arrival time is further delayed 
    by the angular delay time in addition to the radial delay. At $t=t_{0-}$ (see Eq.~\ref{eq:critical-times}), high-latitude emission arrives from the 
    nearer edge of the jet when the shell is at $R=R_0$. Subsequently, high latitude emission arrives from the nearer edge at $R=R_f$ at $t=t_{f-}$, 
    then from the farther edges at $R=R_0$ at $t=t_{0+}$ and from $R=R_f$ at $t=t_{f+}$. 
    While the relative orderings $t_0\leq\, t_{0-}\leq t_{0+}$, $t_0<t_f\leq\,t_{f-}\leq t_{f+}$ (the left and right equalities holding for 
    $\theta_{\rm obs}=\theta_j$ and $\theta_{\rm obs}=0$, respectively), $t_{0-}<t_{f-}$ and $t_{0+}<t_{f+}$ always hold, 
    the relative orderings of $t_f,t_{0-},t_{0+},t_{f-}$ can be different from the particular case shown here, i.e. (1a) $t_0<t_f<t_{0-}<t_{f-}<t_{0+}<t_{f+}$. 
    Other possible relative time orderings for different $\hat R_f$ and $q$ are: (1b) $t_0<t_f<t_{0-}<t_{0+}<t_{f-}<t_{f+}$, 
    (2a) $t_0<t_{0-}<t_f<t_{f-}<t_{0+}<t_{f+}$, (2b) $t_0<t_{0-}<t_f<t_{0+}<t_{f-}<t_{f+}$, and (2c) $t_0<t_{0-}<t_{0+}<t_f<t_{f-}<t_{f+}$. 
    For on-beam emission, all of these angular timescales are particularly important for the calculation of polarization (see Fig.~\ref{fig:Pol-Map-Bperp-Btor}). 
    \textbf{Bottom}: For off-beam emission, the arrival time of first photons from radius $R_0$ is determined by both the radial and angular delay time so 
    that first photons arrive at $t=t_{0-}$ and the same from radius $R_f$ is given by $t=t_{f-}$. The arrival time of last photons, corresponding to the 
    farther edge of the jet, from $R=R_0$ is given by $t=t_{0+}$ and the same from $R=R_f$ is given by $t=t_{f+}$.
    }
    \label{fig:EATS}
\end{figure*}

\begin{figure*}
    \centering
    \includegraphics[width=0.35\textwidth]{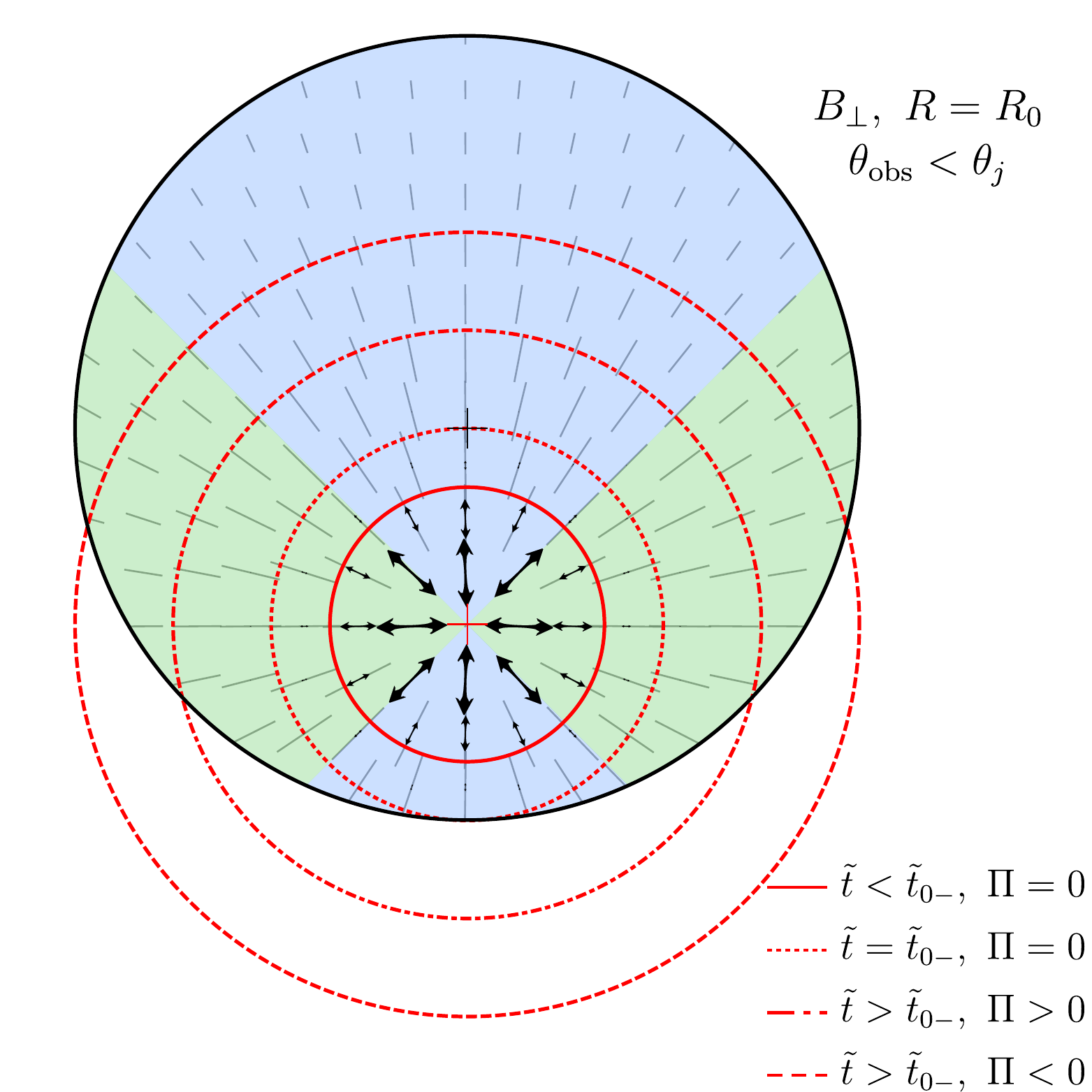}\hspace{10em}
    \includegraphics[width=0.35\textwidth]{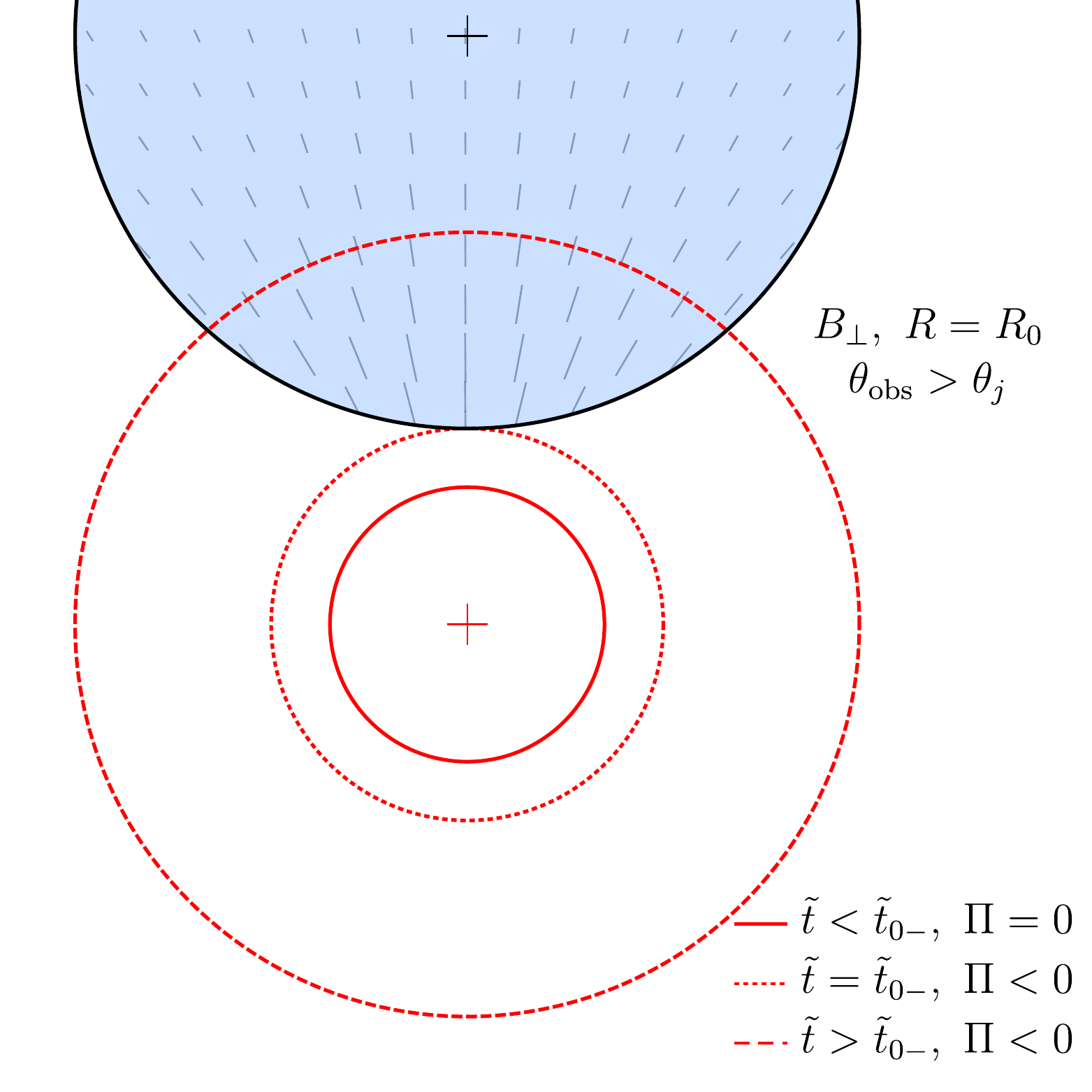}\hspace{10em}
    \includegraphics[width=0.35\textwidth]{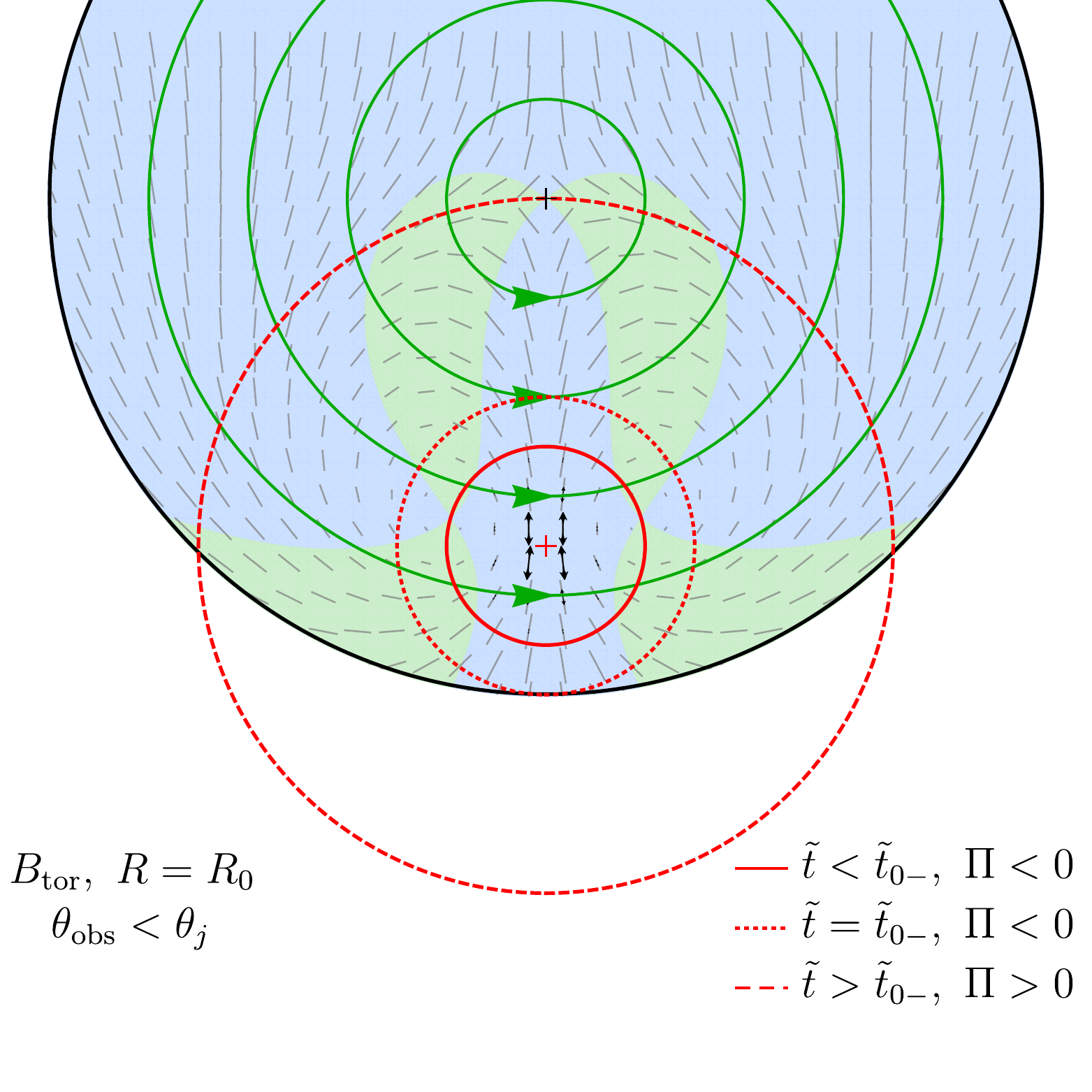}\hspace{10em}
    \includegraphics[width=0.35\textwidth]{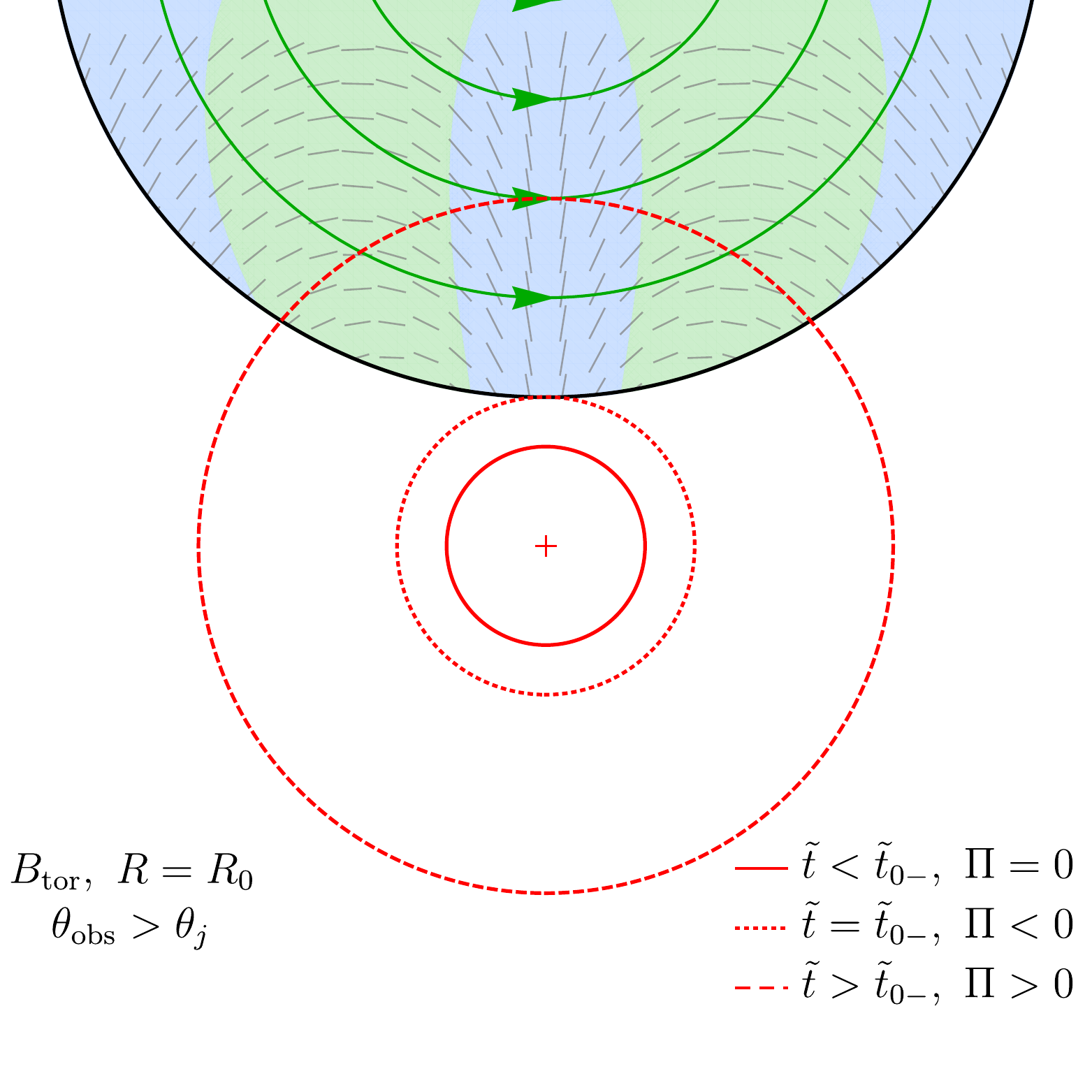}
    \caption{Schematics showing the observed area on the plane of the sky of a top-hat jet coasting at constant bulk $\Gamma$ and 
    its implications for the temporal evolution of polarization. The aperture of the jet is indicated by a black circle with black 
    plus symbol denoting the jet symmetry axis. The aperture of the beaming cone around the observer's LOS, shown with a red plus 
    symbol, with angular extent $\Gamma\tilde\theta=1$ is shown with a solid red circle. The cross-section of the EATS that 
    contributes to the high-latitude emission is shown with a dotted red circle. 
    \textbf{Top}: The polarization map for the $B_\perp$ field configuration for a LOS with $q=0.5$ (left) and $q=1.5$ (right) is 
    shown over the entire aperture of the flow with gray line segments. The size of the black arrows in the beaming cone reflects 
    the polarized intensity, whereas the gray line segments are normalized by $\delta^{3+\beta}$, with $\beta$ being the spectral 
    index, for clarity. The blue shaded regions contribute a net polarization along the line on the plane of the sky that connects 
    the jet symmetry axis and the LOS ($dQ_\nu\propto\cos(2\theta_p)<0$), whereas the green shaded regions contribute a net polarization 
    in the transverse direction ($dQ_\nu\propto\cos(2\theta_p)>0$), which is also the direction from which $\theta_p$ is 
    measured. The predominance of these regions decides the orientation of the 
    net polarization which switches sign going from initially green dominated to blue dominated. In the right panel, the PA remains 
    constant. For the $B_\perp$ (and also for $B_\parallel$) configuration, the polarization starts to grow above zero at 
    $\tilde t\geq \tilde t_{0-}$ due to missing flux that breaks the symmetry. This polarization `break' can be used to 
    constrain the jet half-opening angle. For $q>1+\xi_j^{-1/2}$, the emission starts maximally polarized at $\tilde t=\tilde t_{0-}$ after which time the polarization starts to decline. 
    \textbf{Bottom}: Polarization map for the $B_{\rm tor}$ field configuration with $q=0.7$ (left) and $q=1.3$ (right) is shown at three 
    different times where the PA undergoes a $90^\circ$ change at $\tilde t > \tilde t_{0-}$ at which point the green shaded region dominates 
    over the blue. Another $90^\circ$ change in PA occurs at late times when the blue region becomes dominant. The field lines, shown in green, 
    are symmetric around the jet axis. The emission is maximally polarized at $\tilde t=1$ and shows a sharp decline at $\tilde t>\tilde t_f$ 
    at which point high-latitude emission becomes dominant.
    }
    \label{fig:Pol-Map-Bperp-Btor}
\end{figure*}

The temporal evolution of polarization is shown in Fig.~\ref{fig:Sph-shell-pol} for different values of $x_0$ and $\Delta R/R_0$, and it is broadly 
similar for the two different outflows. Initially, at $\tilde t\approx 1$, since the emission is dominated by photons along the LOS, the polarization 
shows the maximum degree ($\Pi=\Pi_{\rm max}$) for the local spectral index sampled for a given $x_0$. Due to the curvature of the spectrum near the 
spectral peak, the local spectral index varies as $x_0$ is decreased until the asymptotic index is reached. At $\tilde t > 1$, the observer sees photons 
emerging from within the entire beaming cone which causes the polarization to decline and saturate due to partial cancellation. At $\tilde t > \tilde t_f$ 
high latitude emission dominates and the polarization shows a steep decline. For all the cases shown, the polarization curves corresponding to $x_0<1$ 
merge with that for $x_0=1$ at $\tilde t\approx\tilde t_{\rm cross}(x_0)$ due to crossing of the break frequency across the observed frequency.

Here we have only considered a locally ordered magnetic field configuration since it naturally breaks the symmetry in an axisymmetric 
flow and yields non-vanishing polarization. Similar results can be obtained for a globally ordered toroidal field, $B_{\rm tor}$. On the other hand, 
a uniform spherical flow would always yield zero net polarization for the $B_\perp$ and $B_\parallel$ fields due to complete cancellation of polarization 
after averaging over the GRB image over the plane of the sky. Therefore, an inhomogeneous flow or jet geometry with sharp/smooth edges, as treated in 
the next section, is needed to break the symmetry.

\begin{figure*}
    \centering
    \includegraphics[width=0.48\textwidth]{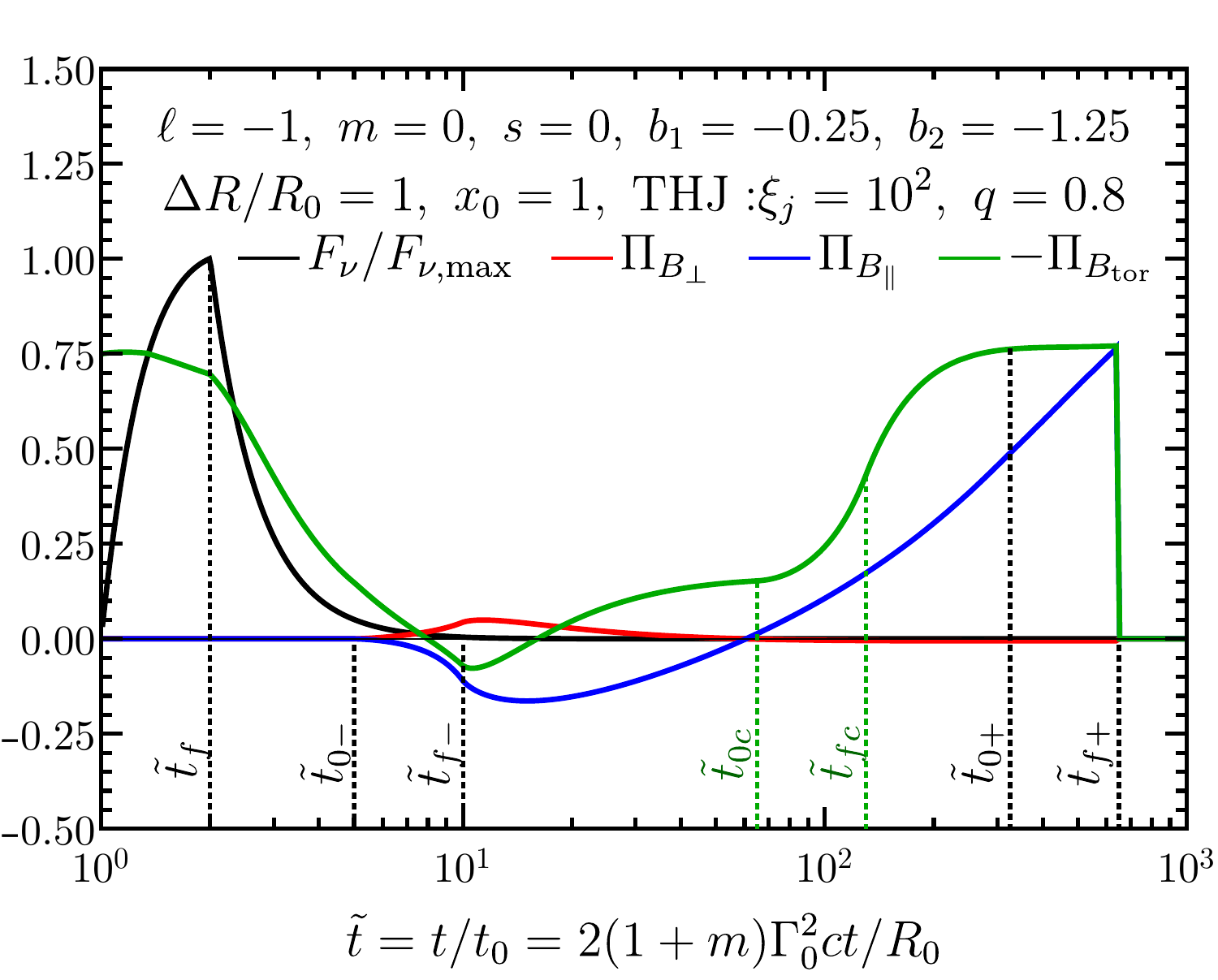}\quad\quad
    \includegraphics[width=0.48\textwidth]{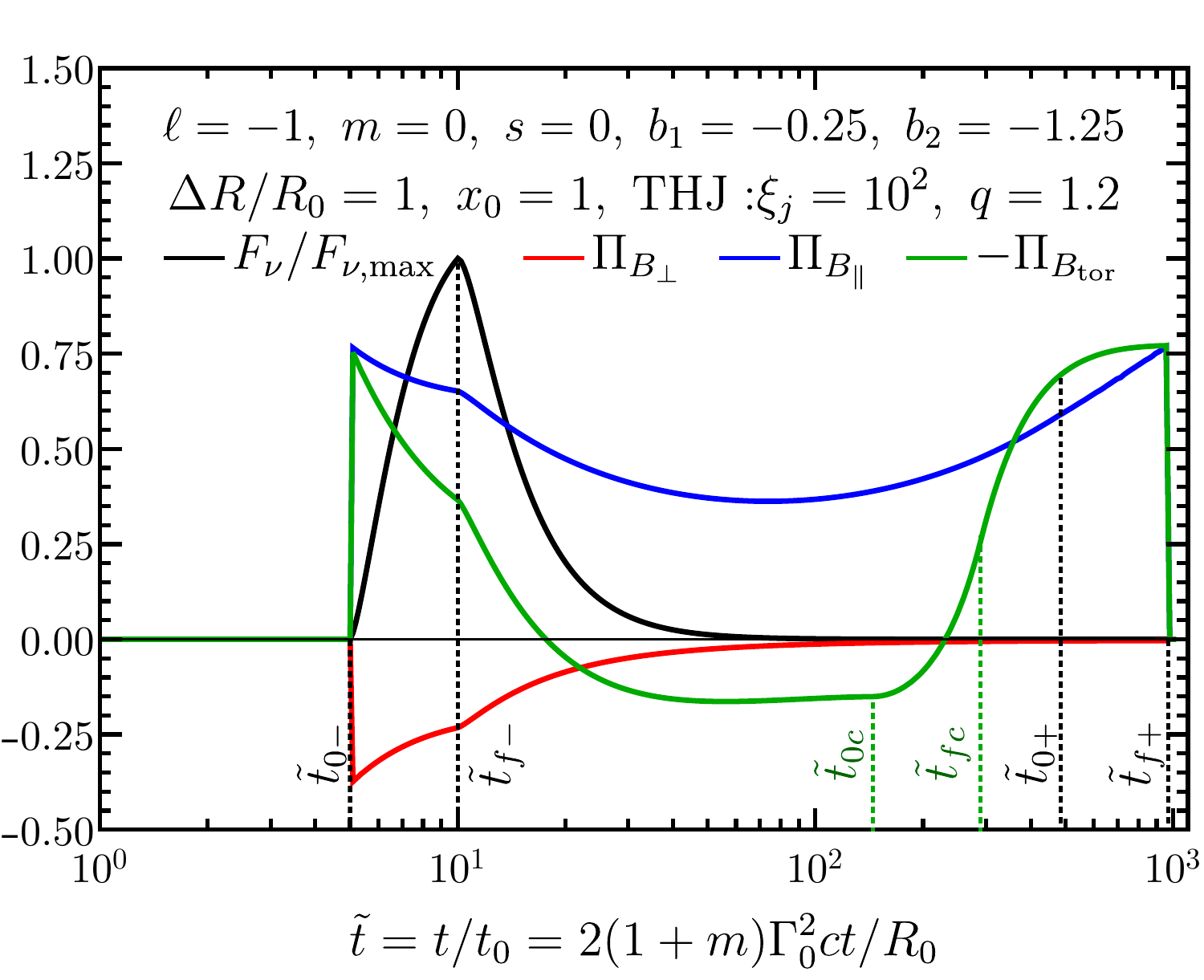}
    \includegraphics[width=0.48\textwidth]{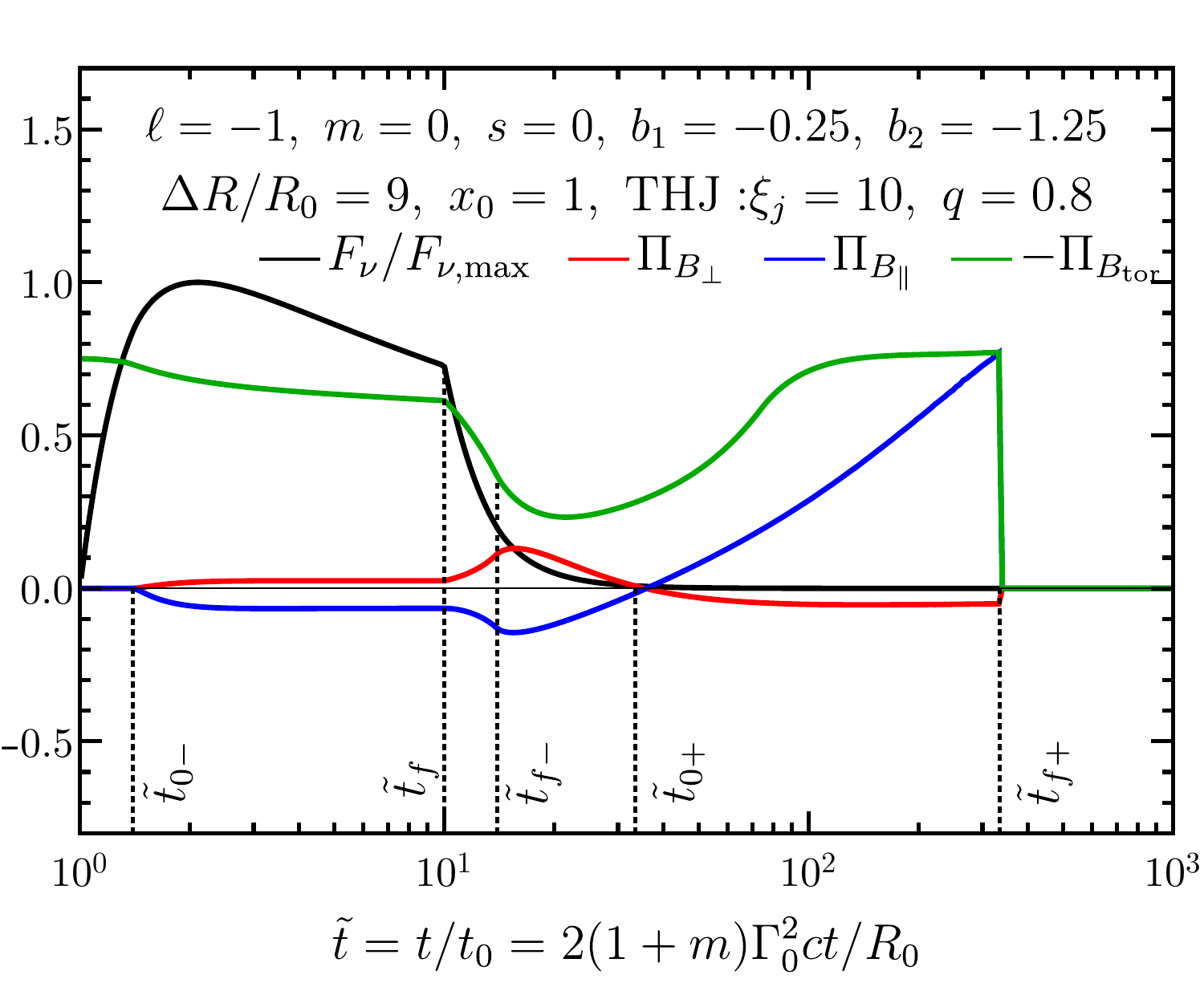}\quad\quad
    \includegraphics[width=0.48\textwidth]{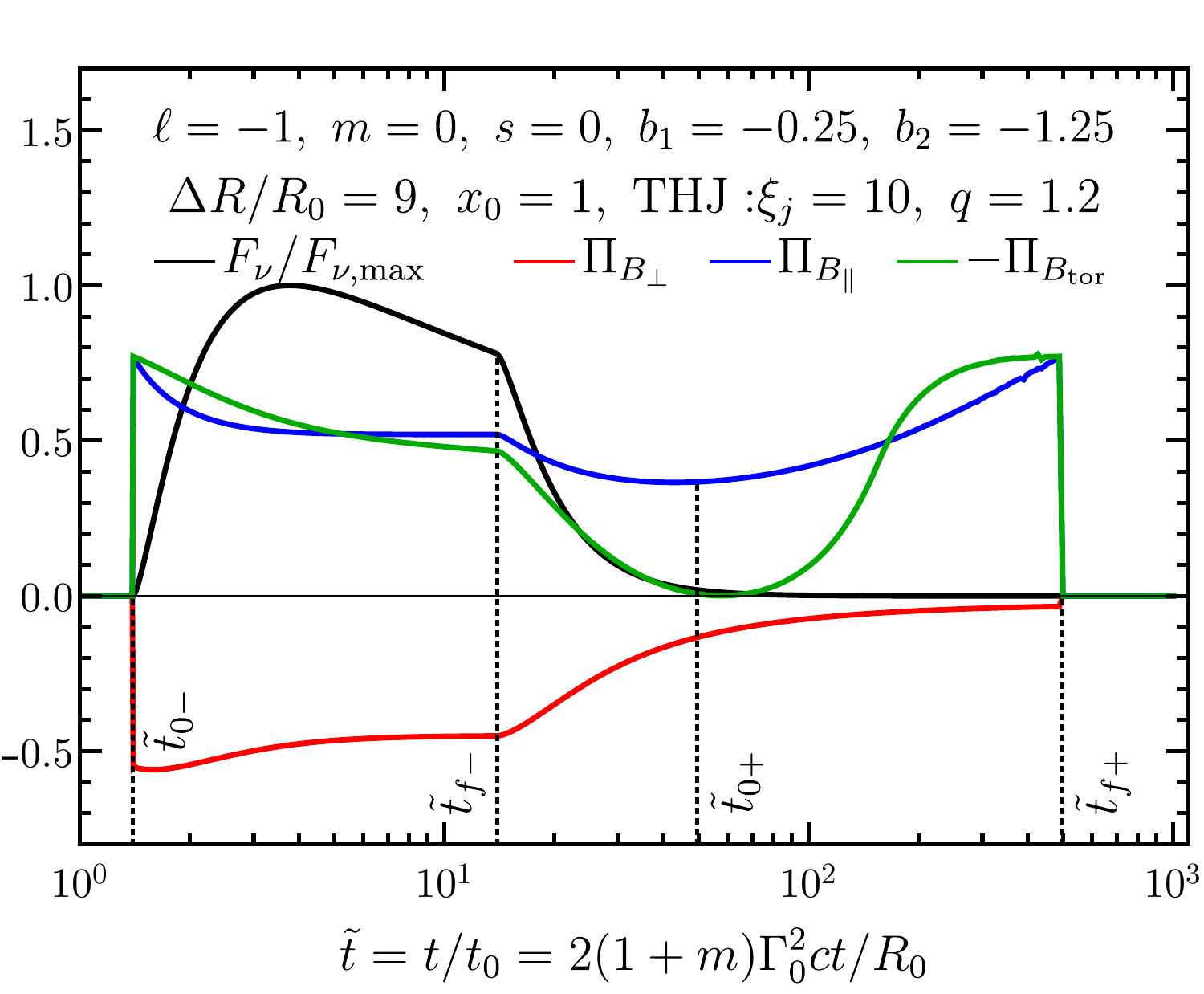}
    \caption{Pulse profile and polarization temporal evolution over a single pulse shown for different magnetic field configurations. 
    The outflow dynamics are that of a KED flow with a top-hat jet geometry with $\xi_j=(\Gamma\theta_j)^2=100$ and $\Delta R/R_0=1$ (top) and 
    $\xi_j=(\Gamma\theta_j)^2=10$ and $\Delta R/R_0=9$ (bottom), both shown for $q=\theta_{\rm obs}/\theta_j=0.8$ (left; on-beam emission) 
    and $q=1.2$ (right; off-beam emission). The critical timescales corresponding to the EATS shown in Fig.~\ref{fig:EATS} 
    are shown with vertical dotted lines. The angular timescales $\tilde t_{0c}$ and $\tilde t_{fc}=\hat R_f\tilde t_{0c}$ correspond to the instant when 
    the EATS reaches the center of the jet for $R=R_0$ and $R=R_f$, respectively. These two angular timescales are relevant for polarization curves 
    for $B_{\rm tor}$ in the top panel. See Table~\ref{tab:symbols} and caption of Fig.~\ref{fig:Sph-shell-pulse} for definition of various symbols.
    }
    \label{fig:time-pulse-pol}
\end{figure*}

\begin{figure*}
    \centering
    \includegraphics[width=0.48\textwidth]{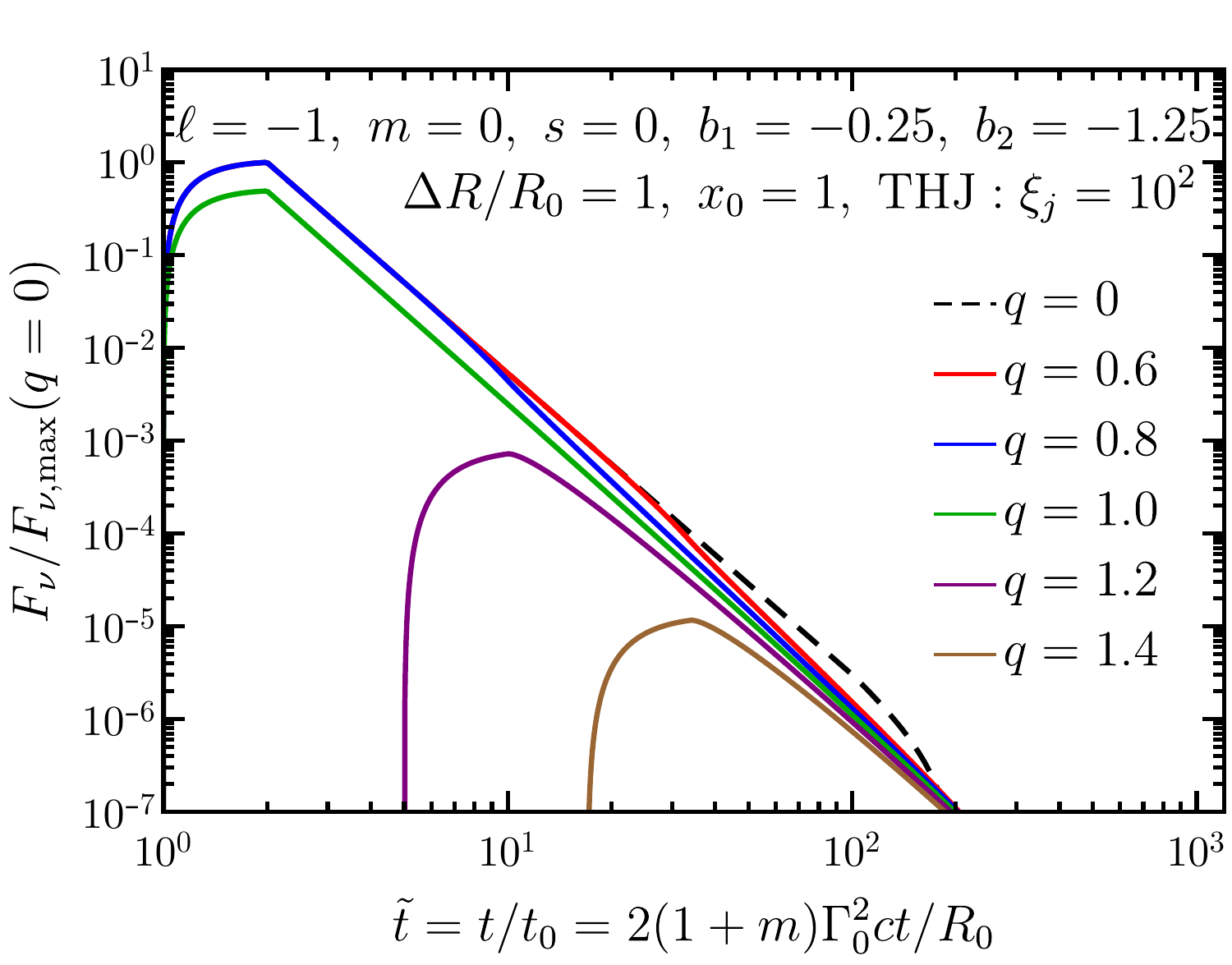}\quad\quad
    \includegraphics[width=0.48\textwidth]{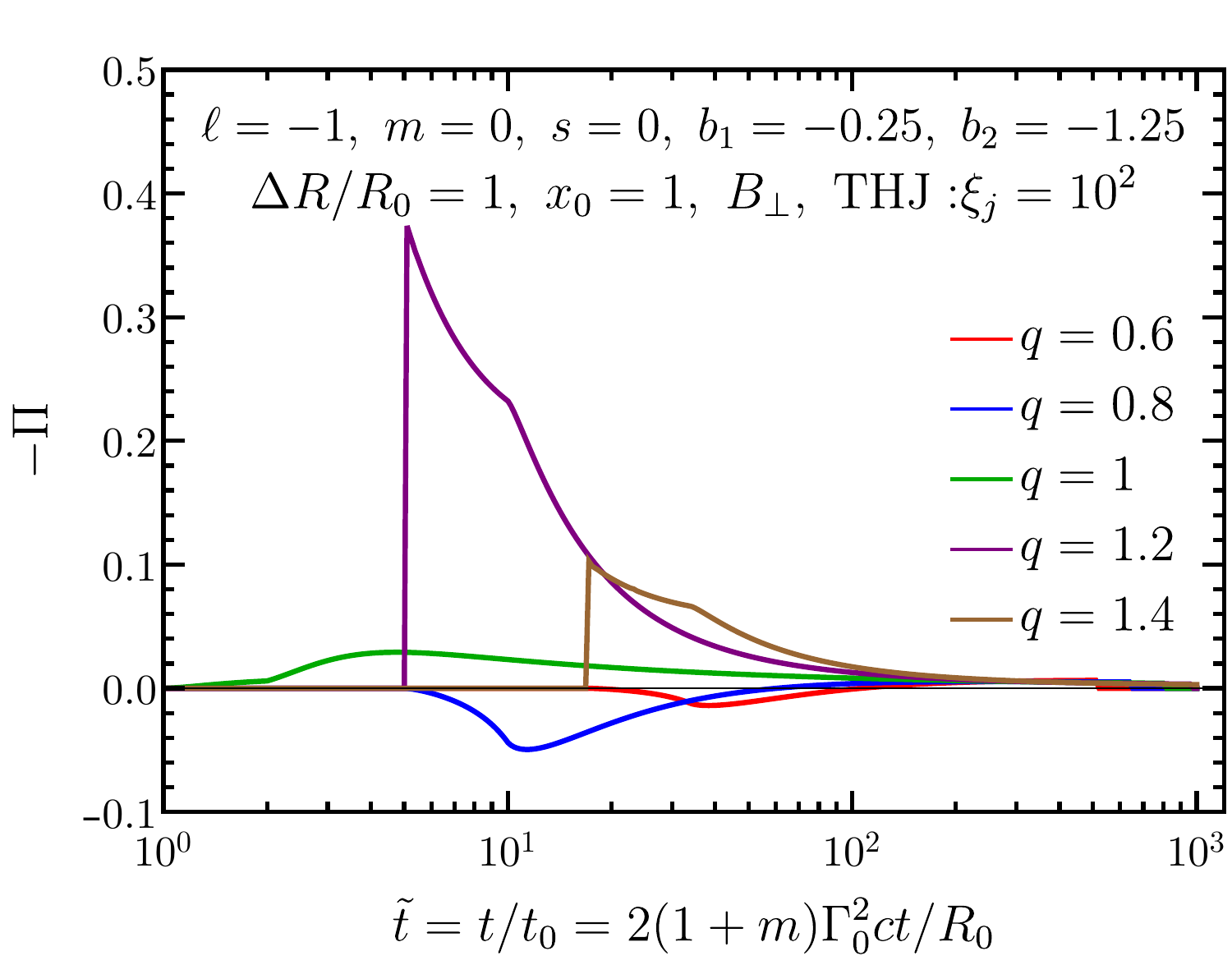}
    \includegraphics[width=0.48\textwidth]{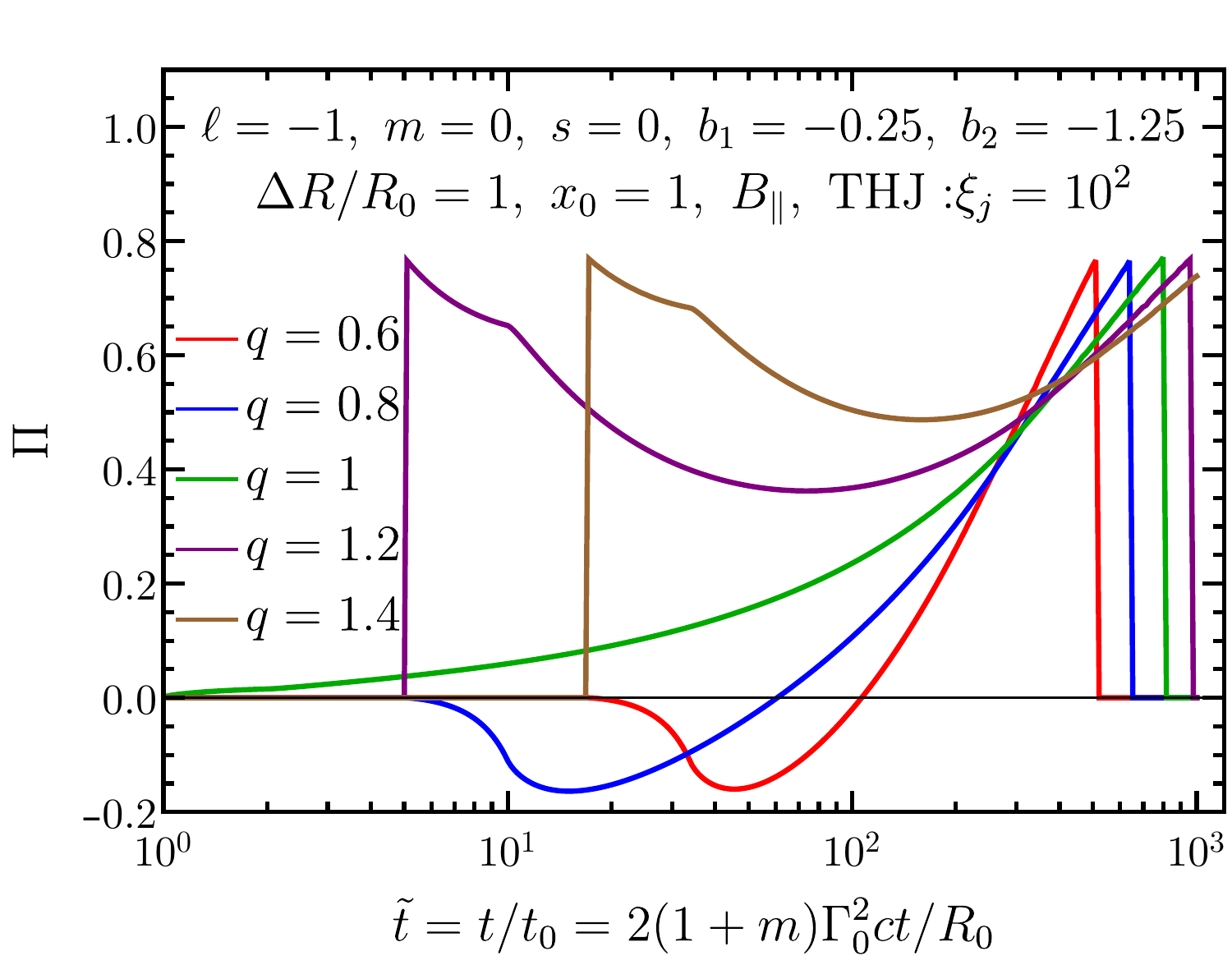}\quad\quad
    \includegraphics[width=0.48\textwidth]{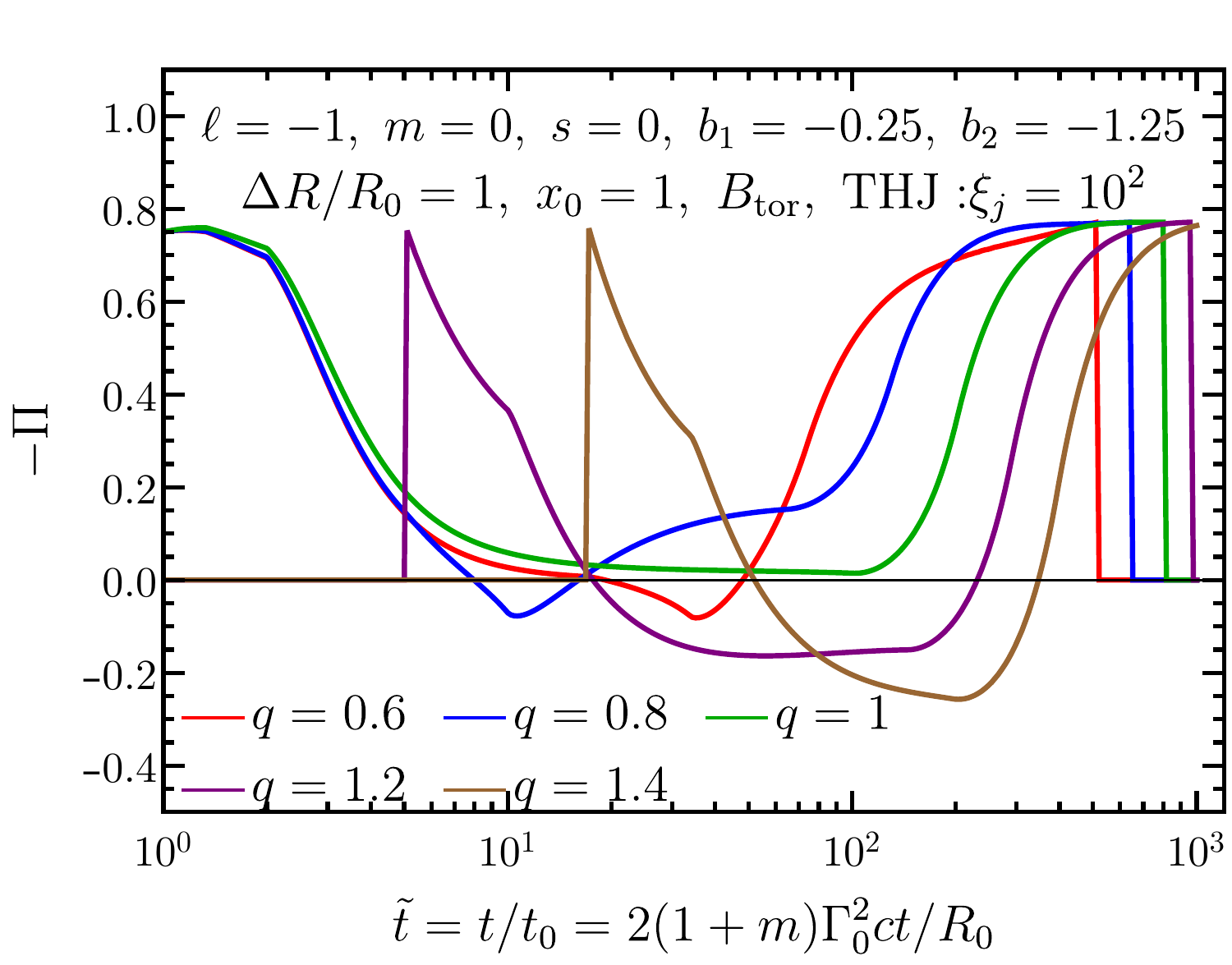}
    \caption{Pulse profile (top-left) and temporal evolution of polarization ($\Pi$) over a single pulse shown for different magnetic field 
    configurations and for different viewing angles $q=\theta_{\rm obs}/\theta_j$. The outflow profile is that of a KED flow with geometry 
    of a top-hat jet (THJ) with $\xi_j=(\Gamma\theta_j)^2=100$. Similar results are obtained for a PFD flow. See Table~\ref{tab:symbols} 
    and caption of Fig.~\ref{fig:Sph-shell-pulse} for definition of various symbols. }
    \label{fig:THJ-Bperp-pulse-pol-diff-q-diff-x0}
\end{figure*}

%%%%%%%%%%%%%%%%%%%%%%%%%%%%%%%%%%%%%%%%%%%%%%%%%%%%%%%%%%%%%%%%%%%%%%
\subsection{Polarization from a top-hat jet}\label{sec:top-hat-jet-pol}
%%%%%%%%%%%%%%%%%%%%%%%%%%%%%%%%%%%%%%%%%%%%%%%%%%%%%%%%%%%%%%%%%%%%%%

For a top-hat jet the emissivity vanishes for $\theta>\theta_j$, where $\theta_j$ is the jet half-opening angle, such that 
$L'_{\nu'}\to L'_{\nu'}H(\xi_{0,j}-\xi_0)$ where $H(x)$ is the Heaviside function, $\xi_{0,j}=(\Gamma_0\theta_j)^2$, and 
$\xi_0 = (\Gamma_0\theta)^2$. The polar angle $\theta$ can be related to $\tilde\theta$ and $\tilde\varphi$ using the 
general relation 
\begin{equation}
    \cos\theta=\mu=\mu_{\rm obs}\tilde\mu-\cos\tilde\varphi\sqrt{(1-\tilde\mu^2)(1-\mu_{\rm obs}^2)}\,,
\end{equation}
with $\tilde\mu=\cos\tilde\theta$ and $\mu_{\rm obs}=\cos\theta_{\rm obs}$. For an ultra-relativistic flow, this relation simplifies to 
\begin{equation}
    \xi_0=\tilde\xi_0+q^2\xi_{0,j}+2q\cos\tilde\varphi\sqrt{\tilde\xi_0\xi_{0,j}}\,,
\end{equation}
where we define $q\equiv\theta_{\rm obs}/\theta_j$ and where the temporal and radial dependence of 
$\tilde\xi_0 = (\Gamma_0\tilde\theta)^2 = (\tilde t\hat R^{-1}-\hat R^m)/(1+m)$ 
is obtained from the EATS condition. 

The top-hat jet case presents significant differences from the spherical flow geometry for different viewing angles $q$. Therefore, we 
briefly establish the appropriate terminology here. The observer is said to be `on-beam' when emission is received within the $1/\Gamma$ 
beaming cone of the emitting material. For a spherical shell this is always the case and the observer is on-beam regardless of the 
viewing angle. In the case of a top-hat jet, the observer receives on-beam emission only when 
$q\lesssim1+\xi_j^{-1/2}\Leftrightarrow\theta_{\rm obs}\lesssim\theta_j+1/\Gamma$. Otherwise, the observer is `off-beam' when the 
emission is received from outside of the $1/\Gamma$ beaming cone of the emitting material. This situation arises when 
$q>1+\xi_j^{-1/2}\Leftrightarrow\theta_{\rm obs}>\theta_j+1/\Gamma$. For an on-beam observer the pulse profile should be similar to that 
obtained for a spherical flow. More often the terms `on-axis' and `off-axis' are used in the literature when discussing top-hat jets, 
where these are synonymous with on-beam and off-beam cases. Care should be taken when describing structured jets, in which both the 
emissivity and bulk-$\Gamma$ decline with increasing $\theta$ outside of a quasi-uniform core. In this case, the observer can be off-axis 
(now with viewing angle outside of the core) and still receive on-axis or on-beam emission since there's always material that could emit 
along the LOS. If the gradient in emissivity and/or bulk-$\Gamma$ is sufficiently steep, the emission is dominated by brighter regions 
at angles $\theta<\theta_{\rm obs}$, in which case the observer would only receive off-beam emission.

% When $q=0\Leftrightarrow\theta_{\rm obs}=0$, the observer is `on-axis', just like 
% in the spherical case, and always receives `on-beam' emission, i.e. the observer's LOS is always located within the beaming cone of the 
% emitter, such that $\tilde\theta\leq1/\Gamma$. When $q>0\Leftrightarrow\theta_{\rm obs}>0$, the observer is `off-axis' where it can 
% receive either on-beam emission, when $q\lesssim1+\xi_j^{-1/2}\Leftrightarrow\theta_{\rm obs}\lesssim\theta_j+1/\Gamma$, or `off-beam' emission, when $q>1+\xi_j^{-1/2}\Leftrightarrow\theta_{\rm obs}>\theta_j+1/\Gamma$ and the LOS 
% is outside the beaming cone of the emitter at $\theta=\theta_j$. For an observer receiving on-beam emission the pulse profile should 
% be similar to that obtained for a spherical flow.

In the case of a top-hat jet, there are several critical timescales at which both the pulse profile as well as the polarization curve undergo 
a change, regardless of the fact that the received emission is on-beam or off-beam. These timescales are 
demonstrated using two dimensional (2D) cross-sections out of the 3D EATSs in Fig.~\ref{fig:EATS} for a coasting jet with $\Gamma=\Gamma_0$. 
The top-panel shows the EATSs for an on-beam observer. Similar to the case of a spherical flow discussed earlier, the observer starts receiving emission along 
the LOS at $\tilde t=1$ when the thin shell is at $\hat R\equiv R/R_0 = 1$. Emission along the LOS terminates at $\tilde t=\tilde t_f$ when the 
shell reaches $\hat R=\hat R_f$.  Next, at $\tilde t>\tilde t_f$, the observer continues to receive emission from increasingly larger $\tilde\theta>0$ 
where the emission becomes dominated by that arising from higher latitudes, i.e. when $\tilde\theta > 1/\Gamma$. Any emission arising from 
$\tilde\theta>0$ suffers an angular delay in addition to the radial one, and the arrival time of photons is readily obtained from the EATS condition 
in Eq.~(\ref{eq:EATS2}) that yields
\begin{equation}\label{eq:t-theta-R0}
    \tilde t(\tilde\theta) = \hat R^{1+m}\left[1 + (1+m)\fracb{\tilde\theta}{\theta_j}^2\frac{\xi_{0,j}}{\hat R^m}\right]\,,
\end{equation}
where we made use of the fact that $\tilde\xi_0=(\Gamma_0\tilde\theta)^2=(\tilde\theta/\theta_j)^2\xi_{0,j}$. 
The radial and angular times are equal when
\begin{equation}
    \tilde\theta = \frac{\hat R^{m/2}}{(1+m)^{1/2}}\frac{1}{\Gamma_0}\,,
\end{equation}
which for $m=0$ yields the well known result, $\tilde\theta=1/\Gamma_0$. 
A critical change both in the pulse profile and polarization curve occurs whenever the observer becomes aware of the edge of the jet at both 
$R=R_0$ and $R=R_f$. The arrival time of photons emitted at any $\hat R$ from the edges of the jet, corresponding to the condition 
$\tilde\theta/\theta_j=1\pm q$, is given by
\begin{equation}
    \tilde t_\pm = \hat R^{1+m}[1 + (1+m)(1\pm q)^2\hat R^{-m}\xi_{0,j}]\,,
\end{equation}
where the negative and positive solutions correspond to the nearer and farther edges from the LOS, respectively. For a coasting flow, the case 
shown in Fig.~\ref{fig:EATS}, $m=0$ for which we get
\begin{equation}
    \tilde t_\pm = \hat R[1 + (1\pm q)^2\xi_{0,j}]\hspace{5em}(m=0)\,.
\end{equation}
From here we define the following critical timescales
\begin{align}\label{eq:critical-times}
    & \tilde t_{0-} \equiv \tilde t_-(\hat R=1) = 1 + (1-q)^2\xi_{0,j}\ , \nonumber\\
    & \tilde t_{f-} \equiv \tilde t_-(\hat R=\hat R_f) = \hat R_f[1 + (1-q)^2\xi_{0,j}]\ , \\
    & \tilde t_{0+} \equiv \tilde t_+(\hat R=1) = 1 + (1+q)^2\xi_{0,j}\ , \nonumber\\
    & \tilde t_{f+} \equiv \tilde t_+(\hat R=\hat R_f) = \hat R_f[1 + (1+q)^2\xi_{0,j}]\ . \nonumber
\end{align}
For a more general $m$ we have
\begin{align}\label{eq:critical-times-general-m}
    & \tilde t_{0\pm} =
    1 + (1+m)(1\pm q)^2\xi_{0,j}\ ,
    \\ \nonumber
    & \tilde t_{f\pm} = \tilde{t}_f+ (1+m)(1\pm q)^2\xi_{0,j}\hat{R}_f\ .
\end{align}
Notice, when $q=0$ the observer receives photons from both edges of the jet at the same time, such that $\tilde t_-=\tilde t_+$. 
For $\tilde t > \tilde t_{f+}$ the flux goes to zero since this marks the arrival time of the last photons from the jet.

The bottom panel of Fig.~\ref{fig:EATS} depicts the situation for an off-beam observer with $q\gtrsim1+\xi_j^{-1/2}$, which implies 
that there is no on-beam emission and the observer only receives high-latitude (off-beam) emission. Therefore, the first photons arrive from the nearer 
edge of the jet at $\tilde t=\tilde t_{0-}$ and the last photons emitted at $R=R_0$ arrive from the farther edge at $\tilde t=\tilde t_{0+}$. 
Likewise, photons from the nearer and farther edges arrive at $\tilde t=\tilde t_{f-}$ and $\tilde t=\tilde t_{f+}$, respectively, 
when the shell is at $R = R_f$.

The timescales at which the observer becomes aware of the edge of the jet and starts noticing a deficit in flux has important implications for 
the polarization. We demonstrate this schematically for two magnetic field configurations, $B_\perp$ and $B_{\rm tor}$, in Fig.~\ref{fig:Pol-Map-Bperp-Btor}, 
again for a uniform coasting flow ($m=0$). 
The expectation for the change in polarization across these timescales is different for the two field configurations. The top panel of the figure 
shows the polarization map for the $B_\perp$ field on the surface of the outflow for an on-beam observer. The blue 
and green shaded regions show the area of the flow that are predominantly polarized in the direction along the line connecting the LOS and the 
jet symmetry axis ($\Pi<0$) and that are polarized in the transverse direction ($\Pi>0$), respectively. Due to the 
inherent symmetry around the LOS $\Pi=0$ at all times $\tilde t\leq\tilde t_{0-}$. As soon as the edge of the jet becomes apparent the missing flux from 
$R=R_0$ breaks this symmetry which should yield a net non-zero polarization at $\tilde t > \tilde t_{0-}$. Initially, the green shaded regions dominate 
that yield $\Pi>0$, and as the observer receives emission from larger angles the blue shaded regions start to dominate. This produces $\Pi<0$ and, as a 
result, the PA changes by $90^\circ$. The observer receives the last photon from the 
jet at $\tilde t=\tilde t_{f+}$, after which time the flux goes to zero and naturally so does the polarization. When the observer is off-beam 
symmetry is broken from the beginning of the pulse as the observer receives the first photons from the 
nearer edge of the jet. In this case, the polarization is expected to be maximal  at the start of the pulse at $\tilde t=\tilde t_{0-}$ 
(with $\Pi=\Pi_{\rm max}$ for $B_{\rm tor}$ or $B_\parallel$ but $\Pi<\Pi_{\rm max}$ for $B_\perp$, as the flux comes from a single point of the 
jet without any canceling between different parts) followed by a decline as increasingly larger area of the outflow comes into view. In addition, since 
in this case only the blue shaded region remains dominant, the PA remains constant with $\Pi<0$ at all times.

The bottom panel of Fig.~\ref{fig:Pol-Map-Bperp-Btor} shows the polarization map for the $B_{\rm tor}$ configuration\footnote{In \citet{Granot-03} 
$\theta_p$ is measured from the direction transverse to the line connecting the jet symmetry axis and the LOS, the same is adopted in the present 
work, however, in \citet{Granot-Taylor-05} $\theta_p$ is measured from the line connecting the jet axis and LOS instead. The former parameterization 
would yield $\Pi<0$ when integrated over the pulse for the $B_{\rm tor}$ case, leading to an opposite sign from what's shown in \citet{Gill-Granot-20} 
who adopted the latter parameterization for $\theta_p$ only for the $B_{\rm tor}$ case.}, 
again for an on-beam (left-panel) and off-beam (right-panel) observer. However, in contrast to the $B_\perp$ or even 
$B_\parallel$ case, since the field is 
ordered the polarization is maximal right from the reception of the first photon. As was seen in the locally ordered field case, the polarization 
would start to decline rapidly when high-latitude emission becomes dominant. As discussed below, the PA changes by $90^\circ$ twice in this case 
depending on which (blue/green) region dominates at any given time. 

While the polarization maps provide a qualitative understanding of how the polarization should evolve over the pulse, a more quantitative picture is 
shown in Fig.~\ref{fig:time-pulse-pol} where we show the polarization curves for the different field configurations and viewing angles\footnote{To calculate the pulse profile and polarization we use the general expression valid for a spherical shell (given by 
Eq.~\ref{eq:general-pol}), but use the Heaviside function to suppress the emissivity for $\xi_0>\xi_{0,j}$. While this technique is more suitable 
for a structured flow, or a smooth top-hat jet as treated in later sections, it may be slightly computationally expensive for a top-hat jet due to 
the sharp edge in emissivity. More suitable expressions for calculating the lightcurve and polarization for a top-hat jet are given in 
\citet{Ghisellini-Lazzati-99} for $q<1$ and in \citet{Granot-03} for a general $q$.}. In the 
figure the critical timescales are presented to allow comparison with the polarization maps. Two cases are highlighted, one with $\xi_j=10^2$ and 
$\Delta R/R_0=1$ (top panel) and the other with $\xi_j=10$ and $\Delta R/R_0=\blue{9}$ (bottom panel), where they have interesting differences. 
In both the left and right panels, corresponding to an 
on-beam and off-beam observer, respectively, the polarization behavior matches the expectation from the polarization maps. 
In the top panel, the polarization curve for the $B_{\rm tor}$ case undergoes a $90^\circ$ change in PA twice over the duration of the pulse in 
both the on-beam and off-beam cases, whereas both the $B_\perp$ and $B_\parallel$ fields only show a single $90^\circ$ change in the PA in the on-beam 
case; no such PA change occurs in the off-beam case. The polarization abruptly vanishes at $\tilde t=\tilde t_{f+}$ 
when the flux vanishes. A noteworthy point is that some of the interesting behavior, e.g. the change in PA, occurs deep in the tail of the pulse profile 
when high-latitude emission dominates. At this point, the decline in flux is so severe that it renders detecting these changes in $\Pi$ very challenging.
This is less of an issue for the scenario shown in the bottom panel. In this case the symmetry breaking timescale $\tilde t_{0-}<\tilde t_f$, and 
therefore, $\vert\Pi\vert>0$ can be measured for both the $B_\perp$ and $B_\parallel$ fields near the pulse peak and thereafter. Still, the degree of 
polarization reaches a more appreciable value only during the tail of the pulse in the lightcurve. While for these two fields the change in PA is similar 
to that in the top-panel, the behaviour is different for $B_{\rm tor}$ which shows no PA change at all.

\begin{figure*}
    \centering
    \includegraphics[width=0.48\textwidth]{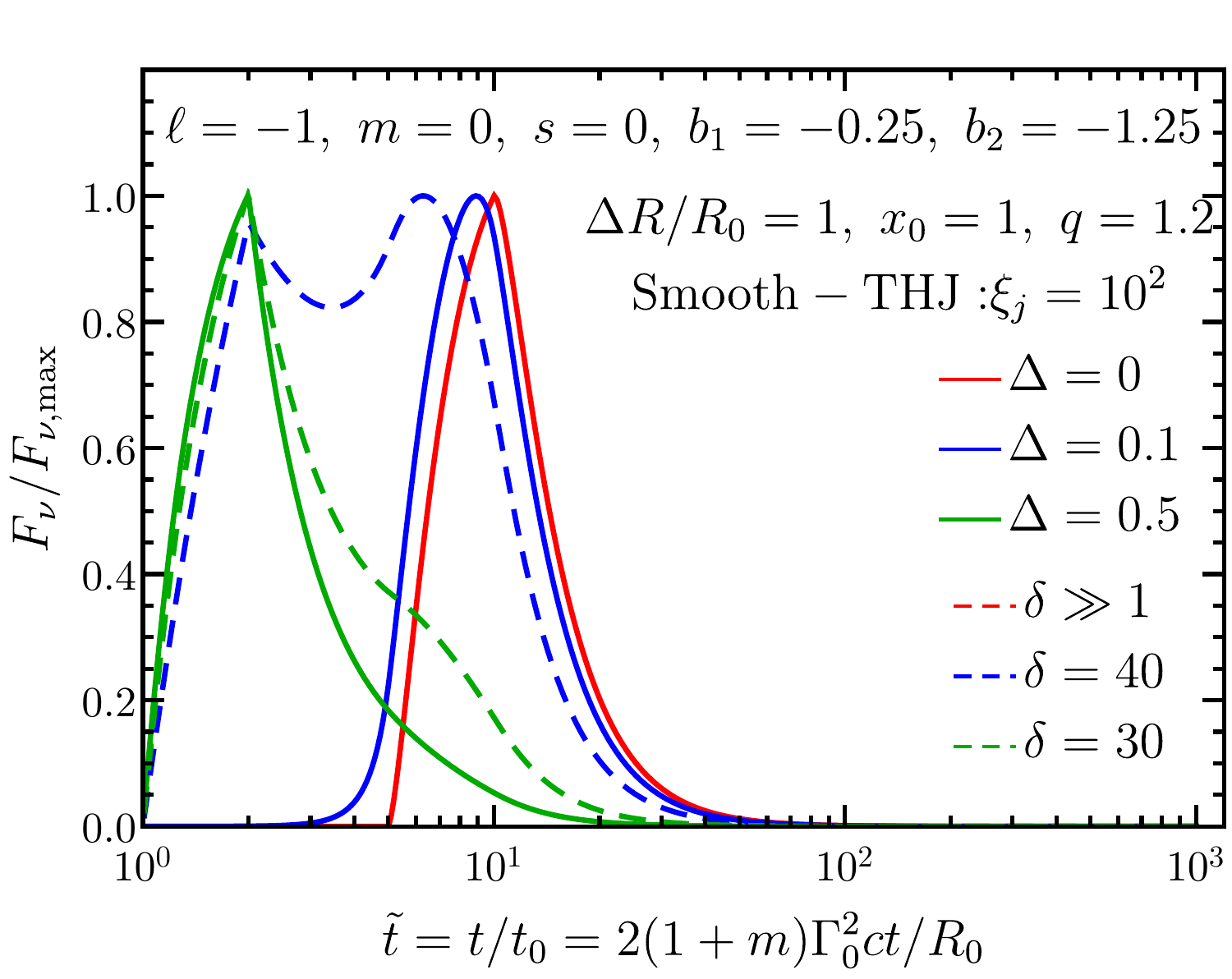}\quad\quad
    \includegraphics[width=0.48\textwidth]{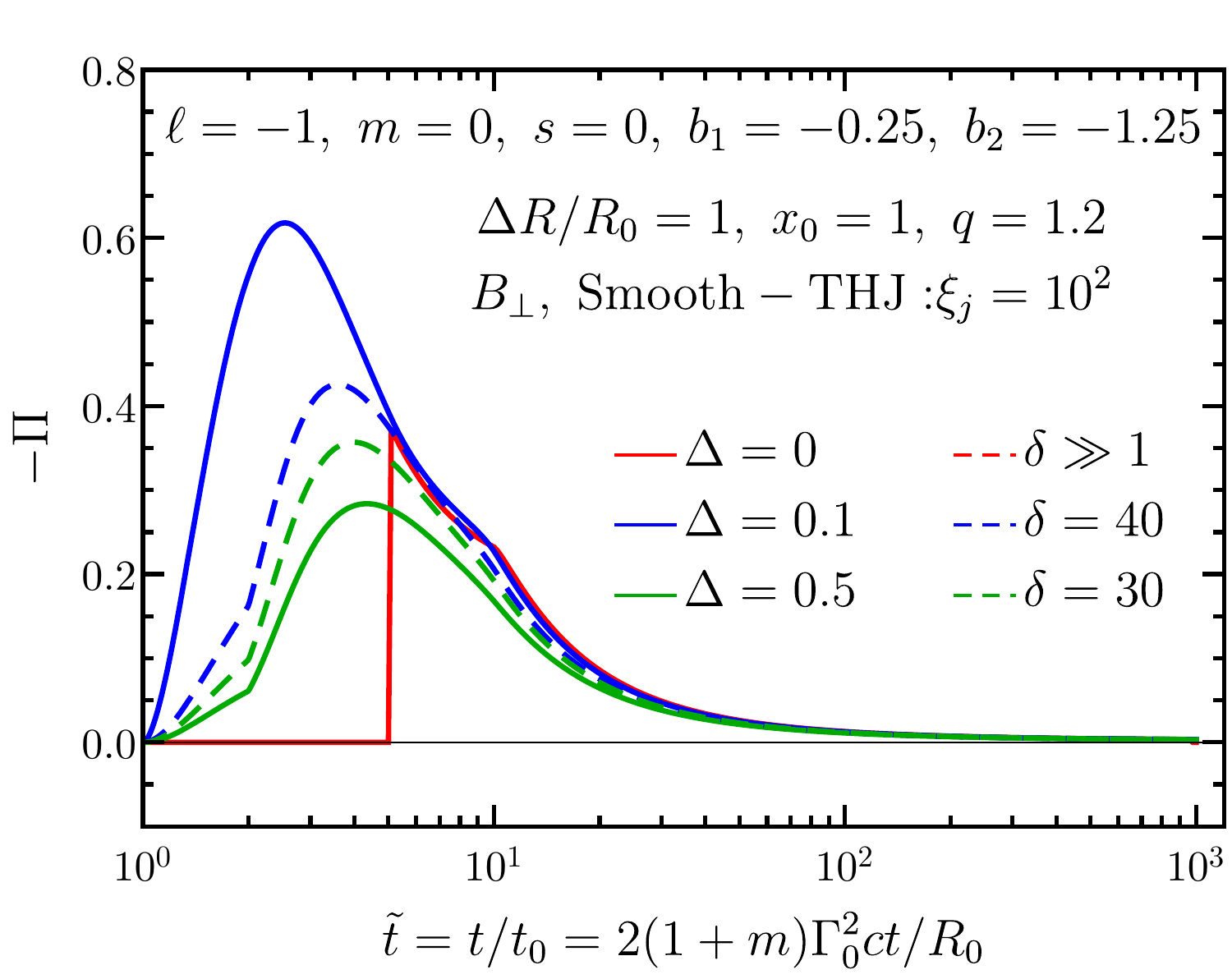}
    \includegraphics[width=0.48\textwidth]{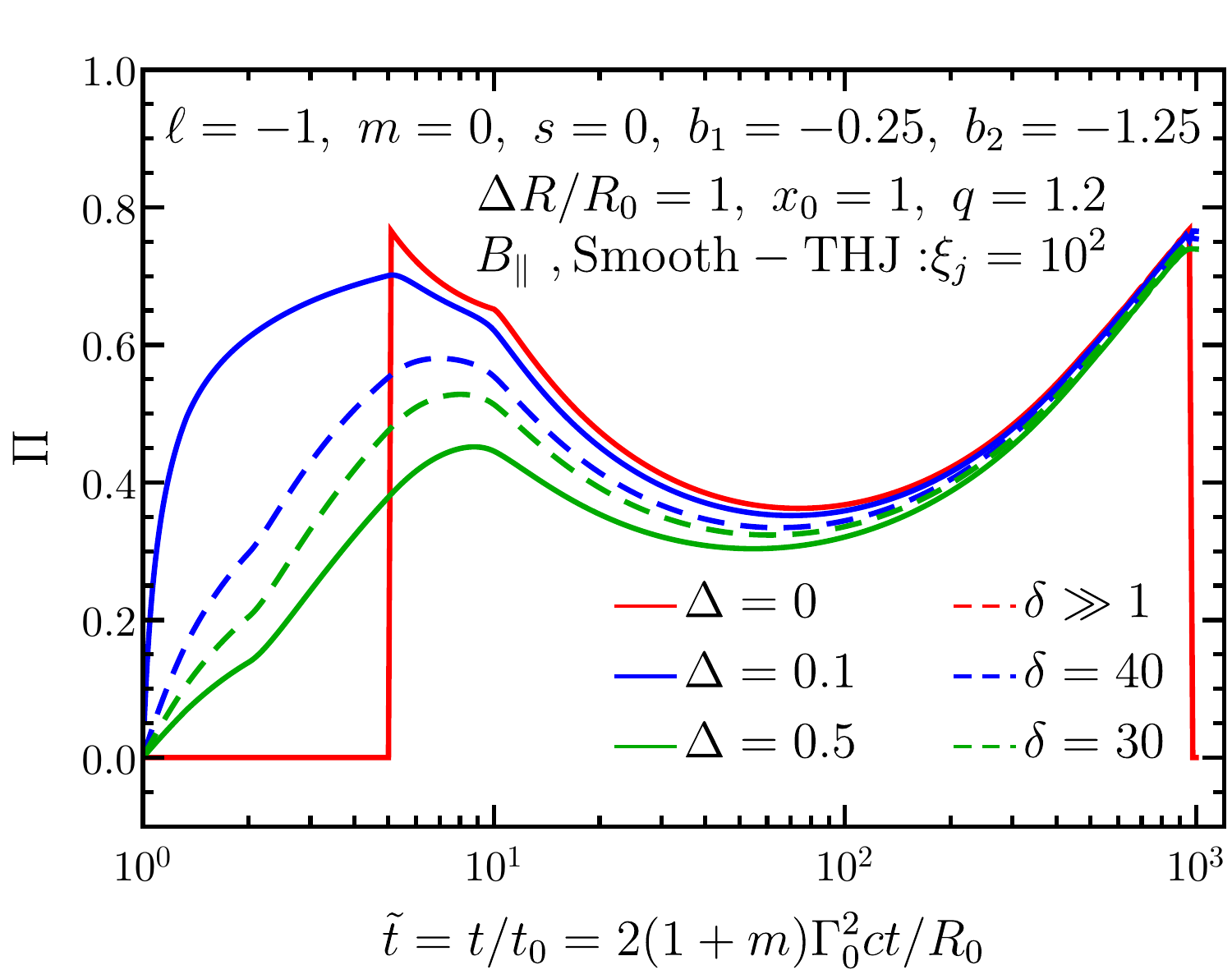}\quad\quad
    \includegraphics[width=0.48\textwidth]{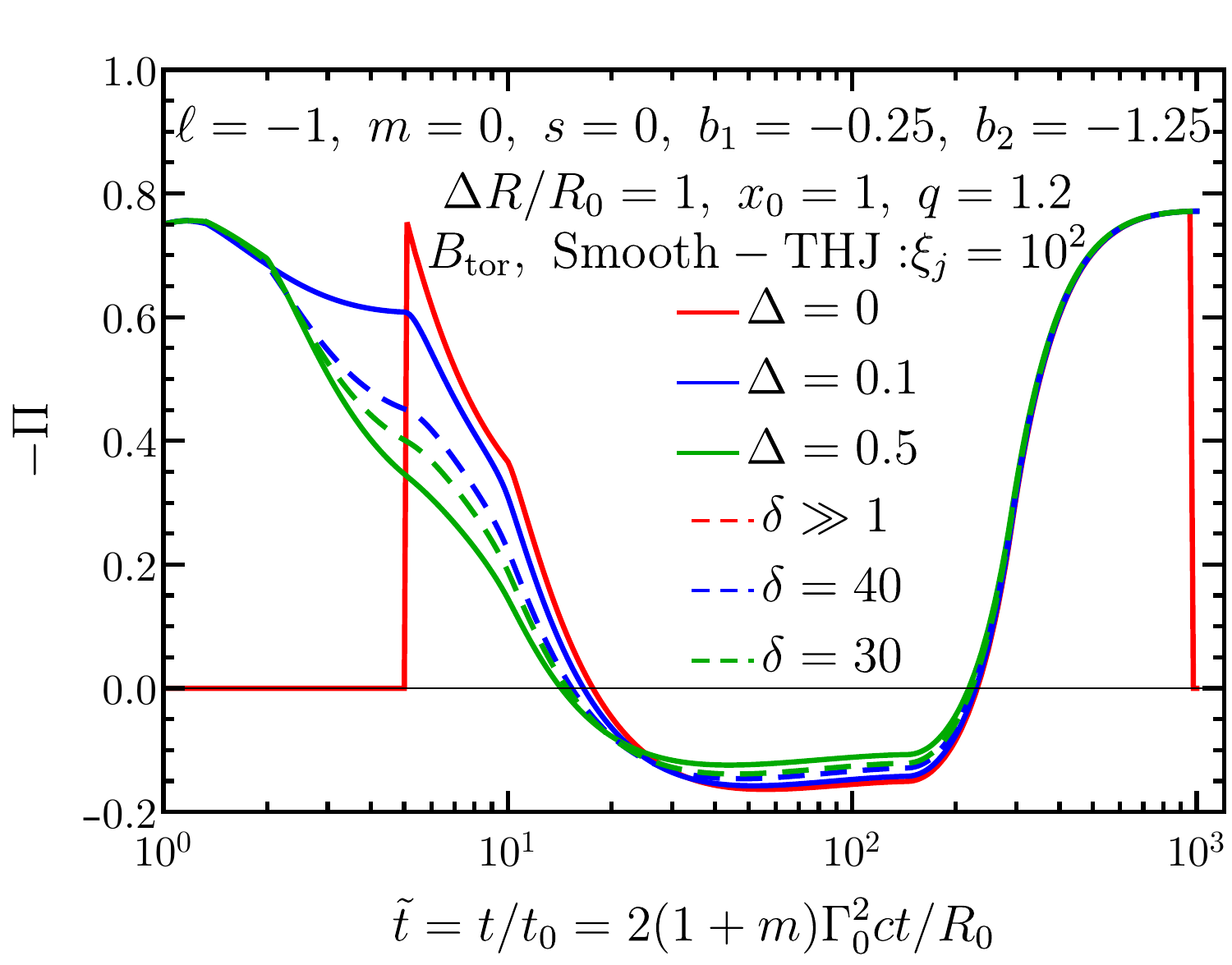}
    \caption{Pulse profile (top-left) and temporal evolution of polarization over a single pulse shown for a smooth top-hat jet (uniform core and 
    decaying profile of $L'_{\nu'}$ for $\theta>\theta_j$) for different smoothing parameter $\Delta$ ($\delta$) for an exponential (power-law decay; 
    see Eq.~\ref{eq:Smooth-THJ-Exponential} \& \ref{eq:Smooth-THJ-PowerLaw}). 
    Polarization curves are shown for random $B_\perp$ (top-right), $B_\parallel$ (bottom-left), and $B_{\rm tor}$ (bottom-right) fields. 
    See Table~\ref{tab:symbols} and caption of Fig.~\ref{fig:Sph-shell-pulse} for definition of various symbols.
    }
    \label{fig:pulse-pol-SMTH-diff-Delta}
\end{figure*}

In the top panel, there are two additional timescales that are relevant for the $B_{\rm tor}$ field. These are shown using a green dotted line in Fig.~\ref{fig:time-pulse-pol}, 
and they correspond to the arrival time of photons from the center of the outflow at $\tilde t=\tilde t_{0c}$ and $\tilde t = \tilde t_{fc}$ when the 
shell is, respectively, at $R=R_0$ and $R=R_f$, where
\begin{align}
    \tilde t_{0c} & = 1 + q^2\xi_{0,j}\hspace{5em}(m=0)\,, \\
    \tilde t_{fc} & = \tilde t_f + q^2\xi_{0,j}\hat R_f\,.
\end{align}
For a more general $m$ we have
\begin{align}
    & \tilde t_{0c} =
    1 + (1+m)q^2\xi_{0,j}\ ,
    \\ \nonumber
    & \tilde t_{fc} = \tilde{t}_f+ (1+m)q^2\xi_{0,j}\hat{R}_f\ .
\end{align}
These times reflect inflexion points across which the temporal behavior of the polarization curve changes. While these can be more clearly seen 
in the top panel, they are not that obvious in the bottom panel due to integration over larger $\Delta R$.

\begin{figure*}
    \centering
    \includegraphics[width=0.48\textwidth]{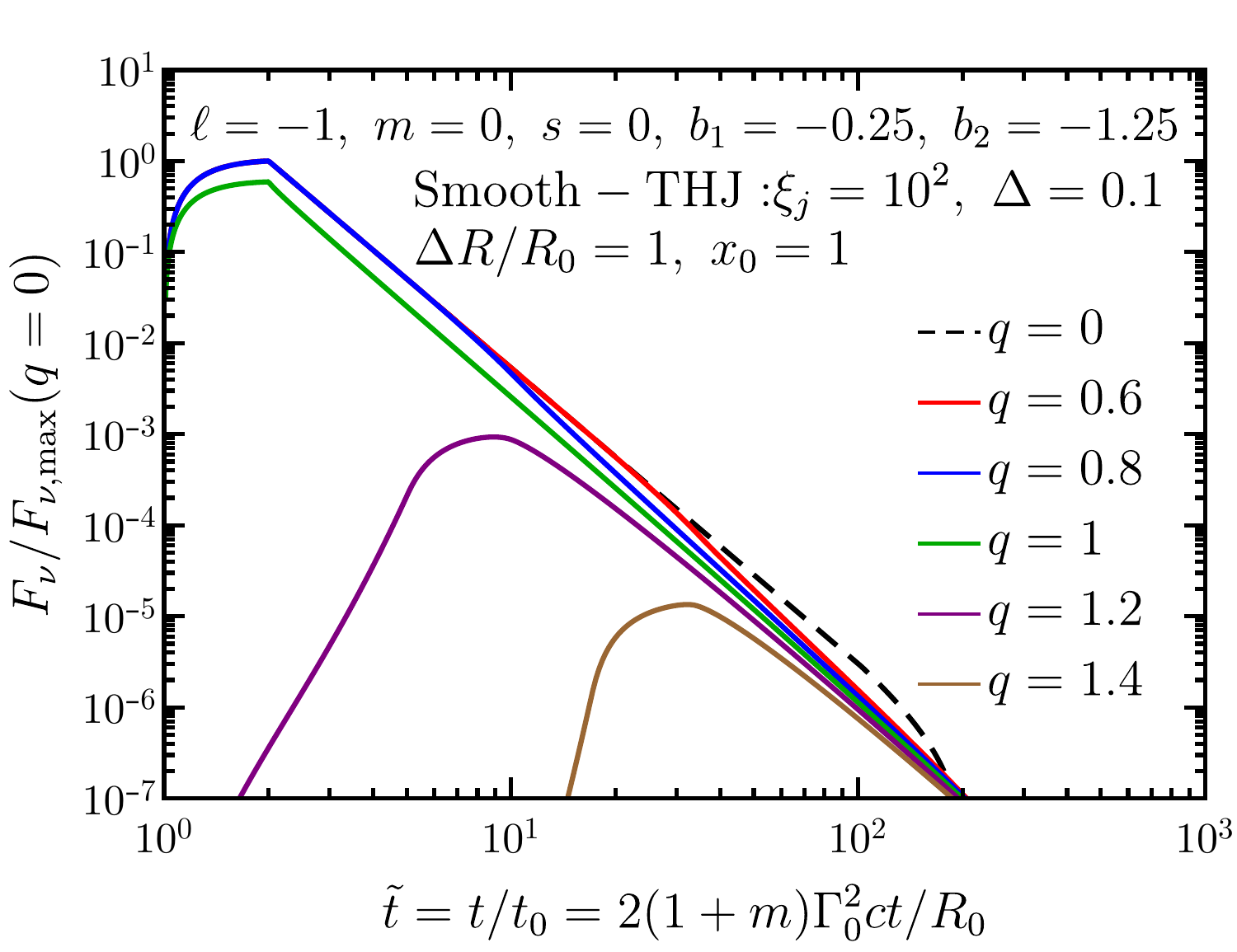}\quad\quad
    \includegraphics[width=0.48\textwidth]{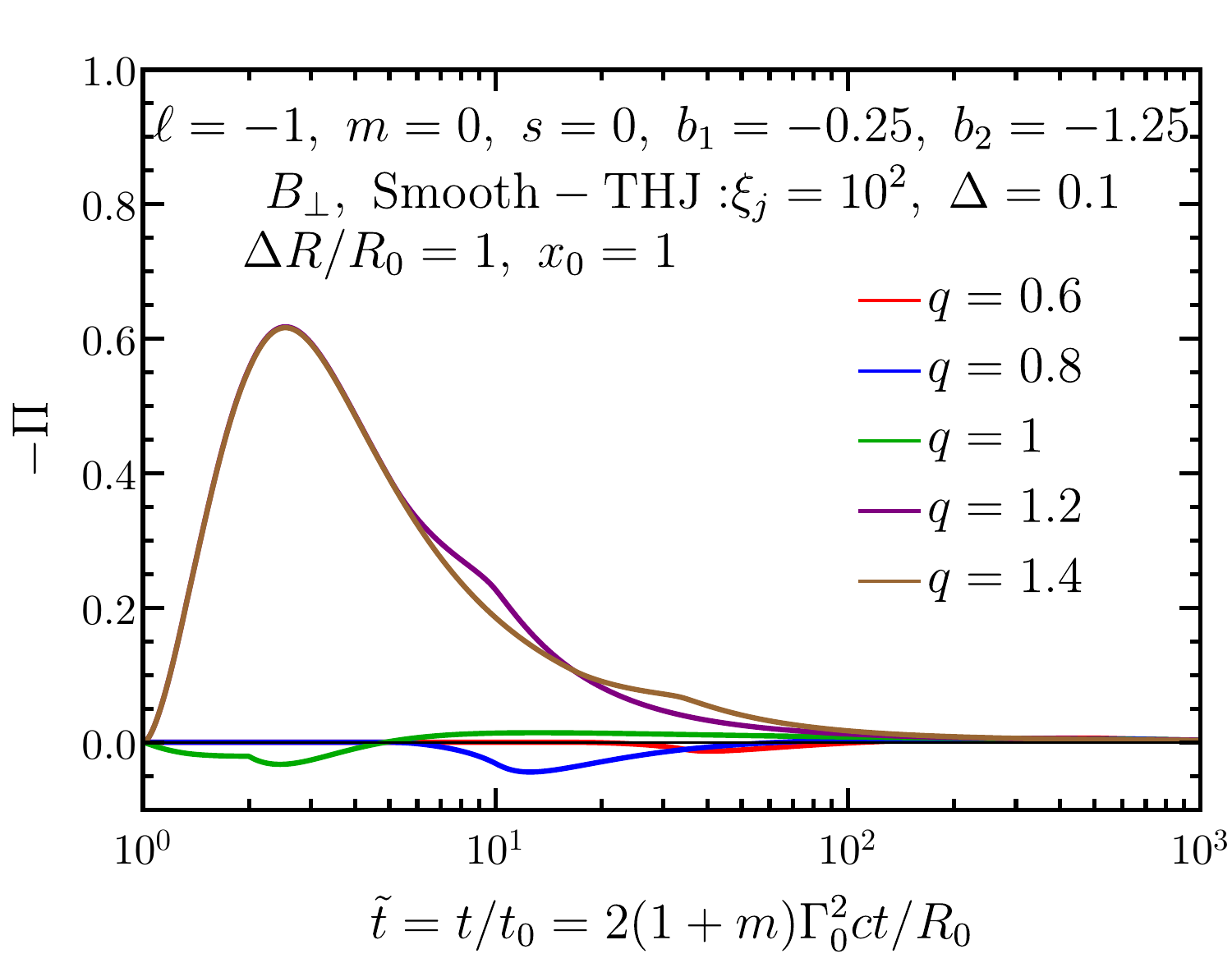}
    \includegraphics[width=0.48\textwidth]{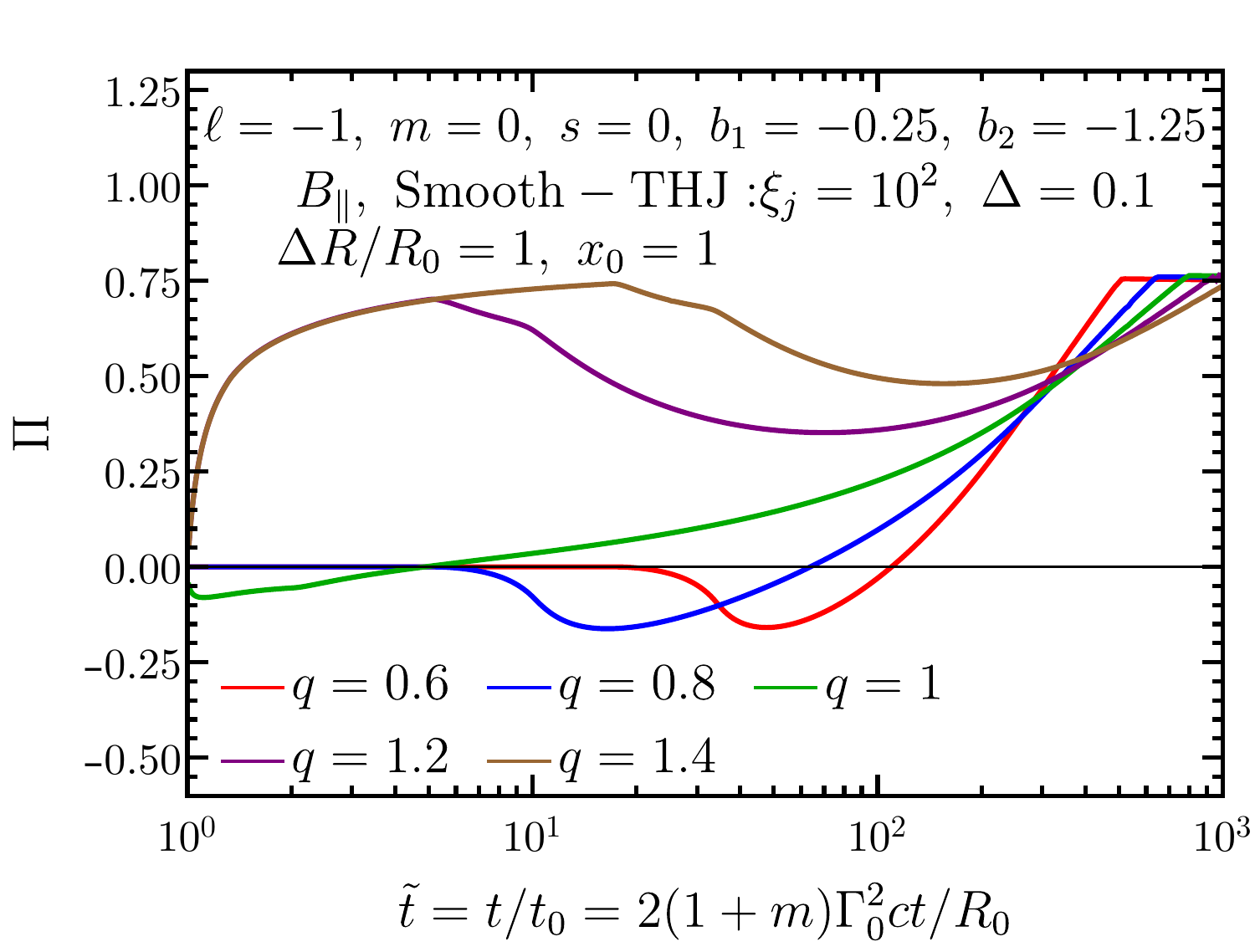}\quad\quad
    \includegraphics[width=0.48\textwidth]{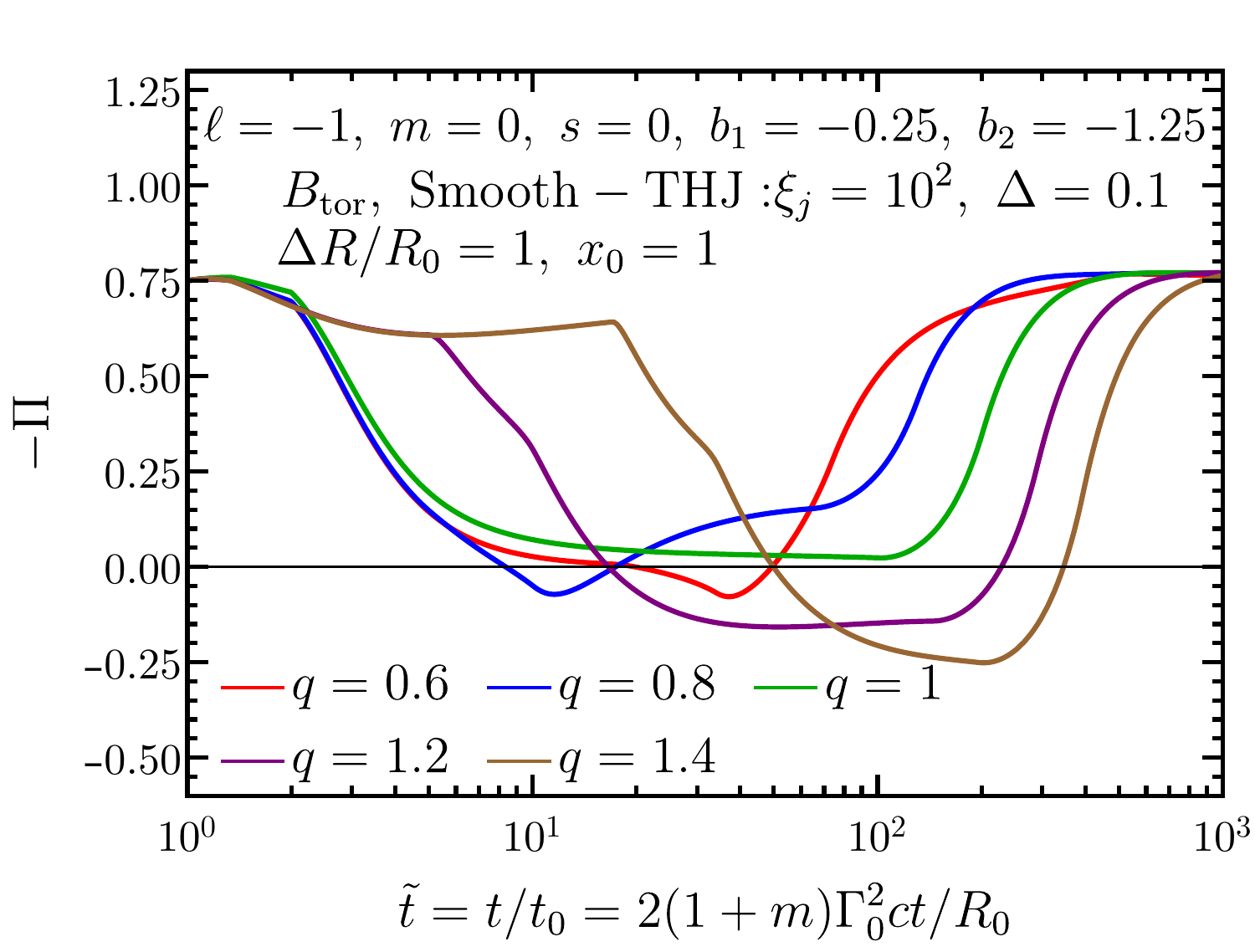}
    \caption{Pulse profile (top-left) and temporal evolution of polarization over a single pulse shown for a smooth top-hat jet (uniform core 
    and exponentially decaying wings in $L_{\nu'}'$) with random $B_\perp$ (top-right), $B_\parallel$ (bottom-left), and $B_{\rm tor}$ (bottom-right) 
    for different $q = \theta_{\rm obs}/\theta_j$. 
    See Table~\ref{tab:symbols} and caption of Fig.~\ref{fig:Sph-shell-pulse} for definition of various symbols.
    }
    \label{fig:pulse-pol-SMTH-diff-q}
\end{figure*}

The polarization `break' obtained at $\tilde t=\tilde t_{0-}$ for the $B_\perp$ and $B_\parallel$ configurations can potentially be used to infer $\theta_j$, 
the half-opening angle of the jet, if $\Gamma$ can be estimated independently from, e.g. high-energy spectral cutoff due to $\gamma\gamma$-annihilation 
\citep[e.g.,][]{Granot+08,Gill-Granot-18a}. In total there are four unknowns, 
namely $\theta_{\rm obs}$, $\theta_j$, $\Gamma$, and the spectral index, where the first three combine to form two variables $q$ and $\xi_j$ that can 
be constrained using polarization while the spectral index is obtained from the observed spectrum. The polarization break time provides one constraint and 
the pulse-integrated polarization, which depends on both $q$ and $\xi_j$, provides another. Of course, in practice this is a challenging exercise since 
it requires an extremely bright GRB that shows a single pulse over which high signal-to-noise measurements for the polarization are obtained.

In the top-left panel of Figure~\ref{fig:THJ-Bperp-pulse-pol-diff-q-diff-x0}, we show the pulse profiles for different viewing angles normalized by 
$F_{\nu,\rm max}$ at $q=0$, the pulse profile for which is shown with a black dashed line. The $q=0$ observer receives 
photons from both edges of the jet at the same time at $\tilde t = \tilde t_{0-} = \tilde t_{0+}$, and the deficit in flux can be seen 
at $\tilde t = 1+(1+m)\xi_{0,j} = 101$ for $\xi_{0,j} = 100$ and $m=0$. When $q<1-\xi_j^{-1/2}$ the pulse profiles overlap with that for $q=0$ until 
$\tilde t=\tilde t_{0-}(q)$ 
due to missing emission. For $q>1-\xi_j^{-1/2}$ the instant at which a deficit in flux occurs moves to smaller times until $q=1$, when the observer's LOS is directly 
along the edge of the jet. In this case, the deficit in flux becomes precisely by half and it is apparent from $\tilde t=1$ since the edge of the jet 
is visible to the observer from the arrival time of the very first photons. The arrival time of the first photons shifts to $\tilde{t}_{0-}>1$ for 
$q>1$ and the flux normalization also declines very rapidly with $q$ since the emission is coming from higher latitudes. The suppression 
in flux is particularly severe for a top-hat jet for $q>1+\xi_j^{-1/2}$ which means that distant GRBs are always observed through (at least nearly) 
on-beam emission, $q\lesssim1+\xi_j^{-1/2}$. In addition, the width of the pulse for off-beam emission also increases in comparison to that for 
on-beam emission, such that 
\begin{equation}
    \frac{\Delta t_{\rm off}}{\Delta t_{\rm on}} = \frac{t_{0-}}{t_0} = \tilde t_{0-} = 1+(1+m)(1-q)^2\xi_{0,j}\,.
\end{equation}

The other panels of Figure~\ref{fig:THJ-Bperp-pulse-pol-diff-q-diff-x0} show the temporal evolution of polarization for different B-field configurations. 
The polarization curves are broadly similar for the $B_\perp$ and $B_\parallel$ configurations, with the only difference being the high level of polarization 
for the latter case due to the field being ordered in the radial direction as compared to it being random in the plane transverse to the radial direction, 
which also reflects in an opposite sign of the polarization (i.e. a $90^\circ$ difference in the PA). 
The feature that is common to polarization curves of both $B_\perp$ and $B_\parallel$ cases is that when $q<1$ the smaller the $q$ the later the occurrence 
of the polarization break and the lower the maximum level of polarization. Therefore, as argued in e.g. \citet{Gill+20} using pulse-integrated polarization, most 
GRBs with such field configurations will not show any discernible levels of polarization, unless the outflow has a $B_{\rm tor}$ field or a locally ordered 
field when emission comes from incoherent patches (or more generally, a field ordered on angular scales $\gtrsim1/\Gamma$). 

In the case of a $B_{\rm tor}$ field, the polarization evolution is completely different from that obtained for the $B_\perp$ and $B_\parallel$ fields. 
Not only the polarization is always maximal at the start of the emission for all $q$ values, it also shows two $90^\circ$ changes in the PA in all 
cases except for $q=1$. Of course, the second flip will be very difficult to detect as it always occurs deep in the tail of the pulse. The abrupt drop in flux 
at $\tilde t = \tilde t_{f+}$ simply reflects the instance when the last photons are received from the flow.

\begin{figure*}
    \centering
    \includegraphics[width=0.48\textwidth]{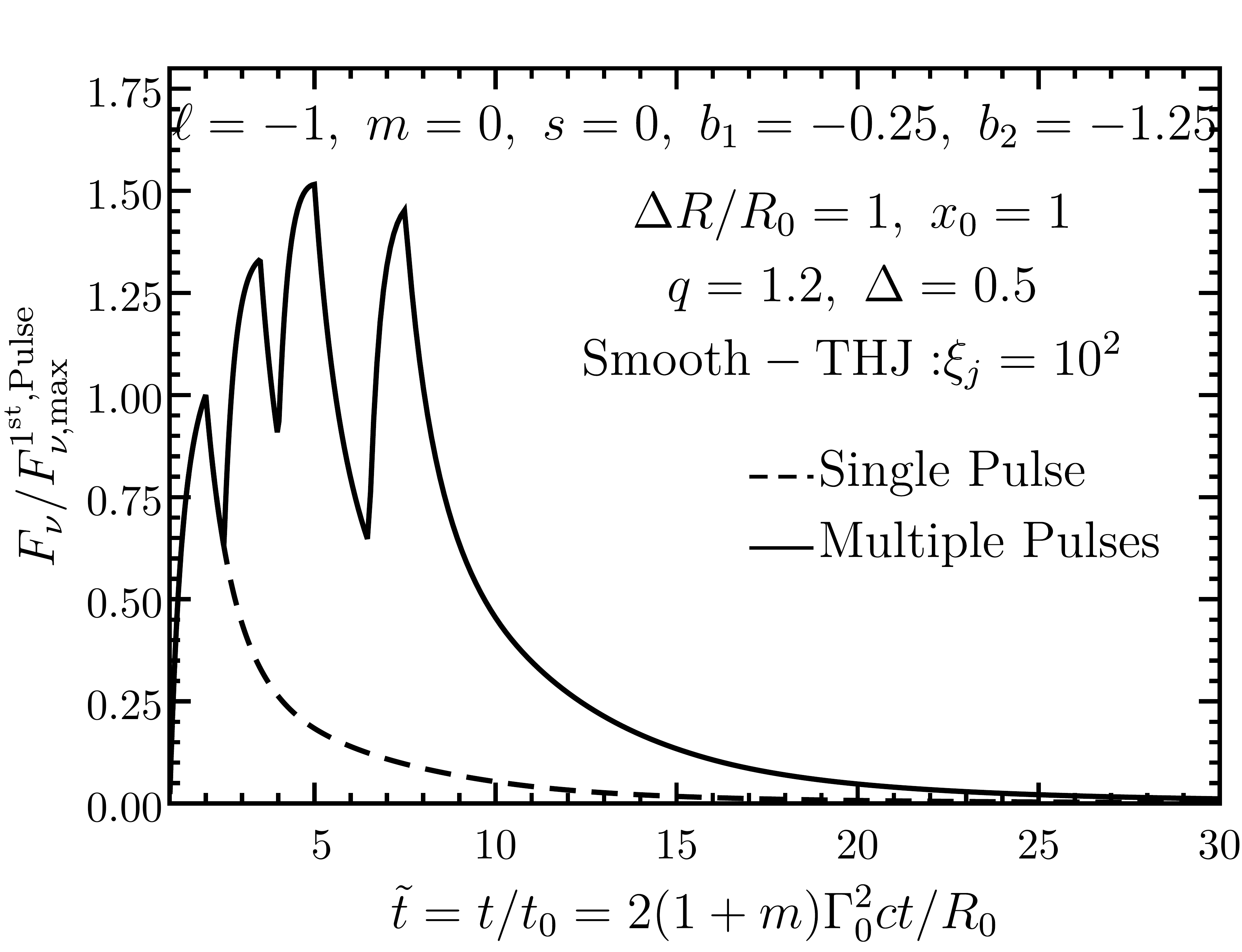}\quad\quad
    \includegraphics[width=0.48\textwidth]{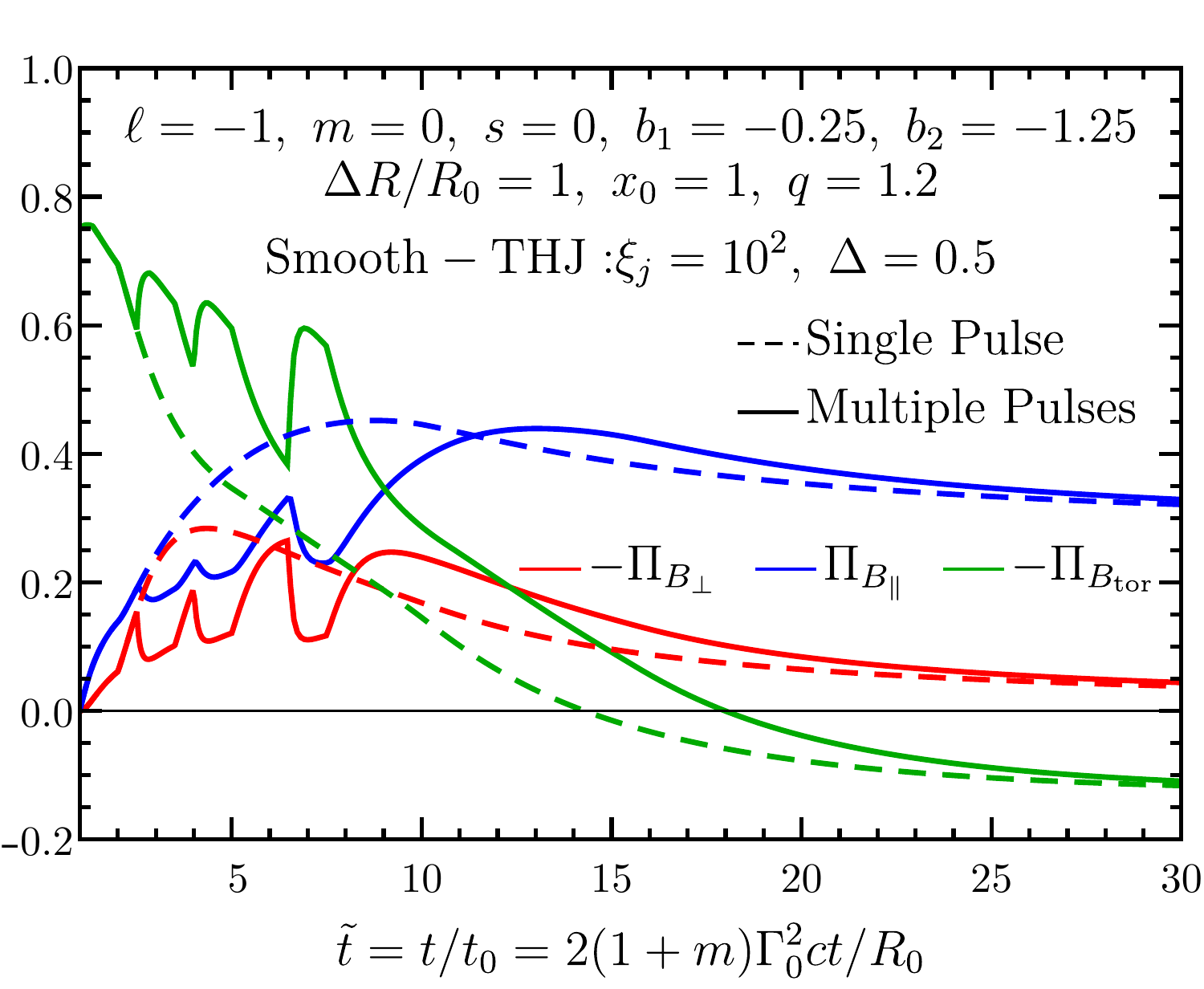}
    \includegraphics[width=0.48\textwidth]{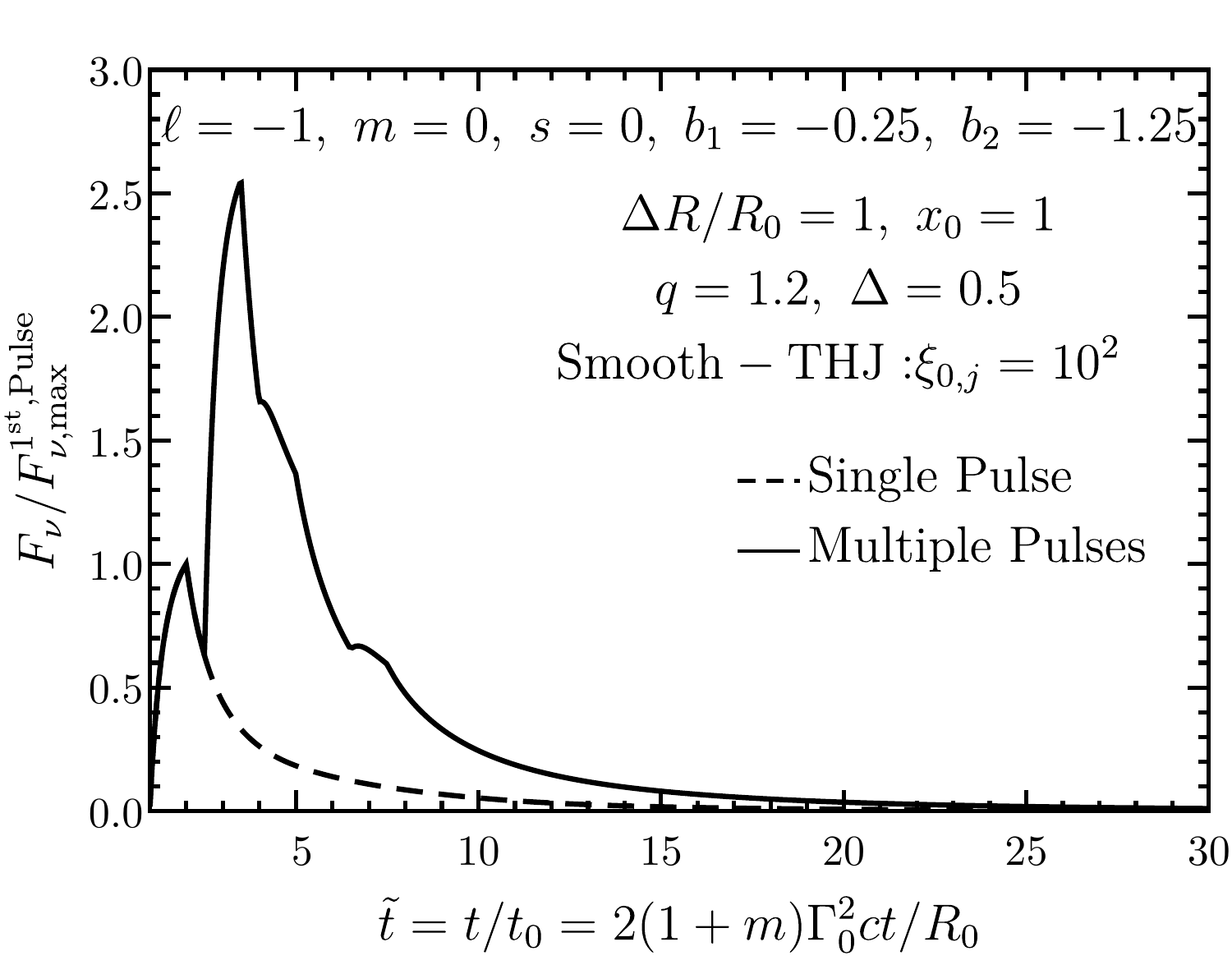}\quad\quad
    \includegraphics[width=0.48\textwidth]{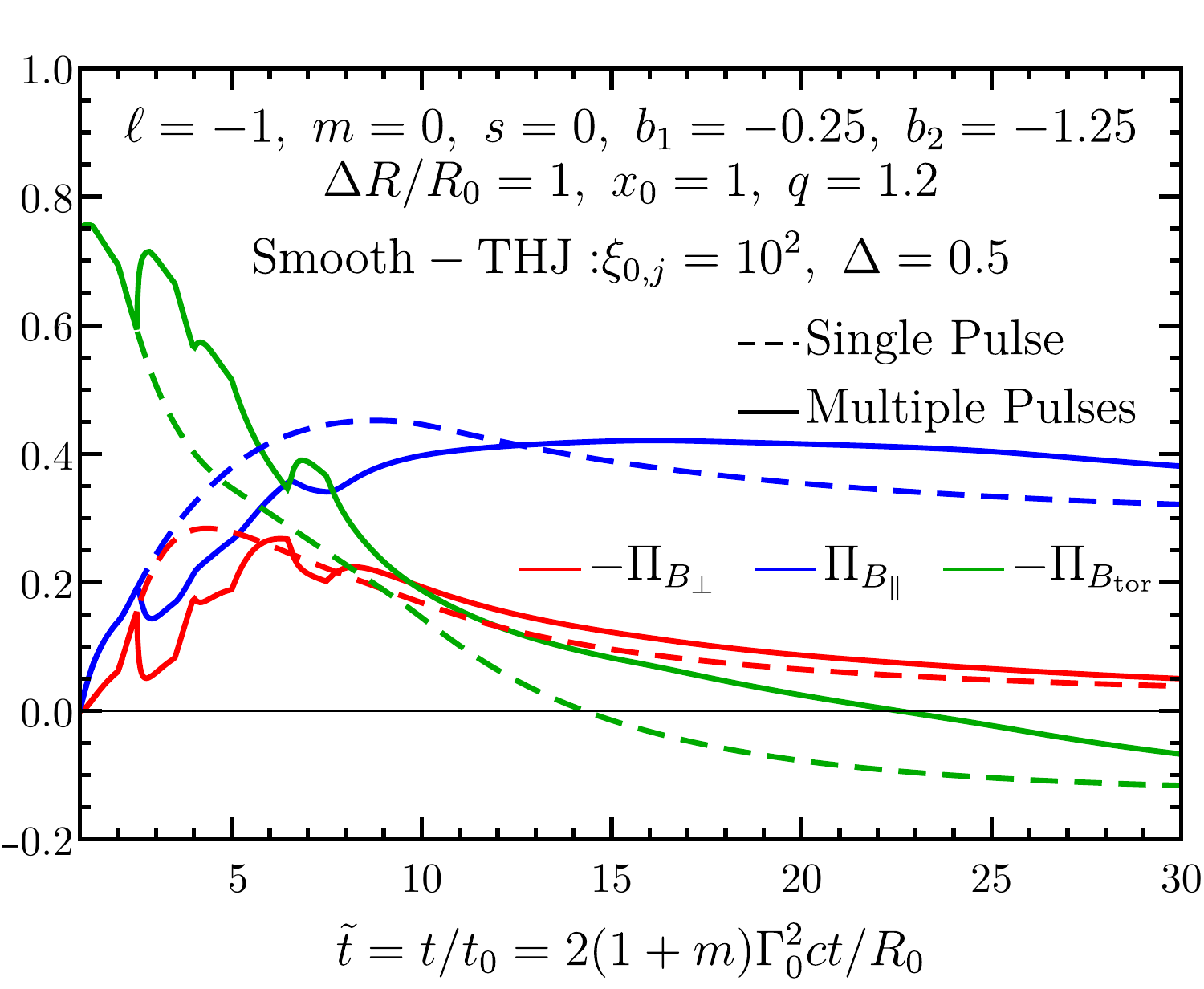}
    \caption{Pulse profile (top-left; normalized by $F_{\nu,\rm max}$ of the first pulse) and temporal evolution of polarization over 
    multiple pulses (top-right) shown for a KED smooth top-hat jet (uniform core 
    and exponentially decaying wings in $L_{\nu'}'$) with different B-field configurations. The pulses are temporally separated 
    due to different ejection times of multiple thin shells from the central engine. In the top-row the bulk-$\Gamma$ of the shells is 
    kept the same whereas in the bottom row shells have different bulk-$\Gamma$ with the $i^{\rm th}$ shell having 
    $\Gamma_{0,i} = \lambda_i\Gamma_0$, and correspondingly $\xi_{j,i}=\lambda_i^2\xi_{0,j}$, with $\lambda_i = 1, 0.8, 1.2, 1.5$. 
    See Table~\ref{tab:symbols} and caption of Fig.~\ref{fig:Sph-shell-pulse} for definition of various symbols.
    }
    \label{fig:multi-pulse-pol-SMTH}
\end{figure*}

%%%%%%%%%%%%%%%%%%%%%%%%%%%%%%%%%%%%%%%%%%%%%%%%%%%%%%%%%%%%%%%%%%%%%%
\subsection{Polarization from a uniform jet with smooth edges}\label{sec:smooth-top-hat-jet-pol}
%%%%%%%%%%%%%%%%%%%%%%%%%%%%%%%%%%%%%%%%%%%%%%%%%%%%%%%%%%%%%%%%%%%%%%
The top-hat jet geometry is an idealization of the narrowly beamed relativistic outflow in GRBs. In reality, the jet can have \textit{smooth} edges both in 
comoving emissivity and bulk $\Gamma$. Such flows are referred to as structured jets. Here, for simplicity, we only consider a smooth top-hat jet 
for which the comoving emissivity is uniform in the core and drops off either exponentially with some smoothing factor $\Delta$ for polar angles $\theta>\theta_j\Leftrightarrow\xi_0>\xi_{0,j}$, such that
\begin{equation}\label{eq:Smooth-THJ-Exponential}
    f_{\rm exp}(\theta) =
    \begin{cases}
        1, & \xi_0 \leq \xi_{0,j} \\ 
        \exp\left(\displaystyle\frac{\sqrt{\xi_{0,j}}-\sqrt{\xi_0}}{\Delta}\right), & \xi_0 > \xi_{0,j}\,,
    \end{cases}
\end{equation}
or as a power law with smoothing factor $\delta$ with
\begin{equation}\label{eq:Smooth-THJ-PowerLaw}
    f_{\rm pl}(\theta) =
    \begin{cases}
        1, & \xi_0 \leq \xi_{0,j} \\
        \displaystyle\fracb{\xi_0}{\xi_{0,j}}^{-\delta/2}, & \xi_0 > \xi_{0,j}\,,
    \end{cases}
\end{equation}
while the bulk $\Gamma$ is uniform everywhere. 

In a flow where the bulk-$\Gamma$ also depends on the polar angle $\theta$, the temporal evolution of polarization will be significantly affected. 
Similar to the comoving emissivity, a typical and physically motivated assumption is that $\Gamma$ also 
decreases away from the jet-symmetry axis with $\theta$. 
In that case, the angular size of the beaming cone, $\sim\Gamma^{-1}$, must grow with $\theta$, which means that observers with $q>1$ that were significantly 
off-beam should now be able to see brighter regions at $\theta<\theta_{\rm obs}$. Not only that, the EATS is strongly modified due to the 
angular structure of $\Gamma$ and as a result the two-dimensional cross-sections of the EATS shown in Fig.~\ref{fig:EATS} for a uniform jet don't hold. 
Bulk-$\Gamma$ angular profiles that drop off steeply with $\theta$ yield larger polarization than shallower ones \citep[see, e.g.,][]{Gill+20}. 
Further discussion of these complexities are out of the scope of the present work and will be presented in a future work. Pulse-integrated polarization 
curves as a function of $q$ and for both smooth top-hat jets as well as structured jets are given in \citet{Gill+20}.

In Fig.~\ref{fig:pulse-pol-SMTH-diff-Delta} we show the pulse profile (top-left panel) and polarization curves for $q=1.2$ and 
for different levels of smoothing with $\Delta > 0$ ($\delta={\rm few}\times10$) for a smooth top-hat jet with exponetial (power-law) wings; 
$\Delta = 0$ and $\delta\gg1$ yields a top-hat jet with sharp edges. As the jet is made smoother the amount of flux 
along the observer's LOS gradually increases since now there is non-vanishing emissivity at $\theta_{\rm obs}>\theta_j$ for sufficiently smoother profiles, 
which means that the observer can receive on-beam emission. As a result, when $\Delta$ ($\delta$) increases (decreases) the arrival time of first photons 
emitted at $R=R_0$ becomes dominated by the radial time delay while the angular delay becomes smaller. Correspondingly, the pulse now shows non-vanishing 
polarization at $\tilde t<\tilde t_{0-}$. In general, the shape of the polarization curve is similar to that obtained for a top-hat geometry, but now only 
smoother and/or broader. In the case of a smooth power-law wing, the pulse profile in some cases shows two peaks/bumps where the second bump at 
$\tilde t>\tilde t_f$ becomes increasingly dominant for larger values of $\delta$ (therefore sharper edged jets).

In Fig.~\ref{fig:pulse-pol-SMTH-diff-q} we again show the pulse profiles and polarization curves, now for a fixed $\Delta=0.1$ and different values of $q$. 
This figure should be compared with Fig.~\ref{fig:THJ-Bperp-pulse-pol-diff-q-diff-x0} with which it shares many features. The pulse profiles still show 
a deficit in flux after the nearer edge first becomes visible, but the off-beam emission pulses now show a shallower rise to the pulse peak due to the structure 
of the outflow. In addition, the peak of the curve corresponding to $q=1$ normalized by that obtained for $q=0$ is now larger than half, again due to the 
angular structure of the jet that contributes additional flux. The polarization curves with $q<1$ show a similar trend as before for all the field configurations, 
but the $q=1$ case now shows a $90^\circ$ change in PA for both the $B_\perp$ and $B_\parallel$ fields. In contrast with the top-hat jet case, where the 
polarization abruptly vanishes due to vanishing flux, 
the polarization curves for a smooth top-hat jet and for the $B_\parallel$ and $B_{\rm tor}$ fields now show a plateau at the maximum level after the whole 
uniform core comes into view, i.e. at $\tilde t>\tilde t_{f+}$ (but the flux is vanishingly small so deep into the tail of the pulse making it extremely 
difficult to measure such a polarization plateau).

%%%%%%%%%%%%%%%%%%%%%%%%%%%%%%%%%%%%%%%%%%%%%%%%%%%%%%%%%%%%%%%%%%%%%%%%%%
\section{Linear Polarization Over Multiple Pulses}\label{sec:multi-pulse}
%%%%%%%%%%%%%%%%%%%%%%%%%%%%%%%%%%%%%%%%%%%%%%%%%%%%%%%%%%%%%%%%%%%%%%%%%%
GRB observations are typically photon starved unless the source is particularly bright or relatively nearby. 
This presents a challenge for polarization measurements that require high photon counts to be able to yield 
statistically significant results. To increase the number of photons integration over multiple pulses is 
generally performed, which alters the polarization evolution from what is expected for a single pulse. Here 
we consider multiple pulses that originate due to episodic internal dissipation in the relativistic flow as 
it expands. In the case of internal shocks, this can occur due to the intermittent ejection of multiple shells 
by the central engine that collide at different radii. On the other hand, a Poynting flux dominated flow can 
suffer multiple dissipation episodes due to magnetic reconnection of a stochastic magnetic field and/or MHD 
instabilities.

To describe the pulse structure and linear polarization from multiple pulses, the same formalism as described 
earlier for a single pulse can be used. Here we follow the formalism introduced in \citet{Genet-Granot-09} to 
describe multiple pulses. In a KED flow, the pulses are now separated temporally due to ejection 
of multiple shells that have different ejection times $t_{{\rm ej},z} = t_{\rm ej}/(1+z)$, such that the onset of 
the $i^{\rm th}$ pulse is given by $t_{{\rm onset},z} = t_{{\rm ej},z,i}+t_{0,z,i}$. Variation in emission radii 
and bulk $\Gamma$ for the different shells can change $t_{0,z,i} = R_{0,i}/2(1+m)\Gamma_{0,i}^2c$ as well as 
having different $(\Delta R)_i/R_{0,i}$, resulting in varied $t_{f,z,i} = t_{0,z,i}[1+(\Delta R)_i/R_{0,i}]$, 
can lead to a variety of pulse structures. For simplicity, here we fix $R_{0,i}=R_0$, $(\Delta R)_i=\Delta R$, and 
$\theta_j$ for all the pulses, where it's understood that $\theta_{\rm obs}$ cannot change between the different pulses. 
We allow $\Gamma$ to vary, such that $\Gamma_{0,i}=\lambda_i\Gamma_0$ where we define the bulk $\Gamma$ of 
the first pulse $\Gamma_{0,1}=\Gamma_0$, which yields $\tilde t_{0,i} \equiv t_{0,z,i}/t_{0,z,1} = t_{0,z,i}/t_0 = \lambda_i^{-2}$. 
Finally, the onset and peak times of the $i^{\rm th}$ pulse normalized by the onset time of the first pulse, 
$t_{0,z,1}=t_0$, is 
\begin{align}
    & \tilde t_{{\rm onset},i} = \tilde t_{{\rm ej},i} + \lambda_i^{-2} \\
    & \tilde t_{{\rm peak},i} = \tilde t_{{\rm ej},i} + \lambda_i^{-2}(1+\Delta R/R_0)\,.
\end{align}
We note that the peak times of the subsequent pulses after the first one can be different from $\tilde t_{{\rm peak},i}$ due 
to overlapping flux of neighbouring pulses.

In Fig.~\ref{fig:multi-pulse-pol-SMTH}, we show the pulse profiles (left-panel) for multiple pulses and the corresponding 
polarization (right-panel) for a KED smooth top-hat jet with three different magnetic field configurations. The single 
pulse case is also shown for comparison and to isolate the effect of having multiple pulses on polarization. In the top row 
we consider pulses with the same fixed bulk $\Gamma$, i.e. $\lambda_i=1$, for which the different timescales are: 
$\tilde t_{{\rm ej},i}=0,~1.5,~3,~5.5$, and $\tilde t_{{\rm onset},i} = \tilde t_{{\rm ej},i} + 1 = 1,~2.5,~4,~6.5$. In the bottom 
row different pulses have different $\Gamma_{0,i}$, with $\lambda_i = 1,~0.8,~1.2,~1.5$ and $\tilde t_{{\rm ej},i}=0,~0.94,~3.3,~6.1$, 
which yields the same onset time as for the $\lambda_i=1$ case. 
For incoherent radiation the Stokes parameters are additive, and therefore the polarization at any given time is obtained from 
\begin{equation}
    \Pi(\tilde t) = \frac{\sum_i Q_i(\tilde t)}{\sum_i I_i(\tilde t)}\,,
\end{equation}
where $I_i$ and $Q_i$ are the Stokes parameters for the individual pulses. Here again, due to symmetry, $U_i(\tilde t)=0$. 
The polarization evolution also shows multiple 
spikes, mirroring the behavior of the lightcurve. This occurs due to the fact that the increase in total intensity $I(\tilde t)$ 
is not matched equally with the increase in the polarized intensity $Q(\tilde t)$ as contribution from other pulses are 
added. As a result, a sharp drop in polarization occurs at the time of the onset of the next pulse. The late-time level of 
polarization, however, coincides with that obtained for a single pulse when $\lambda_i=1$. Here we have only explored a scenario where all the 
pulses have the same properties except variation in bulk $\Gamma$. A variety of pulse shapes \citep[see, e.g.,][]{Genet-Granot-09}, 
and their corresponding polarization curves, due to variations in peak intensity, duration, as well as the spectrum can be 
produced using the given formalism in a straightforward manner. In order to observe these rapid changes in polarization a very 
bright GRB as well as a sensitive gamma-ray polarimeter that can obtain such time-resolved observations is needed. The latter 
requirement is still not feasible and therefore when integration is carried out over multiple pulses it becomes critical that 
measurements are compared with theoretical models that account for multiple pulses.

\begin{figure}
    \centering
    \includegraphics[width=0.48\textwidth]{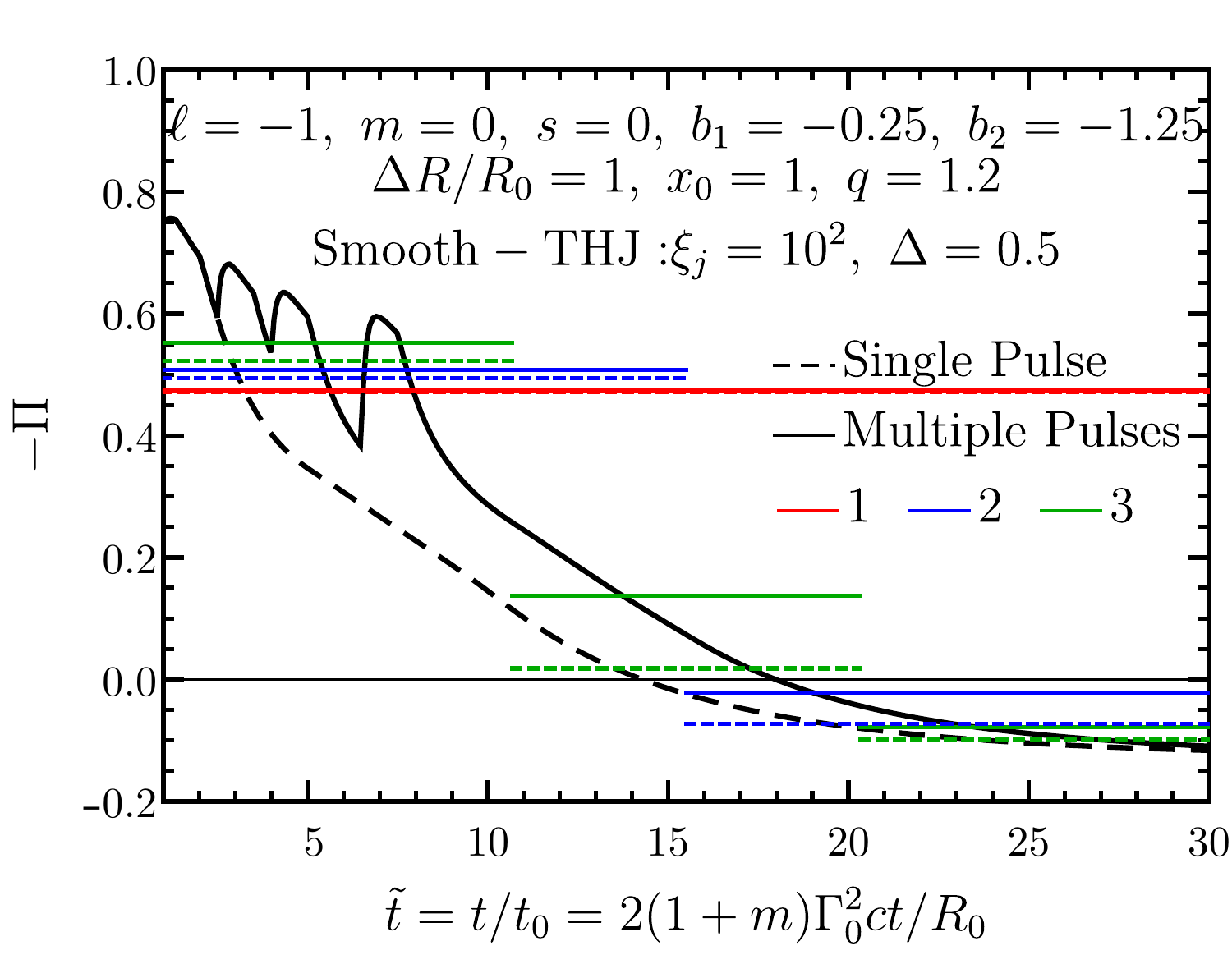}
    \caption{Time-integrated polarization for a KED smooth top-hat jet (uniform core 
    and exponentially decaying wings in $L_{\nu'}'$) with $B_{\rm tor}$ field and with bulk-$\Gamma$ the same for different pulses. 
    Temporal segments over which polarization is obtained are calculated by dividing the pulse into one (red), two (blue), or three (green) part(s). 
    The assumed parameters are the same as in the top row of Fig.~\ref{fig:multi-pulse-pol-SMTH}. See Table~\ref{tab:symbols} and caption of Fig.~\ref{fig:Sph-shell-pulse} for definition of various symbols.}
    \label{fig:time-integrated-pol}
\end{figure}

When time integration is performed, whether over a single bright pulse or over an emission episode comprising of multiple pulses, it is equally critical to compare 
the measurements with time-integrated model predictions. Time-integrated polarization over a temporal segment $\tilde t_1<\tilde t<\tilde t_2$ is obtained from
\begin{equation}
    \Pi(\tilde t_1,\tilde t_2) = \frac{\int_{\tilde t_1}^{\tilde t_2}Q(\tilde t)d\tilde t}{\int_{\tilde t_1}^{\tilde t_2}I(\tilde t)d\tilde t}\,.
\end{equation}
In Fig.~\ref{fig:time-integrated-pol}, we show the time-integrated polarization for a smooth top-hat jet with $B_{\rm tor}$ field. 
Two cases are highlighted, one comprising of a single pulse and the other with multiple pulses. In both, the time-integrated 
polarization is significantly different, as should be expected, from the instantaneous polarization within the temporal segment. The $90^\circ$ change in PA can 
still be obtained in this particular case, however, with an obvious side-effect that time-integration washes out the information on the exact time when the 
PA changes relative to the arrival time of the first photons.

%%%%%%%%%%%%%%%%%%%%%%%%%%%%%%%%%%%%%%%
\section{Summary \& Discussion}\label{sec:Summary-Discussion}
%%%%%%%%%%%%%%%%%%%%%%%%%%%%%%%%%%%%%%%
The temporal dependence of linear polarization in GRB prompt emission is an important tool that can elucidate the composition and structure of ultrarelativistic 
flows that power GRBs and, when synchrotron emission is dominant, help us understand the magnetic field configuration in the emission 
region. In this work, we present a comprehensive treatment of time-resolved polarization of synchrotron emission from different magnetic field configurations, 
jet structures, and outflow compositions with different dynamics. 

In particular, we consider a kinetic-energy-dominated (KED) flow, in which internal shocks efficiently dissipate the kinetic energy of baryons 
when the flow is coasting, and a Poynting-flux-dominated (PFD) flow, where magnetic energy is dissipated due to magnetic reconnection or MHD 
instabilities which also accelerates the flow. The pulse profiles and polarization curves are calculated for a radially expanding ultrarelativistic (with 
bulk $\Gamma\gg1$) thin outflow that continuously radiates over radii $R_0\leq R\leq R_0+\Delta R$. Both of these are broadly similar for the two outflows. 
The main difference between the two cases can be attributed to the fact that the PFD flow is accelerating, with $\Gamma\propto R^{1/3}$, which causes all critical 
timescales, including the width of the pulses, to become shorter in comparison to a KED flow.

In both types of outflows power-law electrons gyrating in either shock-generated small-scale or globally ordered large-scale magnetic fields produce synchrotron 
emission. In a KED flow, the magnetic field configuration, in general, can resemble a tangled field that lies entirely in the plane transverse to the radial 
direction ($B_\perp$) or an ordered field aligned with the radial direction ($B_\parallel$) at every point of the outflow. 
We note that although $B_\perp$ is considered in many works as it's physically motivated, achieving such an anisotropic field configuration in reality 
may be challenging \citep[see, e.g, the discussion in][for shock-generated field structure in relativistic collisionless shocks]{Gill-Granot-20}.  
Moreover, $B_\parallel$ is likely even harder to achieve physically, its main appeal for a shock produced field being that it is (trivially) symmetric 
w.r.t the local shock normal. For a PFD flow, the most relevant field 
configuration is that of an ordered toroidal field ($B_{\rm tor}$), although other global field configurations are possible, and significant magnetic 
reconnection can create a substantial field component that has random orientations on small angular scales. Alternatively, it's possible that the emission 
region is pervaded by a locally ordered field 
whose coherence length is similar to or larger than the angular size of the beaming cone, such that $\theta_B\gtrsim1/\Gamma$. For observers with 
viewing angle $\theta_{\rm obs}=0$ only the locally ordered field would yield non-vanishing polarization. All the other field configurations considered here, 
which are all axisymmetric, require the observer to have $\theta_{\rm obs}>0$ and for the $B_\perp$ and $B_\parallel$ fields the flow to be 
inhomogeneous in some way to break the symmetry and yield net non-zero polarization.

The symmetry is broken by means of the uniform flow either having a sharp edge (at $\theta=\theta_j$), as in a narrowly beamed top-hat jet, or a smooth edge, as 
in a smooth top-hat jet with smoothly decaying emissivity wings outside of the uniform narrow core, and the observer having $q\equiv\theta_{\rm obs}/\theta_j > 0$. Magnetic field configurations in which the polarization vectors are symmetric around the observer's LOS, e.g. $B_\perp$ 
and $B_\parallel$, the polarization only begins to grow when the observer first `sees' the edge of the jet nearer to their LOS. The time at which this happens, 
$\tilde t=\tilde t_{0-}$ (see Eq.~\ref{eq:critical-times}), is exactly the same when the received emission is on-beam ($q < 1+\xi_{0,j}^{-1/2}$) or off-beam 
($q > 1+\xi_{0,j}^{-1/2}$) from a narrowly beamed top-hat jet with $\xi_{0,j}\equiv(\Gamma_0\theta_j)^2\gg1$. In the case of off-beam emission this time also 
marks the arrival time of the first photons, however, since the flux is dominated by high latitude emission, the GRB becomes too dim making it harder to reliably 
detect any polarization.

In a top-hat jet, the three magnetic field configurations, namely $B_\perp$, $B_\parallel$, and $B_{\rm tor}$, show distinct polarization evolution over a single 
pulse with some common features. For an on-beam observer both $B_\perp$ and $B_\parallel$ always show a $90^\circ$ change in PA, 
but the $B_{\rm tor}$ field case either shows two such changes over the pulse duration or none depending on $\xi_j$, $\Delta R/R_0$, and $q$.
When the received emission is off-beam, only the $B_{\rm tor}$ shows $90^\circ$ PA changes whereas both $B_\perp$ and $B_\parallel$ fields show 
a steady PA.    

In a smooth top-hat jet, the general features of the polarization curves as obtained for the top-hat jet still remain. As the jet becomes more smooth in 
emissivity around the edges, by developing e.g. exponentially/power-law decaying wings, the observer starts receiving on-beam emission now that there's some material 
that emits along the LOS. Therefore, significantly high levels of polarization are obtained for $q>1+\xi_{0,j}^{-1/2}$ in a smooth top-hat jet at 
$\tilde t<\tilde t_{0-}$ when there was none in the case of a top-hat jet. 

Due to the paucity of $\gamma$-ray photons, time-integration over smaller (compared to pulse duration) temporal segments of a bright single pulse or an 
emission episode comprising of multiple pulses is routinely performed to gain high signal-to-noise measurements. We show how multiple pulses can be modeled 
using the same single pulse framework and how their polarization can be obtained. In the case of a single bright pulse, time-integrated polarization over 
several temporal segments yields a coarse trend that can only be interpreted accurately by modeling the time-resolved polarization for a given field 
configuration and outflow dynamic and structure. The same is true for multiple pulses.

Below we collect the main points of this work:
\begin{itemize}
    \item In a top-hat jet, when $q=\theta_{\rm obs}/\theta_j<1$, an ordered B-field, e.g. $B_{\rm tor}$, always yields the highest polarization ($\Pi\lesssim75\%$) over most of the pulse duration starting from the arrival time of the first photon. 
    \item For the $B_\perp$ and $B_\parallel$ fields $\Pi$\,$=$\,$0$ until the observer sees the nearest edge of the jet at $\tilde 
    t = \tilde t_{0-}$. Then the polarization grows to $\Pi\lesssim15\%$ before the low flux level in the pulse tail makes it challenging to 
    measure even higher $\Pi$ values, especially for $B_\parallel$ where $\Pi$ is highest deep in the pulse tail.
    \item All three B-field cases show a $90^\circ$ change in PA, sometimes even twice over the pulse duration for the $B_{\rm tor}$ case. The PA can only change by $90^\circ$ if the global flow and magnetic field configuration is axisymmetric.
    %Changes in the PA other than $90^\circ$ cannot occur if the flow is axisymmetric.
    Furthermore, the PA only changes in the pulse tail for all three field 
    configurations which makes it difficult to measure.
    \item For $q>1$ high $\Pi$ can be measured for the three field configurations right from the start of the pulse. However, 
    the fluence sharply drops with $q$ for $q>1$, making such measurements challenging. %in practice.
    \item Pulse-integrated polarization is significantly different from time-resolved polarization, both when the emission consists of a truly single pulse or multiple overlapping pulses. Therefore, time-resolved measurements should only be compared with time-resolved theoretical models.
\end{itemize}

%%%%%%%%%%%%%%%%%%%%%%%%%%%%%%%%%%%%%%%%%%%%%%
\subsection{Time-Varying Polarization Angle}\label{sec:Time-Dependent-PA}
%%%%%%%%%%%%%%%%%%%%%%%%%%%%%%%%%%%%%%%%%%%%%%
Most measurements of prompt GRB polarization report a fixed PA \citep[see, e.g., Table 1 in][and references therein]{Gill+20}. In many cases, this may 
simply be a result of integrating over the entire pulse, which naturally gives a single PA, and not carrying out a time-resolved analysis due to low photon 
counts. Where the latter has been possible, albeit only for a few GRBs, a change in PA was noted. A change in PA was reported for \textit{IKAROS}-GAP detected 
GRB 100826A \citep{Yonetoku+11} where the prompt emission was divided into two $\sim50\,$s intervals, each containing multiple spikes, with the two intervals 
having polarization $\Pi_1=25\pm15$ per cent and $\Pi_2=31\pm21$ per cent and PAs $\phi_1=159^\circ\pm18^\circ$ and $\phi_2=75^\circ\pm20^\circ$. Time-resolved 
analysis of a single pulse GRB 170114A detected by POLAR showed a large change in PA between two $2\,$s intervals with $\phi_1=122^\circ$ and $\phi_2=17^\circ$ 
\citep{Zhang+19}. A more refined time-resolved analysis \citep{Burgess+19} that simultaneously fitted both the spectrum and polarization for the same GRB showed 
a gradual change in PA over the pulse while the polarization showed an increase towards the peak of the pulse reaching levels of $\Pi\sim30$ per cent at the peak. 
\citet{Chand+19} reported an energy dependent gradually changing polarization and PA for GRB 171010A but the statistical significance of the polairzation was 
low ($\lesssim2.5\sigma$). A varying degree of polarization and PA was also reported in \citet{Sharma+19} for GRB 160821A that was observed by multiple 
instruments, namely \textit{Fermi}-GBM/LAT, \textit{AstroSat}-CZTI, and \textit{Swift}-BAT. The polarization was obtained in three time intervals and it remained 
high with $\Pi_1=71_{-41}^{+29}$ per cent, $\Pi_2=58_{-30}^{+29}$ per cent, and $\Pi_3=61_{-46}^{+39}$ per cent with detection significance for all intervals 
between $\sim3\sigma$ and $\sim4\sigma$. The PA showed a remarkable evolution where it changed by $\Delta\phi_{1,2}=81^\circ\pm13^\circ$ and 
$\Delta\phi_{2,3}=80^\circ\pm19^\circ$.

For the magnetic field configurations considered in this work, the polarization angle can only change exactly by $\Delta\phi=90^\circ$ and a gradual change of 
the PA is not possible. There are tantalizing hints of a $90^\circ$ change in the PA in some of the GRBs, as discussed above, but the results are not yet conclusive. 
The result presented by \citet{Sharma+19} where the PA changes by $90^\circ$ twice over the emission is again very exciting as such a change over 
a single pulse can only occur for the $B_{\rm tor}$ field configuration. The only difficulty, according to the modeling done here, is that both $90^\circ$ changes 
occur in the decaying tail of the pulse when high latitude emission dominates the flux. In the measurement presented by \citet{Sharma+19} the PA shows a change 
close to the peak of the emission. Another scenario in which a $90^\circ$ PA change can be obtained includes contribution from multiple pulses and when the LOS 
is close to the edge of the jet, such that $\theta_{\rm obs}\approx\theta_j$, along with a change in bulk $\Gamma$ between the pulses which would change 
$\xi_j=(\Gamma\theta_j)^2$. Alternatively, such a change in the PA can be obtained due to magnetic reconnection, e.g. in the ICMART model \citep{Zhang-Yan-11}, 
where the local magnetic field orientation, which is orthogonal to the wave vector of the emitted photon, itself changes by $90^\circ$ 
as the field lines are destroyed and reconnected in the emission region \citep{Deng+16}. To obtain a change in the PA other than 
$\Delta\phi=90^\circ$ or to get a gradually changing PA the condition for axisymmetry must be relaxed and the magnetic field configuration 
or orientation in the emission region must change. One possibility is that if the different pulses that contribute to the emission arise 
in a `mini-jet' within the outflow 
\citep[e.g.,][]{Shaviv-Dar-95,Lyutikov-Blandford-03,Kumar-Narayan-09,Lazar+09,Narayan-Kumar-09,Zhang-Yan-11}. In this case the different directions of the mini-jets 
or bright patches w.r.t. the LOS \citep[e.g.][]{Granot-Konigl-03,Nakar-Oren-04} would cause the PA to also be different between the pulses even for a 
field that is locally symmetric w.r.t the local radial direction (e.g. $B_\perp$ or $B_\parallel$) as well as for fields that are axisymmetric w.r.t to 
the center of each mini-jet (e.g. a local $B_{\rm tor}$ for each mini-jet). Finally, broadly similar result would follow from an ordered field within 
each mini-jet ($B_{\rm ord}$) which are incoherent between different mini-jets. Time-resolved measurment in such a case would naturally yield 
a time-varying PA. Alternatively, as shown by \citet{Granot-Konigl-03} for GRB afterglow polarization, a combination of an ordered field component 
(e.g. $B_{\rm ord}$) and a random field, like $B_\perp$, can give rise to a time-varying PA between different pulses that, e.g., arise from internal 
shocks. The ordered field component here would be that advected from the central engine and the random field component can be argued to be shock-generated. 
Notice that the ordered field component should not be axisymmetric in order for the position angle to smoothly vary.

%%%%%%%%%%%%%%%%%%%%%%%%%%%%%%%%%%%%%%%%%%%%%%%
\subsection{Energy-Dependent Polarization}\label{sec:Energy-Dependent-Pol}
%%%%%%%%%%%%%%%%%%%%%%%%%%%%%%%%%%%%%%%%%%%%%%%
As we showed in this work, using a thin spherical shell and band-function spectrum that evolves with radius, polarization at any given time is energy dependent. 
In the case of pure synchrotron emission the spectrum would move across the instrument's energy window towards lower energies due to spectral softening caused 
by the radial evolution of the spectral peak energy. This softening can be inferred from Eq.~(\ref{eq:x-t-on-axis}) 
where setting $x=1$ would yield the temporal evolution of the spectral peak, such that $x_0\equiv\nu/\nu_0=\tilde t^{(2d-m)/2(1+m)}$, which for $m=0$ yields 
$x_0=\tilde t^{d}$ where $d=-1$ for a KED flow. Consequently, the level of polarization at a fixed observed energy would change not only due to its temporal 
evolution, as in an infinite power-law spectrum, but also due to the drift of the spectral peak since the synchrotron polarization depends on the local spectral 
index.

Energy dependent polarization is a powerful tool particularly when multiple spectral components are present. These can arise, e.g., in photospheric emission 
models where the peak of the spectrum may be dominated by a quasi-thermal component while a non-thermal spectrum develops both below and above the spectral peak 
energy \citep[see, e.g.,][and references therein]{Beloborodov-Meszaros-17}. Detailed numerical simulations of a PFD flow in \citet{Gill+20b} showed that 
depending on how particles are accelerated/heated, e.g, into a power-law energy distribution or where they form a monoenergetic distribution due to distributed 
heating and cooling, the non-thermal component of the spectrum arises either due to synchrotron emission or Comptonization, respectively. In such a scenario, 
energy dependent polarization can be used to discriminate between the two emission mechanisms. The spectral peak is expected to be unpolarized as the photons 
there suffer multiple Compton scatterings causing the polarization to average out nearly to zero. In the case of synchrotron emission, significant polarization 
is expected away from the spectral peak energy \citep{Lundman+18}. However, if Comptonization dominates the non-thermal component, again negligible polarization 
is expected. Detailed exploration of energy dependent polarization is deferred to future work (Gill \& Granot, 2021, in prep.).

Upcoming missions, e.g, POLAR-II \citep{Kole-19}, LEAP \citep{McConnell+16}, and eXTP \citep{eXTP}, with the ability to carry out simultaneous 
spectro-polarimetric measurements will be able to provide much needed insights 
into the composition of ultrarelativistic jets and radiation processes that power GRB prompt emission.

%\vspace{0.10cm}
\section*{Acknowledgements}
We thank the anonymous referee for useful comments. 
This research was supported by the ISF-NSFC joint research program (grant No. 3296/19).

\section*{Data Availability}
The data underlying this article will be shared on reasonable request to the corresponding author.

%%%%%%%%%%%%%%%%%%%%%%%%%%%%%%%%%%%%%%%%%%%
\bibliographystyle{mnras}
\bibliography{refs.bib}
%%%%%%%%%%%%%%%%%%%%%%%%%%%%%%%%%%%%%%%%%%%
% Don't change these lines
\bsp	% typesetting comment
\label{lastpage}
\end{document}